\documentclass[twocolumn,times]{aastex63}
\usepackage{xspace}
\usepackage{amsmath}
\usepackage{graphicx}
\usepackage{bm}
\usepackage{tikz}
\usetikzlibrary{shapes}
\usetikzlibrary{fit}
\usetikzlibrary{chains}
\usetikzlibrary{arrows}
\let\tablenum\relax
\usepackage{siunitx}
\usepackage{tikz}

\tikzstyle{latent} = [circle, fill=white, draw=black, inner sep=1pt,
minimum size=25pt, font=\fontsize{10}{10}\selectfont, node distance=1]
\tikzstyle{obs} = [latent,fill=blue!10]
\tikzstyle{const} = [rectangle, inner sep=0pt, node distance=1]
\tikzstyle{factor} = [rectangle, fill=black, minimum size=5pt, inner
sep=0pt, node distance=0.4]
\tikzstyle{det} = [latent, diamond]
\tikzstyle{input} = [rectangle, fill=blue!10, draw=black, inner sep=1pt,
minimum size=25pt, font=\fontsize{10}{10}\selectfont, node distance=1]

\tikzstyle{plate} = [draw, rectangle, rounded corners, fit=#1]
\tikzstyle{wrap} = [inner sep=0pt, fit=#1]
\tikzstyle{gate} = [draw, rectangle, dashed, fit=#1]

\tikzstyle{caption} = [font=\normalsize, node distance=0] %
\tikzstyle{plate caption} = [caption, node distance=0, inner sep=0pt,
below left=-2pt and 3pt of #1.south east] %
\tikzstyle{factor caption} = [caption] %
\tikzstyle{every label} += [caption] %

\newcommand{\factoredge}[4][]{ %
  \foreach \f in {#3} { %
    \foreach \x in {#2} { %
      \path (\x) edge[-,#1] (\f) ; %
    } ;
    \foreach \y in {#4} { %
      \path (\f) edge[->, >={latex}, #1] (\y) ; %
    } ;
  } ;
}

\newcommand{\edge}[3][]{ %
  \foreach \x in {#2} { %
    \foreach \y in {#3} { %
      \path (\x) edge [->, >={latex}, #1] (\y) ;%
    } ;
  } ;
}

\newcommand{\factor}[5][]{ %
  \node[factor, label={[name=#2-caption]#3}, name=#2, #1,
  alias=#2-alias] {} ; %
  \factoredge {#4} {#2-alias} {#5} ; %
}

\newcommand{\plate}[4][]{ %
  \node[wrap=#3] (#2-wrap) {}; %
  \node[plate caption=#2-wrap] (#2-caption) {#4}; %
  \node[plate=(#2-wrap)(#2-caption), #1] (#2) {}; %
}

\newcommand{\TESS}{\emph{TESS}\xspace}
\makeatletter
\newcommand\notsotiny{\@setfontsize\notsotiny\@vipt\@viipt}
\newcommand{\changes}[1]{\textrm{#1}}
\makeatother

\usepackage{natbib}
\defcitealias{brah19}{Br19}
\defcitealias{brah20}{Br20}
\defcitealias{niel19}{Ni19}
\defcitealias{rodr19}{Ro19}
\defcitealias{newt19}{Ne19}
\defcitealias{daws19}{Da19}
\defcitealias{kipp19}{Ki19}
\defcitealias{addi20}{Ad21}
\defcitealias{huan20}{Hu20}
\defcitealias{jord20}{Jo20}
\defcitealias{lend14}{Le14}
\defcitealias{quel10}{Qu10}
\defcitealias{hell19}{He19}
\defcitealias{hell17}{He17}
\defcitealias{smit14}{Sm14}
\defcitealias{carm20}{Ca20}
\defcitealias{mont20}{Mo20}
\defcitealias{mire20}{Mi20}
\defcitealias{carm20a}{Ca20a}
\defcitealias{carm20b}{Ca20b}
\defcitealias{boum20}{Bo20}
\defcitealias{rodr21}{Ro21}
\defcitealias{hobs21}{Ho21}
\defcitealias{daws21}{Da21}

\received{January 25, 2021}
\revised{April 2, 2021}
\accepted{}
\submitjournal{AAS Journals}

\shorttitle{Warm Jupiters in Year 1 \TESS FFIs}
\shortauthors{Dong, Huang, \& Dawson et al.}
\graphicspath{{./}{figures/}}

\begin{document}

\title{Warm Jupiters in \TESS Full-Frame Images: A Catalog and Observed Eccentricity Distribution for Year 1}

\correspondingauthor{Jiayin Dong}
\email{jdong@psu.edu}

\author[0000-0002-3610-6953]{Jiayin Dong} 
\affiliation{Department of Astronomy \& Astrophysics, The Pennsylvania State University, University Park, PA 16802, USA}
\affiliation{Center for Exoplanets \& Habitable Worlds, 525 Davey Laboratory, The Pennsylvania State University, University Park, PA 16802, USA}
\affiliation{Center for Computational Astrophysics, Flatiron Institute, 162 Fifth Avenue, New York, NY 10010, USA}

\author[0000-0003-0918-7484]{Chelsea X. Huang} 
\altaffiliation{Juan Carlos Torres Fellow}
\affiliation{Department of Physics and Kavli Institute for Astrophysics and Space Research, Massachusetts Institute of Technology, Cambridge, MA 02139, USA}

\author[0000-0001-9677-1296]{Rebekah I. Dawson} 
\affiliation{Department of Astronomy \& Astrophysics, The Pennsylvania State University, University Park, PA 16802, USA}
\affiliation{Center for Exoplanets \& Habitable Worlds, 525 Davey Laboratory, The Pennsylvania State University, University Park, PA 16802, USA}

\author[0000-0002-9328-5652]{Daniel Foreman-Mackey} 
\affiliation{Center for Computational Astrophysics, Flatiron Institute, 162 Fifth Avenue, New York, NY 10010, USA}

\author[0000-0001-6588-9574]{Karen A.\ Collins} 
\affiliation{Center for Astrophysics \textbar \ Harvard \& Smithsonian, 60 Garden Street, Cambridge, MA 02138, USA}

\author[0000-0002-8964-8377]{Samuel N. Quinn} 
\affiliation{Center for Astrophysics \textbar \ Harvard \& Smithsonian, 60 Garden Street, Cambridge, MA 02138, USA}

\author[0000-0001-6513-1659]{Jack J. Lissauer} 
\affiliation{Space Science \& Astrobiology Division, MS 245-3, NASA Ames Research Center, Moffett Field, CA 94035, USA}

\author[0000-0002-9539-4203]{Thomas Beatty} 
\affiliation{Department of Astronomy and Steward Observatory, University of Arizona, Tucson, AZ 85721, USA}

\author[0000-0002-9644-8330]{Billy Quarles} 
\affiliation{Center for Relativistic Astrophysics, School of Physics, Georgia Institute of Technology, Atlanta, GA 30332 USA}

\author[0000-0001-5401-8079]{Lizhou Sha} 
\affiliation{Department of Astronomy, University of Wisconsin--Madison, Madison, WI 53706, USA}

\author[0000-0002-1836-3120]{Avi Shporer} 
\affiliation{Department of Physics and Kavli Institute for Astrophysics and Space Research, Massachusetts Institute of Technology, Cambridge, MA 02139, USA}

\author{Zhao Guo} 
\affiliation{Department of Applied Mathematics and Theoretical Physics (DAMTP), University of Cambridge, UK}

\author[0000-0002-7084-0529]{Stephen R. Kane} 
\affiliation{Department of Earth and Planetary Sciences, University of California, Riverside, CA 92521, USA}

\author[0000-0002-0856-4527]{Lyu Abe} 
\affiliation{Universit\'e C\^ote d'Azur, Observatoire de la C\^ote d'Azur, CNRS, Laboratoire Lagrange, Bd de l'Observatoire, CS 34229, 06304 Nice cedex 4, France}

\author[0000-0003-1464-9276]{Khalid Barkaoui} 
\affiliation{Astrobiology Research Unit, Université de Liège, 19C Allée du 6 Août, 4000 Liège, Belgium}
\affiliation{Oukaimeden Observatory, High Energy Physics and Astrophysics Laboratory, Cadi Ayyad University, Marrakech, Morocco}

\author[0000-0001-6285-9847]{Zouhair Benkhaldoun} 
\affiliation{Oukaimeden Observatory, High Energy Physics and Astrophysics Laboratory, Cadi Ayyad University, Marrakech, Morocco}

\author[0000-0002-9158-7315]{Rafael Brahm} 
\affiliation{Facultad de Ingenier\'ia y Ciencias, Universidad Adolfo Ib\'{a}\~{n}ez, Av. Diagonal las Torres 2640, Pe\~{n}alol\'{e}n, Santiago, Chile}
\affiliation{Millennium Institute for Astrophysics, Chile}

\author{Fran\c{c}ois Bouchy} 
\affiliation{Observatoire de l'Universit\'e de Gen\`eve, 51 chemin des Maillettes,1290 Versoix, Switzerland}

\author[0000-0001-6416-1274]{Theron W. Carmichael} 
\altaffiliation{National Science Foundation Graduate Research Fellow}
\affiliation{Department of Astronomy, Harvard University, Cambridge, MA 02138, USA}
\affiliation{Center for Astrophysics \textbar \ Harvard \& Smithsonian, 60 Garden St, Cambridge, MA 02138, USA}

\author[0000-0003-2781-3207]{Kevin I.\ Collins} 
\affiliation{George Mason University, 4400 University Drive, Fairfax, VA, 22030 USA}

\author[0000-0003-2239-0567]{Dennis M.\ Conti} 
\affiliation{American Association of Variable Star Observers, 49 Bay State Road, Cambridge, MA 02138, USA}

\author[0000-0001-7866-8738]{Nicolas Crouzet} 
\affiliation{European Space Agency (ESA), European Space Research and Technology Centre (ESTEC), Keplerlaan 1, 2201 AZ Noordwijk, The Netherlands}

\author[0000-0002-3937-630X]{Georgina Dransfield} 
\affiliation{School of Physics \& Astronomy, University of Birmingham, Edgbaston, Birmingham B15 2TT, United Kingdom}

\author[0000-0002-5674-2404]{Phil Evans} 
\affiliation{El Sauce Observatory, Coquimbo Province, Chile}

\author[0000-0002-4503-9705]{Tianjun Gan} 
\affiliation{Department of Astronomy and Tsinghua Centre for Astrophysics, Tsinghua University, Beijing 100084, China}

\author{Mourad Ghachoui} 
\affiliation{Oukaimeden Observatory, High Energy Physics and Astrophysics Laboratory, Cadi Ayyad University, Marrakech, Morocco}

\author[0000-0003-1462-7739]{Micha\"el Gillon} 
\affiliation{Astrobiology Research Unit, Universit\'e de Li\`ege, 19C All\`ee du 6 Ao\^ut, 4000 Li\`ege, Belgium}

\author[0000-0001-8105-0373]{Nolan Grieves} 
\affiliation{Observatoire de Gen\`eve, Universit\'e de Gen\`eve, Chemin Pegasi 51b, 1290 Sauverny, Switzerland}

\author[0000-0002-7188-8428]{Tristan Guillot} 
\affiliation{Universit\'e C\^ote d'Azur, Observatoire de la C\^ote d'Azur, CNRS, Laboratoire Lagrange, Bd de l'Observatoire, CS 34229, 06304 Nice cedex 4, France}

\author{Coel Hellier} 
\affiliation{Astrophysics Group, Keele University, Staffordshire ST5 5BG, U.K.} 

\author{Emmanu\"el Jehin} 
\affiliation{Space Sciences, Technologies and Astrophysics Research (STAR) Institute, Universit\'e de Li\`ege, 19C All\`ee du 6 Ao\^ut, 4000 Li\`ege, Belgium}

\author[0000-0002-4625-7333]{Eric L.\ N.\ Jensen} 
\affiliation{Department of Physics \& Astronomy, Swarthmore College, Swarthmore PA 19081, USA}

\author[0000-0002-5389-3944]{Andres Jord\'an} 
\affiliation{Facultad de Ingenier\'ia y Ciencias, Universidad Adolfo Ib\'{a}\~{n}ez, Av. Diagonal las Torres 2640, Pe\~{n}alol\'{e}n, Santiago, Chile}
\affiliation{Millennium Institute for Astrophysics, Chile}

\author{Jacob Kamler} 
\affiliation{John F. Kennedy High School, 3000 Bellmore Avenue, Bellmore, NY 11710, USA}

\author[0000-0003-0497-2651]{John F.\ Kielkopf} 
\affiliation{Department of Physics and Astronomy, University of Louisville, Louisville, KY 40292, USA}

\author[0000-0001-5000-7292]{Djamel M\'ekarnia} 
\affiliation{Universit\'e C\^ote d'Azur, Observatoire de la C\^ote d'Azur, CNRS, Laboratoire Lagrange, Bd de l'Observatoire, CS 34229, 06304 Nice cedex 4, France}

\author[0000-0002-5254-2499]{Louise D. Nielsen} 
\affiliation{Observatoire de l'Universit\'e de Gen\`eve, 51 chemin des Maillettes,1290 Versoix, Switzerland}

\author[0000-0003-1572-7707]{Francisco J. Pozuelos} 
\affiliation{Space Sciences, Technologies and Astrophysics Research (STAR) Institute, Universit\'e de Li\`ege, 19C All\`ee du 6 Ao\^ut, 4000 Li\`ege, Belgium}
\affiliation{Astrobiology Research Unit, Universit\'e de Li\`ege, 19C All\`ee du 6 Ao\^ut, 4000 Li\`ege, Belgium}

\author[0000-0002-3940-2360]{Don J. Radford} 
\affiliation{Brierfield Observatory, New South Wales, Australia}

\author[0000-0003-3914-3546]{Fran{\c c}ois-Xavier Schmider} 
\affiliation{Universit\'e C\^ote d'Azur, Observatoire de la C\^ote d'Azur, CNRS, Laboratoire Lagrange, Bd de l'Observatoire, CS 34229, 06304 Nice cedex 4, France}

\author[0000-0001-8227-1020]{Richard P. Schwarz} 
\affiliation{Patashnick Voorheesville Observatory, Voorheesville, NY 12186, USA}

\author[0000-0003-2163-1437]{Chris Stockdale} 
\affiliation{Hazelwood Observatory, Australia}

\author[0000-0001-5603-6895]{Thiam-Guan Tan} 
\affiliation{Perth Exoplanet Survey Telescope, Perth, Western Australia}

\author{Mathilde Timmermans} 
\affiliation{Astrobiology Research Unit, Universit\'e de Li\`ege, 19C All\`ee du 6 Ao\^ut, 4000 Li\`ege, Belgium}

\author[0000-0002-5510-8751]{Amaury H.M.J. Triaud} 
\affiliation{School of Physics \& Astronomy, University of Birmingham, Edgbaston, Birmingham B15 2TT, United Kingdom}

\author[0000-0003-3092-4418]{Gavin Wang} 
\affiliation{Stanford Online High School, 415 Broadway Academy Hall, Floor 2, 8853, Redwood City, CA 94063, USA}
\affiliation{Tsinghua International School, Beijing 100084, China}

\author[0000-0003-2058-6662]{George Ricker} 
\affiliation{Department of Physics and Kavli Institute for Astrophysics and Space Research, Massachusetts Institute of Technology, Cambridge, MA 02139, USA}

\author[0000-0001-6763-6562]{Roland Vanderspek} 
\affiliation{Department of Physics and Kavli Institute for Astrophysics and Space Research, Massachusetts Institute of Technology, Cambridge, MA 02139, USA}

\author[0000-0001-9911-7388]{David W. Latham} 
\affiliation{Center for Astrophysics \textbar \ Harvard \& Smithsonian, 60 Garden St, Cambridge, MA 02138, USA}

\author[0000-0002-6892-6948]{Sara Seager} 
\affiliation{Department of Physics and Kavli Institute for Astrophysics and Space Research, Massachusetts Institute of Technology, Cambridge, MA 02139, USA}
\affiliation{Department of Earth, Atmospheric and Planetary Sciences, Massachusetts Institute of Technology, Cambridge, MA 02139, USA}
\affiliation{Department of Aeronautics and Astronautics, MIT, 77 Massachusetts Avenue, Cambridge, MA 02139, USA}

\author[0000-0002-4265-047X]{Joshua N. Winn} 
\affiliation{Department of Astrophysical Sciences, Princeton University, NJ 08544, USA}

\author[0000-0002-4715-9460]{Jon M. Jenkins} 
\affiliation{NASA Ames Research Center, Moffett Field, CA 94035, USA}

\author[0000-0002-4510-2268]{Ismael Mireles} 
\affiliation{Department of Physics and Astronomy, University of New Mexico, 210 Yale Blvd NE, Albuquerque, NM 87106, USA}

\author[0000-0003-4755-584X]{Daniel A. Yahalomi} 
\affiliation{Department of Astronomy, Columbia University, 550 West 120th Street, New York, NY 10027, USA}
\affiliation{Center for Astrophysics \textbar \ Harvard \& Smithsonian, 60 Garden Street, Cambridge, MA 02138, USA}

\author[0000-0003-1447-6344]{Edward~H.~Morgan} 
\affiliation{Department of Physics and Kavli Institute for Astrophysics and Space Research, Massachusetts Institute of Technology, Cambridge, MA 02139, USA}

\author{Michael~Vezie} 
\affiliation{Department of Physics and Kavli Institute for Astrophysics and Space Research, Massachusetts Institute of Technology, Cambridge, MA 02139, USA}

\author[0000-0003-1309-2904]{Elisa~V.~Quintana} 
\affiliation{NASA Goddard Space Flight Center, 8800 Greenbelt Road, Greenbelt, MD 20771, USA}

\author[0000-0003-4724-745X]{Mark E. Rose} 
\affiliation{NASA Ames Research Center}

\author[0000-0002-6148-7903]{Jeffrey C. Smith} 
\affiliation{NASA Ames Research Center}
\affiliation{SETI Institute}

\author{Bernie Shiao} 
\affiliation{Mikulski Archive for Space Telescopes}

\begin{abstract}
Warm Jupiters -- defined here as planets larger than 6 Earth radii with orbital periods of 8--200 days -- are a key missing piece in our understanding of how planetary systems form and evolve. It is currently debated whether Warm Jupiters form in situ, undergo disk or high eccentricity tidal migration, or have a mixture of origin channels. These different classes of origin channels lead to different expectations for Warm Jupiters' properties, which are currently difficult to evaluate due to the small sample size. We take advantage of the \TESS survey and systematically search for Warm Jupiter candidates around main-sequence host stars brighter than the \TESS-band magnitude of 12 in the Full-Frame Images in Year 1 of the \TESS Prime Mission data. We introduce a catalog of \changes{55} Warm Jupiter candidates, including \changes{19} candidates that were not originally released as \TESS Objects of Interest (TOIs) by the \TESS team. We fit their \TESS light curves, characterize their eccentricities and transit-timing variations (TTVs), and prioritize a list for ground-based follow-up and \TESS Extended Mission observations. Using hierarchical Bayesian modeling, we find the preliminary eccentricity distributions of our Warm-Jupiter-candidate catalog using a Beta distribution, a Rayleigh distribution, and a two-component Gaussian distribution as the functional forms of the eccentricity distribution. Additional follow-up observations will be required to clean the sample of false positives for a full statistical study, derive the orbital solutions to break the eccentricity degeneracy, and provide mass measurements.
\end{abstract}


\section{Introduction} \label{sec:intro}
We do not yet understand the formation of Warm Jupiters. Although intrinsically uncommon relative to other planetary demographics (e.g., \citealt{jone03,udry03,witt10,sant16}), they are a result of physical processes that likely sculpt many planetary systems and are excellent test cases for theories inspired by the more frequently-detected hot Jupiters (i.e., gas giants with orbital periods less than 8 days) and Warm mini-Neptunes/super-Earths (i.e., planets between the sizes of Earth and Neptune with orbital periods of 8--200 days). It is currently debated whether Warm Jupiters form in situ \changes{\citep[e.g.,][]{lee14, lee16, baty16, bole16}} or undergo disk \changes{migration \citep[e.g.,][]{gold80, lin86, lin96, ida08, baru14}} or high-eccentricity tidal migration \changes{(e.g., \citealt{rasi96, wu11, petr15}; see Section 4.3 \citealt{daws18} for a comprehensive review)}. Three proposed origin channels lead to different expectations for Warm Jupiters' masses (e.g., \citealt{ida08,lee14,lee16}), eccentricities (e.g., \citealt{duff14,petr16,ande19,frel19}), host star obliquities (e.g., \citealt{naoz12,li16,petr16}) and metallicities (e.g., \citealt{daws13,tsan14a}), and the presence and properties of other planets in the system (e.g., \citealt{dong14, huan16}). Recently, the diversity in these properties has inspired the hypothesis that multiple origin channels contribute substantially to the Warm Jupiter population (e.g., \citealt{daws18}).

The \textit{Transiting Exoplanet Survey Satellite} \citep[\TESS;][]{rick15} provides an excellent opportunity to examine these hypotheses for Warm Jupiters' origins. Using \TESS data, a handful of Warm Jupiter systems \citep[e.g., TOI-172, TOI-216, TOI-481, TOI-677, TOI-1130;][]{rodr19, daws19, kipp19, brah20, jord20, huan20} have been discovered and characterized, allowing us to examine whether multiple properties of the system tell a consistent story about the system's origin (i.e., whether each property is consistent with the same origin theory). More generally, \TESS is discovering a large sample of Warm Jupiters in the Full-Frame Images (FFIs) that allows for a statistical study of the Warm Jupiter population. Origin theories make different predictions for the eccentricity distribution and occurrence rates of Warm Jupiters \citep[e.g.,][]{petr16}, which are currently difficult to evaluate due to their small sample size.

In this work, we describe a systematic search for Warm Jupiters, defined here as planets larger than 6 Earth radii with orbital periods of 8--200 days, around stars brighter than 12th \TESS-band magnitude (Tmag) in the FFIs in the first year of \TESS data. We note that the Warm Jupiters here can be more accurately termed warm, \emph{large} planet candidates since we do not have their mass measurements. Planets larger than 6 Earth radii are likely to be gas giants, although we cannot rule out the possibility of super-puffs (i.e., large planets with low densities) which likely have different formation process. We construct a southern ecliptic hemisphere catalog of Warm Jupiter candidates in Year 1 of the \TESS FFIs; prioritize a list of Warm Jupiter candidates showing evidence of strong transit-timing variations (TTVs) and high eccentricities for ground-based follow-up and the \TESS Extended Mission; derive the eccentricity distribution of our Warm Jupiter candidate catalog using hierarchical Bayesian modeling to compare to expectations of different origin theories.

In Section \ref{sec:selection}, we describe our pipeline for discovering Warm Jupiter candidates in \TESS FFIs. In Section \ref{sec:lc}, we present our fitting model for \TESS light curves and post fitting analysis. In Section \ref{sec:results}, we catalog the resulting Warm Jupiter candidates discovered in Year 1 of the \TESS FFIs and highlight candidates showing possible TTV signals and evidence of high eccentricities. In Section \ref{sec:ecc_dist}, we infer the eccentricity distribution of the catalog using hierarchical Bayesian modeling. We put our work in the context of \TESS Extended Mission and ground-based follow-up observations and discuss the implications of our results for origins of Warm Jupiters in Section \ref{sec:discussion}. We summarize our findings in Section \ref{sec:summary}.

\section{Transit search} \label{sec:selection}
During the first year of the \TESS Prime Mission (July 25, 2018 -- July 18, 2019), \TESS surveyed almost the entire southern ecliptic hemisphere \citep{rick15}. The Year 1 Prime Mission was divided into thirteen \TESS sectors (Sector 1--13) with each sector monitoring a fraction of sky for $\sim$27 days. The \TESS data release includes 30-minute cadence FFIs and 2-minute cadence for $\sim$200,000 pre-selected target stars. Given the long orbital periods and low occurrence rates of Warm Jupiters, we should expect only a dozen or so Warm Jupiter candidates on the pre-selected list. Here we systematically search for Warm Jupiter candidates in \TESS FFIs, which includes over a million stars brighter than Tmag of 12. Warm Jupiters' large transit depths ($\sim$10$^4$ ppm) make them readily detected in FFIs.

\subsection{Identifying Threshold-Crossing Events}
We identify Threshold-Crossing Events (TCEs) in \TESS Sector 1--13 that would correspond to transits of objects with radii 6--20 $R_\Earth$, orbital periods 8--200 days, and transit signal-to-noise ratios (SNRs) greater than 9 \citep[see the definition of SNRs in][]{hart16}. We focus on TCEs with host stars brighter than the Tmag of 12 to produce a catalog feasible for ground-based follow up. The Tmag and parameter cut off here are based on \TESS Input Catalog v7 \citep[TICv7;][]{stas18}. The TICv8 catalog \citep{stas19} was not yet published when we made the target selection. All the TCEs are required to receive a $\mathtt{triage}$ score $>$ 0.09 \citep{yu19}, which is the threshold from the deep learning algorithm to distinguish eclipse-like signals (e.g., planet candidates and eclipsing binaries) from stellar variability and instrumental noise. After applying these selection criteria, we identify $\sim$2000 TCEs.

\subsection{Vetting TCEs}
We further vet the $\sim$2000 TCEs to remove false positives (e.g., obvious eclipsing binaries, stellar variability, and instrumental artifacts) that were not identified by $\mathtt{triage}$, and single-transit events.\footnote{A list of single-transit events is available from J.D. upon request, although we note that our detection pipeline (i.e., a Box-Least Squares search) is not optimized for single transit events.} The procedures are similar to standard vetting procedures applied to \TESS Objects of Interests (TOIs) \citep{guer21}. This process truncates our sample to $\sim$500 Warm Jupiter candidates.
For the remaining $\sim$500 Warm Jupiter candidates, we examine more closely each candidate to identify those with signatures of false positives. We remove the target from our sample if it shows
\begin{itemize}
    \setlength\itemsep{0.1em}
    \item detectable motion of the center of the light during fading events
    \item $>$ 3$\sigma$ secondary eclipse detection at any phase
    \item $>$ 3$\sigma$ transit depth difference detection between a large aperture (3.5-pixel) and a nominal aperture (2.5-pixel)
    \item transit signal synchronized with stellar variability
    \item matching any known eclipsing binary catalogs \citep{coll18}
    \item impact parameter $b >$ 0.9
\end{itemize}
For giant planets with a typical planet-star radius ratio of $\sim$0.1, an impact parameter greater than 0.9 may lead to difficulty in constraining the planet-star radius ratio and impact parameter from their light curves. To minimize the contamination of eclipsing binaries in our sample, we remove these targets. The impact-parameter cutoff has a trade off between false positives and real planets with high impact parameters. At the time of this writing, we notice the confirmation of a high-impact-parameter Warm Jupiter, TIC 237913194 \citep{schl20}, which is not identified in our catalog due to the impact-parameter cutoff. One may also concern that the impact-parameter cutoff will result in removing candidates on highly elliptical orbits because of their short transit durations. Since giant planets usually have large transit depths and well-resolved transit shapes, we do not expect the cutoff will strongly affect planets with high eccentricities. Lastly, we remove about 15 targets that have been labeled as false positives by \TESS Follow-up Observing Program at the time of vetting. See Section~\ref{subsec:tfop} for more details.

{\changes Our study focuses on host stars that are on or near the main sequence and have TCEs around substantially evolved stars removed (i.e., $R_\star > 3 R_\odot$ or $\log g < 3.5$).} We derive posteriors for stellar parameters (including $\rho_{\star}$ and $R_\star$) from isochrone fitting with the Dartmouth \citep{dott08} stellar evolution models. We use the approach described by \citet{daws15} to fit the stellar effective temperature, metallicity (when available), \emph{Gaia} DR2 parallax, and \emph{Gaia} apparent $g$ magnitude \citep{gaia16,gaia18}. We apply the systematic correction to \emph{Gaia} parallaxes from \citet{stas18b}. The stellar temperature, metallicity, and uncertainties are taken from the TICv8 catalog \citep{stas19}.

After this step, we have left with a sample of 197 Warm Jupiter candidates with 12 candidates on the pre-selected target list.

\subsection{TFOP WG Follow-up}\label{subsec:tfop}
Several candidates that survived vetting have already been dispositioned as false-positives (FPs) or false-alarms (FAs) by \TESS Follow-up Observing Program (TFOP)\footnote{https://tess.mit.edu/followup} Sub Group 1 (SG1) using \emph{Seeing-Limited Photometry} and Sub Group 2 (SG2) using \emph{Recon Spectroscopy}. As ground-based follow-up observations progress, we will continue to drop newly discovered FPs and FAs from the catalog. Table~\ref{tbl:fp} lists the \changes{15} FPs and FA that survived our vetting and 15 FPs that were removed during the TCE vetting discussed in previous subsection.

\begin{deluxetable*}{llDDc}
\tablecaption{Facilities used for SG1 Follow-up Observations\label{table:walesobservatories}}
\tabletypesize{\small}
\tablehead{\colhead{Observatory} & \colhead{Location} & \twocolhead{Aperture} & \twocolhead{Pixel scale} & \colhead{FOV}\\[-2mm]
 & & \twocolhead{(m)} & \twocolhead{(arcsec)} & \colhead{(arcmin)}
}
\decimals
\startdata
Antarctic Search for Transiting ExoPlanets (ASTEP) & Concordia Station, Antarctica & 0.4 & 0.93 & $63 \times 63$\\
Brierfield Private Observatory & Bellingen, New S. Wales, Australia & 0.36 & 1.47 & $50 \times 50$ \\
Chilean-Hungarian Automated Telescope (CHAT) & Las Campanas Observatory, Chile & 0.7 & 0.6 & $21 \times 21$ \\
Evans Private Telescope & El Sauce Observatory, Chile & 0.36 & 1.47 & $19 \times 13$\\
Hazelwood Private Observatory & Churchill, Victoria, Australia & 0.32 & 0.55 & $20 \times 14$\\
Hungarian Automated Tel. Network-South (HATS) & Chile, Namibia, Australia & 0.18 & 3.7 & $492 \times 492$\\
Las Cumbres Observatory Global Telescope (LCOGT) & Chile, South Africa, Australia & 1.0 & 0.39 & $26 \times 26$ \\
MEarth-South & La Serena, Chile & 0.4 & 0.84 & $29 \times 29$ \\
Perth Exoplanet Survey Telescope (PEST) & Perth, Australia & 0.3 & 1.2 & $31 \times 21$ \\
TRAPPIST-North & Oukaimeden Observatory, Morocco & 0.6 & 0.60 & $20 \times 20$ \\
TRAPPIST-South & La Silla, Chile & 0.6 & 0.64 & $ 22 \times 22$ \\
Wide Angle Search for Planets-South (WASP-South) & Sutherland, South Africa & 0.1 & 13.7 & $468 \times 468$ \\
\enddata
\tablerefs{TRAPPIST \citep{trappsit11}}
\end{deluxetable*}

The SG1 follow-up observations attempt to (1) rule out or identify nearby eclipsing binaries (NEBs) as sources of the \TESS data detection, (2) detect the transit-like event on-target to confirm the event depth and thus the \TESS photometric deblending factor, (3) refine the \TESS ephemeris, (4) provide additional epochs of transit center time measurements to supplement transit timing variation (TTV) analysis, and (5) place constraints on transit depth differences across optical filter bands. We used the $\mathtt{TESS\;Transit\;Finder}$, which is a customized version of the $\mathtt{Tapir}$ software package \citep{Jensen:2013}, to schedule SG1 transit observations.

The SG2 follow-up observations attempt to provide spectroscopic parameters that will more precisely constrain the masses and radii of planet host stars, to detect false positives caused by spectroscopic binaries (SBs), and to identify stars unsuitable for precise RV measurements, such as rapid rotators.

Joint analysis of ground-based and \TESS data for planet candidates in our catalog that have been verified to be on-target is beyond the scope of this work. The ground-based data will be presented in follow-up statistical validation or radial velocity confirmation publications. However, in the process of following-up these candidates, we identified several nearby eclipsing binaries (NEBs) that are the cause of their corresponding signals in the \TESS data. We also identified a few systems with strong transit depth chromaticity, which we interpret as a blended eclipsing binary (BEB) in the follow-up photometric aperture. We also identified one candidate that is an FA after it was not detected on or off target in the ground-based follow-up data. Further analysis of the \TESS data led us to conclude that the apparent signal in the \TESS data is a detrending residual due to high scattered light near the beginning of the two corresponding \TESS orbits. 
The follow-up light curve data are available at ExoFOP-TESS\footnote{https://exofop.ipac.caltech.edu/tess}. Table \ref{table:walesobservatories} lists observatories that have participated in SG1 follow-up observations of these targets.

\section{Light Curve Characterization} \label{sec:lc}
\TESS light curves are obtained from the MIT Quick Look Pipeline \citep[QLP;][]{huan19, huan20b, huan20c}. To reduce the computational effort for light curve fitting, we trim the light curves around each transit to roughly 6 times the transit duration. We generally use the QLP detrended light curves (i.e., $\mathtt{KSPSAP\_FLUX}$), which are corrected for systematics using Kepler splines \citep{shal18}. However, we identify some $\mathtt{KSPSAP\_FLUX}$ light curves showing over-detrended transit signals (e.g., transit signals being washed out due to the correction). In such cases, we switch to the Simple Aperture Photometry (SAP) light curves. We also notice that transit signals could be masked by the \TESS quality flag if they are close to momentum dumps. To avoid missing any transit data, we do not apply the \TESS quality flag to produce the trimmed light curves. Instead, we manually remove truly abnormal data points (e.g., a sudden 10\% flux deviation on a single data point) in the trimmed light curves. 

For the pre-selected targets with 2-minute cadence data available, we obtain their \TESS light curves processed with the Science Processing Operations Center (SPOC) pipeline \citep{jenk10}. The SPOC pipeline is a descendant of the Kepler mission pipeline based at the NASA Ames Research Center \citep{jenk02,jenk10}, analyzing target pixel postage stamps that are obtained for pre-selected target stars. We download the publicly available data from the Mikulski Archive for Space Telescopes (MAST) using the $\mathtt{lightkurve}$ package \citep{lightkurve18}. Similar to FFI light curves, we use $\mathtt{PDCSAP}$ light curves in most cases but switch to $\mathtt{SAP}$ light curves if the $\mathtt{PDCSAP}$ light curves are over-detrended.

\subsection{Light curve modeling}\label{subsec:lc_model}
\begin{deluxetable*}{clc}
\tablecaption{Summary of priors used for the light curve fits. \label{tbl:priors}}
\tabletypesize{\normalsize}
\tablehead{\colhead{Parameter} & \colhead{Description} & \colhead{Distribution}} 
\startdata	
\multicolumn{3}{l}{Transit model}\\      
$\rho_{\textrm{circ}}$        & Stellar density assuming a circular orbit (g/cm$^3$)     & logUniform(10$^{-3}$, 10$^{3}$) \\
$b$                           & Impact parameter                                         & Uniform(-2, 2)\\
$r_{\textrm{p}}/r_{\star}$    & Planet-star radius ratio              		             & logUniform(\emph{testval}-5, \emph{testval}+5)\\
$P$                           & Orbital period (days)                                    & Uniform(\emph{testval}-5, \emph{testval}+5) or \emph{constant}\\ 
$T_{1..N}$                    & Mid-transit times (BJD-2457000)                          & Uniform(\emph{testval}-5, \emph{testval}+5)\\ 
$u_0$, $u_1$                  & Quadratic limb-darkening coefficients                    & Adopted from \cite{exoplanet:kipping13}\\ 
\hline
\multicolumn{3}{l}{Gaussian Process model}\\
$s_{\textrm{gp}}$             & Photometric jitter                                       & logUniform($30\times10^{-6}$, 1)\\
$\rho_{\textrm{gp}}$          & Amplitude of the Matern-$3/2$ kernel                     & logUniform($30\times10^{-6}$, 1)\\
$\sigma_{\textrm{gp}}$        & Timescale of the Matern-$3/2$ kernel                     & logUniform(0.001, 1000)\\
\enddata
\tablecomments{\emph{testval}: test values; The absolute value of the impact parameter is used when computing light curves.}
\end{deluxetable*}
We use a quadratic limb darkening transit model \citep{mand02}, along with a Gaussian process (GP) likelihood function, to infer planet properties from \TESS light curves. The free parameters in our transit model are $\big\{\rho_{\textrm{circ}}, b, r_{\textrm{p}}/r_{\star}, P, T_0, u_0, u_1\}$, where $\rho_{\textrm{circ}}$ is the stellar density assuming a circular orbit, $b$ is the impact parameter of the transiting planet, $r_{\textrm{p}}/r_{\star}$ is the planet-star radius ratio, $P$ is the planet orbital period, $T_0$ is the mid-transit time of the first observed transit, and $u_0$ and $u_1$ are the quadratic limb darkening coefficients \citep{exoplanet:kipping13}.
We note that we fit our candidates' stellar densities, $\rho_{\textrm{circ}}$ assuming they have circular orbits. We later compare the marginalized $\rho_{\textrm{circ}}$ to $\rho_{\star}$, the ``true" stellar densities derived from the isochrone fitting, to constrain the candidates' eccentricities (Section \ref{subsec:post_analysis}). 
To allow transit-timing variation (TTV) characterization, we adopt a slightly different approach for candidates with 3$+$ transits. Instead of fitting the orbital period and mid-transit time, we fit mid-transit times for each transit (i.e., $T_{1..N}$). We take the orbital period as a fixed value in the fitting and it is only involved in the computation of the transit duration. Since the orbital period is known with much greater precision than the precision with which the transit duration can be determined, and since the transit duration has a weak scaling with the orbital period ($\tau \propto P^{1/3}$), fixing the orbital period for light curve modeling and computing it later from a linear fit to mid-transit times does not significantly affect the inference on any of the parameters.
Lastly, to account for correlated noise in the light curves, we adopt a GP kernel including a diagonal jitter term ($s_{\textrm{gp}}$) to characterize the light curve white noise and a Matern-$3/2$ term ($\sigma_{\textrm{gp}}$ for the amplitude and $\rho_{\textrm{gp}}$ for the timescale) to take account photometric variability \citep{exoplanet:foremanmackey17, exoplanet:foremanmackey18}. 
In Table \ref{tbl:priors}, we list the priors we put on the free parameters described above.
For the pre-selected targets with both 2-minute and 30-minute cadence data, we perform a joint fit of parameters $\big\{\rho_{\textrm{circ}}, b, r_{\textrm{p}}/r_{\star}, T_{1..N}, u_0, u_1\}$, but give each dataset a different set of GP-kernel parameters to treat the difference on time sampling.

The light curve fits use the $\mathtt{exoplanet}$ package \citep{exoplanet:exoplanet} that implements the quadratic limb darkening transit model using $\mathtt{starry}$ \citep{exoplanet:luger18, exoplanet:agol20} and the Matern-$3/2$ GP model in $\mathtt{celerite}$ \citep{exoplanet:foremanmackey17}. The light curve model is averaged into 30-minute bins over 15 evaluations per bin (i.e., $\mathtt{texp}$ = 0.02083, $\mathtt{oversample}$ = 15) for the 30-minute cadence light curves and averaged into 2-minute bins over 8 evaluations per bin (i.e., $\mathtt{texp}$ = 0.00139, $\mathtt{oversample}$ = 8) for the 2-minute cadence light curves. For each set of light curves, we sample 4 chains using the Markov Chain Monte Carlo (MCMC) technique with gradient-based proposals \citep{neal12, hoff11, beta17}. Each chain contains 5000 tuning steps and 3000 sampling steps. We set the target accept rate universally to 0.99 to reduce divergences caused by the degeneracy between $\rho_{\textrm{circ}}$ and $b$ for some \emph{v}-shaped light curves. We assess MCMC convergence using the Gelman-Rubin diagnostic (i.e., $\hat{\mathcal{R}} < 1.1$ for convergence), trace plots, and corner plots \citep{corner} of the marginal joint distributions. We combine all four chains to get the posterior distribution. For a typical Warm Jupiter candidate with three transits \citep[e.g., TOI-172b;][]{rodr19}, it takes $\sim$30 minutes to analyze the light curves, benefiting from the efficiency of gradient-based proposals and No U-Turns (NUTs) sampling \citep{hoff11}.

\begin{figure*}[t]
    \centering
    \includegraphics{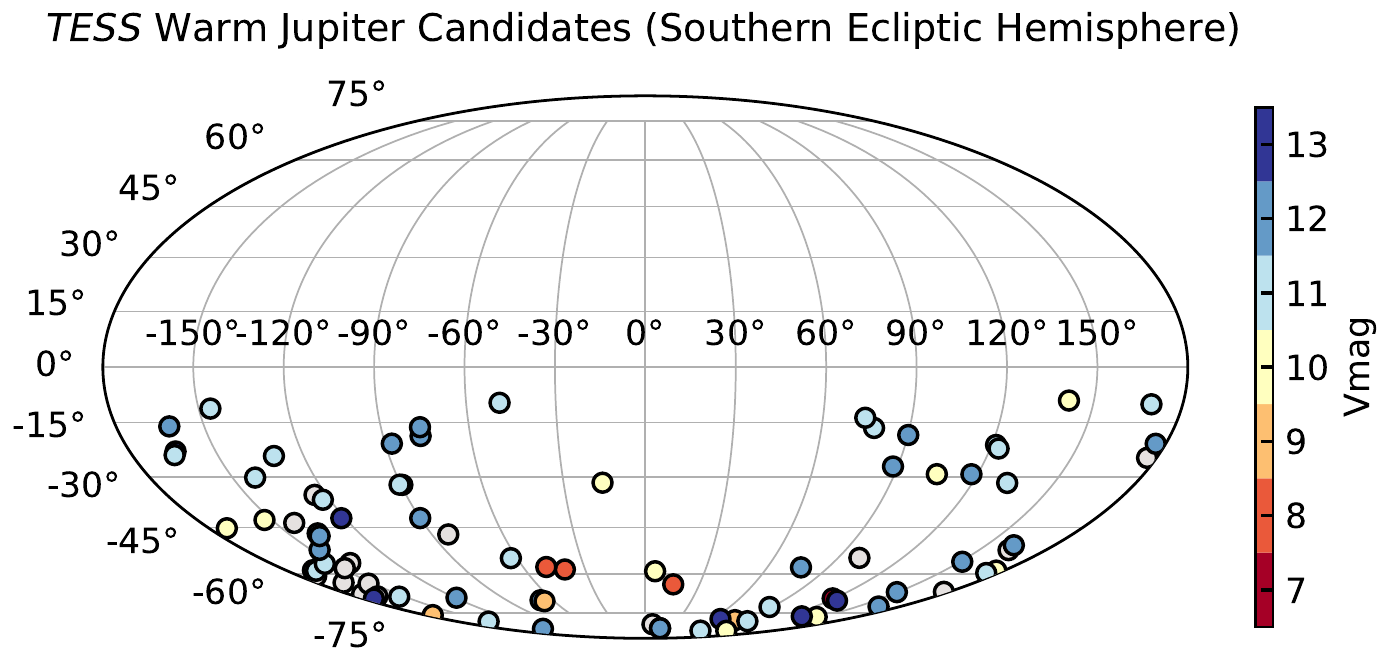}
    \caption{Warm Jupiter candidates discovered in Year 1 \TESS Full-Frame Images around main-sequence host stars brighter than \TESS-band magnitude of 12. The candidates are plotted in ecliptic coordinates ($\lambda$, $\beta$), where $\lambda$ is the ecliptic longitude and $\beta$ is the ecliptic latitude. The \changes{55} candidates with well-constrained parameters, along with 11 candidates with well-fitted light curves but undetermined orbital periods and/or missing stellar parameters, are colored in their V-band magnitudes. The 19 \emph{possible} candidates with unconstrained impact parameters and radii are colored in grey. Most of our targets are bright enough for ground-based follow-up observations.}
    \label{fig:WaLEs_south}
\end{figure*}

\subsection{Post light curve analysis}
\label{subsec:post_analysis}
Our light curve fitting identifies a large group of Warm Jupiter candidates ($\sim$100) with grazing orbits that were not previously identified in Section \ref{sec:selection}. Their impact parameter posteriors have a mode centered around $b=1$ with large uncertainties (e.g., $\sigma_b = 0.3$--0.5).
We remove these candidates from our catalog because we cannot constrain their planet-star radius ratios due to poorly constrained impact parameters. The range of possible radii extends to the stellar companion regime in many cases.

We assess TTVs for Warm Jupiter candidates with $3+$ transits. We calculate a best-fit linear ephemeris to the mid-transit times using the medians and uncertainties on mid-transit time posteriors from light curve fitting. We perform a least-square fitting to
\begin{equation}
    T_n = T_c + nP, 
\end{equation}
where $T_n$ is the mid-transit time for the $n$-th transit, $T_c$ is the conjunction time for reference, and $P$ is the orbital period of the planet.
To compute the $O\mathrm{-}C$ (Observed-minus-Calculated) times, we subtract the linear ephemeris $T_n$ from the observed mid-transit times. To evaluate the significance of the TTV signal, we find the absolute difference between every pair of $O\mathrm{-}C$ data points and normalize it by their uncertainties (i.e., the quadratic error in the two mid-transit times). For example, for a pair of mid-transit data points $T_x$ and $T_y$ with $O\mathrm{-}C$s of $(O\mathrm{-}C)_x$ and $(O\mathrm{-}C)_y$ and uncertainties of $\sigma_{T_x}$ and $\sigma_{T_y}$, respectively, their TTV significance $\sigma_{\mathrm{TTV}|x,y}$ is calculated as
\begin{equation}\label{eqn:ttv}
    \sigma_{\mathrm{TTV}|x,y} = \frac{\lvert (O\mathrm{-}C)_x - (O\mathrm{-}C)_y \rvert}{\sqrt{\sigma_{T_x}^2+\sigma_{T_y}^2}}
\end{equation}
We take the maximum value of the TTV significance of every pair of the mid-transit times as the significance level of the TTV detection of the system.

The orbital eccentricity of a planet can be inferred from its transit light curve, sometimes termed the ``photoeccentric" effect, although the solution is degenerate with the planet's argument of periapse. Using the posterior distribution of $\rho_{\mathrm{circ}}$ from light curve fitting and the posterior distribution of $\rho_\star$ from the isochrone fitting, we compute a joint posterior distribution for $(e, \omega)$ following \cite{daws12}.

\section{Candidate Catalog}
\label{sec:results}
Our vetting process results in a catalog of \changes{55} Warm Jupiter candidates in the FFIs in Year 1 of the \TESS Prime Mission data. Due to the depth of our survey (i.e., Tmag$<$12), \changes{19} of our candidates have not yet been identified as TOIs. We present their light curves in Appendix~\ref{appendix:lc}. A complete figure set (\changes{19} figures) of the FFI light curves of non-TOI candidates is available in the online journal. In Section \ref{subsec:catalog}, we tabulate our candidates with their host star and planet properties and present their TTV and eccentricity analysis. In Section \ref{subsec:catalog_more}, we introduce 11 additional Warm Jupiter candidates with undetermined orbital periods and/or missing stellar parameters and list 19 more \emph{possible} Warm Jupiter candidates with unconstrained impact parameters.

\subsection{A catalog of Warm Jupiter candidates} \label{subsec:catalog}
We present \changes{55} Warm Jupiter candidates discovered in \TESS Sectors 1--13 in FFIs around main-sequence host stars brighter than \TESS magnitude of 12. In Figure \ref{fig:WaLEs_south}, we display our candidates in ecliptic coordinates and color them by their V-band magnitude. In Table \ref{tbl:pc}, we tabulate the \changes{55} candidates with their planet and host star properties. A detailed description of each column can be found in the table caption. The \changes{19} candidates that were not originally released as TOIs by the \TESS team can be identified from the \changes{``TOI Name" column (Column 2)} with no TOI names listed. Some of the non-TOI targets have been independently vetted by other groups \citep[e.g.,][]{mont20} and reported as Community TOIs (CTOIs), labeled in Column \changes{17}.

\begin{figure*}[htb!]
  \centering
  \hspace*{-0.5cm}
  \includegraphics{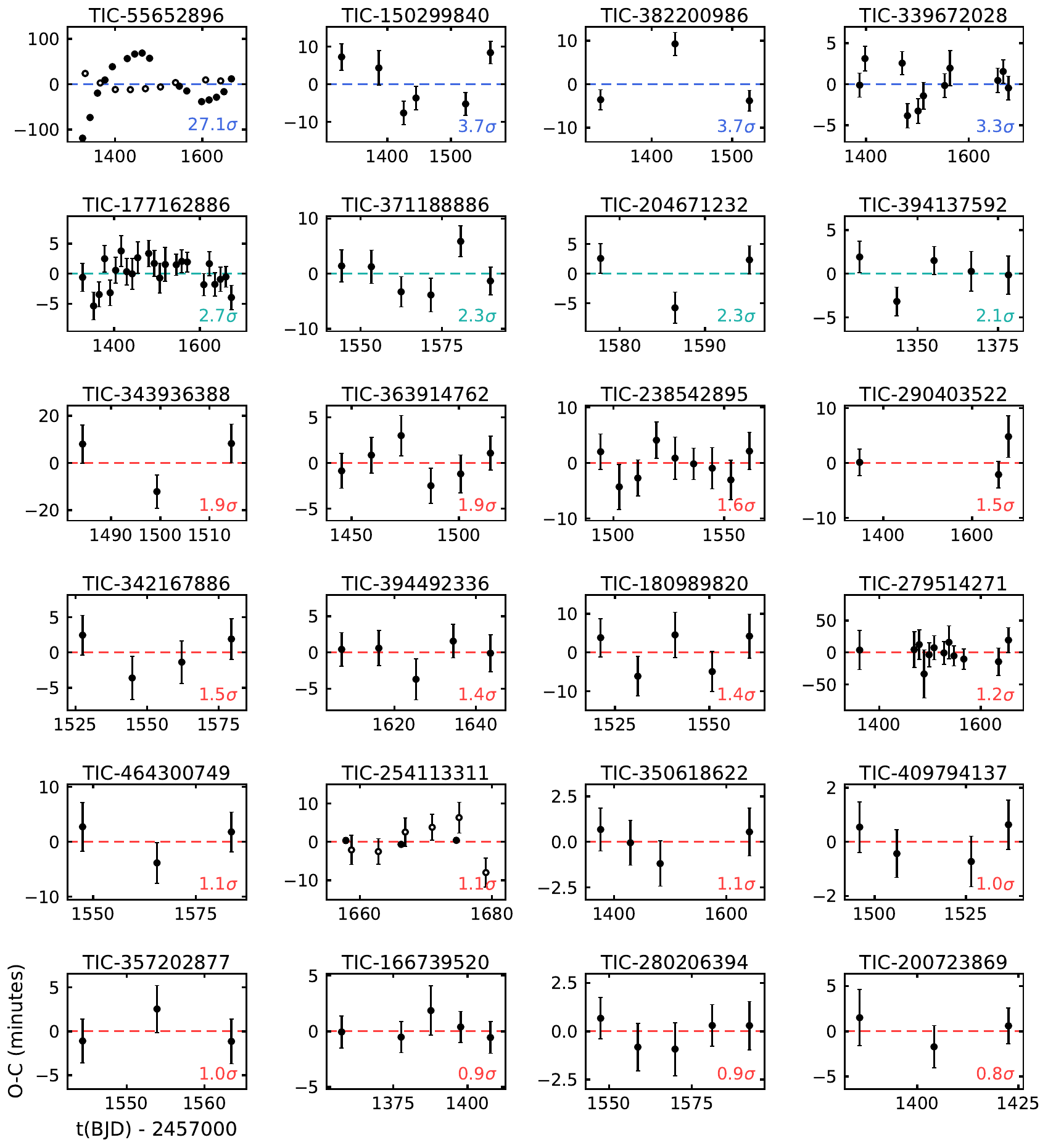}
  \caption{TTV analysis for Warm Jupiter candidates with 3$+$ transits (Part 1). In each panel, we show the $O$-$C$ diagram (Observed-Calculated of mid-transit times versus time in BJD-2457000) for one candidate. We include a horizontal dashed line centered at zero in each panel for reference. The significance levels of the TTV detections are calculated using Equation (\ref{eqn:ttv}) by taking the absolute differences between every pair of $O$-$C$ data points normalized by their quadratic errors. We report the maximum TTV significance level in each panel. Candidates with $>3\sigma$ TTV detections are colored in lavender; with 2--3$\sigma$ detections are colored in green; with $<2 \sigma$ detections are colored in coral. For two-planet systems (i.e., TIC-55652896 and TIC-254113311), we plot one planet using closed circles and the other using open circles.}
  \label{fig:ttv1}
\end{figure*}

\begin{figure*}[htb!]
  \centering
  \hspace*{-0.5cm}
  \includegraphics{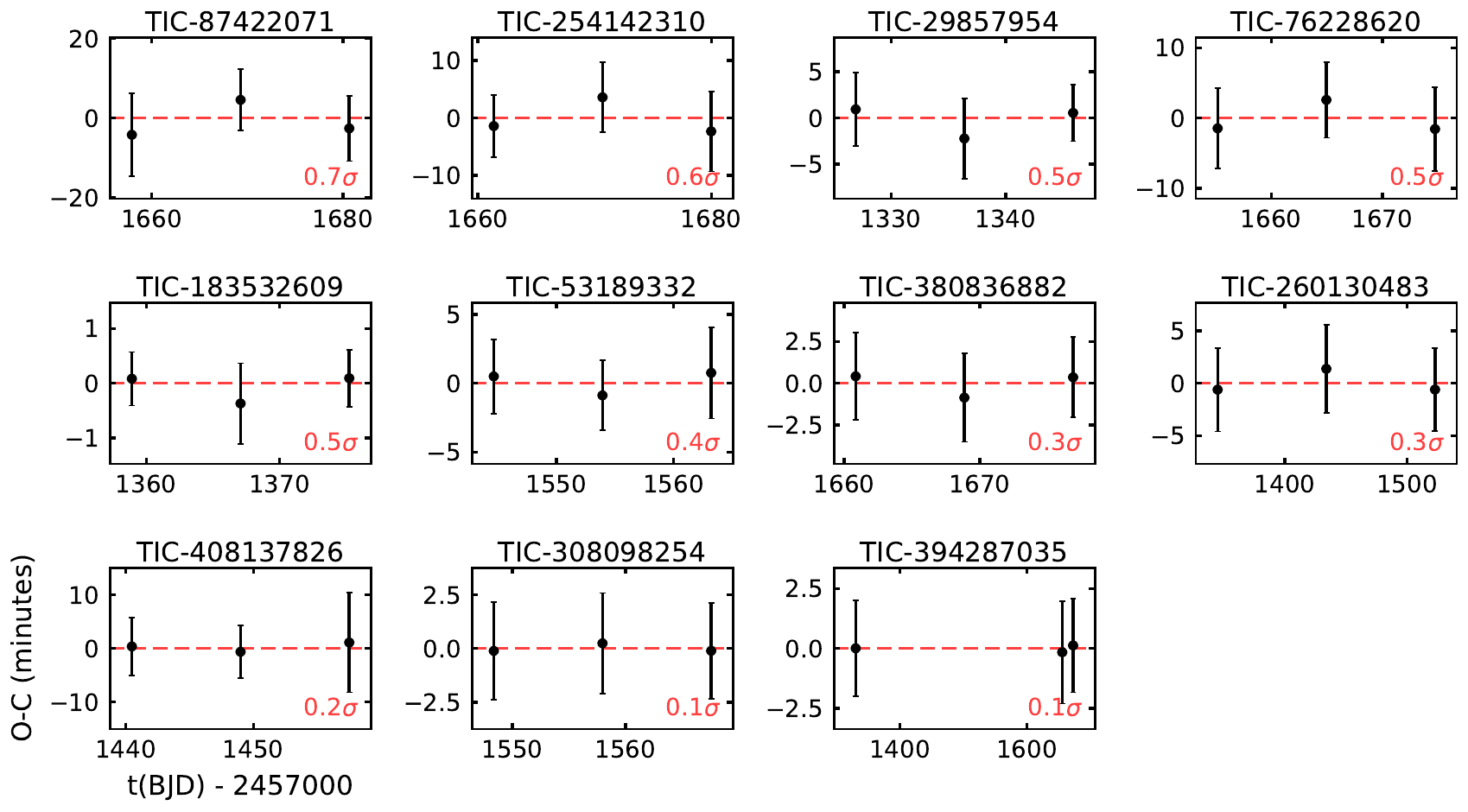}
  \caption{Same as Figure~\ref{fig:ttv1}.}
  \label{fig:ttv2}
\end{figure*}

In Figures \ref{fig:ttv1} and \ref{fig:ttv2}, we present the TTV analysis for the \changes{35} Warm Jupiter systems with 3$+$ transits. In each panel, we show the $O$-$C$ diagram (i.e., the observed mid-transit times with the calculated mid-transit times subtracted) of the planet candidates in the system. The panels are sorted by the TTV significance. About half of the systems are observed in multiple \TESS Sectors. One of the most prominent TTV systems is TOI-216 \citep[TIC-55652896; Dawson et al. AJ in press;][]{daws19, kipp19} with an outer Warm Jupiter (labeled as open circles) in 2:1 mean-motion resonance with the inner Warm Neptune (labeled as black dots). The system has a TTV significance level of 27.1 using our metric discussed in Section \ref{subsec:post_analysis}. We do not identify any other systems with TTV signals as significant as TOI-216's in the rest of the catalog. TOI-1130 \citep[TIC-254113311;][]{huan20} is another \TESS multi-transiting-planet system with an outer 8.4-day Warm Jupiter (TOI-1130c) and an inner 4.1-day hot Neptune (TOI-1130b). Our TTV analysis on the system shows that TOI-1130c (i.e., black dots) has no obvious TTV signals, and TOI-1130b (i.e., open circles) has some tentative TTV signals. We might be able to detect the TTV signals of TOI-1130c when combined with the \TESS Extended Mission data. Meanwhile, the system demonstrates that even with a nearby companion, a target can still show no significant TTV signals due to the short observing baseline.

As shown in Figure \ref{fig:ttv1} and \ref{fig:ttv2}, we evaluate the significance of the TTV signals (i.e., $>$3$\sigma$, 2--3$\sigma$, and $<$2$\sigma$) for each system. \changes{4/35 ($\sim$11\%)} systems show a $>$3$\sigma$ TTV detection, colored and labeled in lavender; \changes{4/36 ($\sim$11\%)} candidates show 2--3$\sigma$ TTV detection, colored and labeled in green; and \changes{27/35 (77\%)} systems show less than 2$\sigma$ TTV signals, colored and labeled in coral. The horizontal dashed line in each panel is centered at zero for reference. Although they do not have the strongest TTV detections, a few systems show TTV patterns that could be sinusoidal, e.g.,
\begin{itemize}
    \item TIC-150299840
    \item TIC-382200986
    \item TIC-371188886
    \item TIC-343936388,
\end{itemize}
which might be worthwhile to explore further. 
Many of the systems (i.e., systems colored in coral) show no significant TTV signals. The lack of detected TTVs does not rule out the presence of other planets and may be due to the lack of precision of mid-transit times (i.e., demonstrated by error bars in Figure \ref{fig:ttv1} and \ref{fig:ttv2}) and/or short observing baselines. With the \TESS Extended Mission observation, which will provide a longer observing baseline and finer cadence data, we expect to improve the TTV analysis for many of the Warm Jupiter systems.

\begin{figure*}[htb!]
    \centering
    \includegraphics{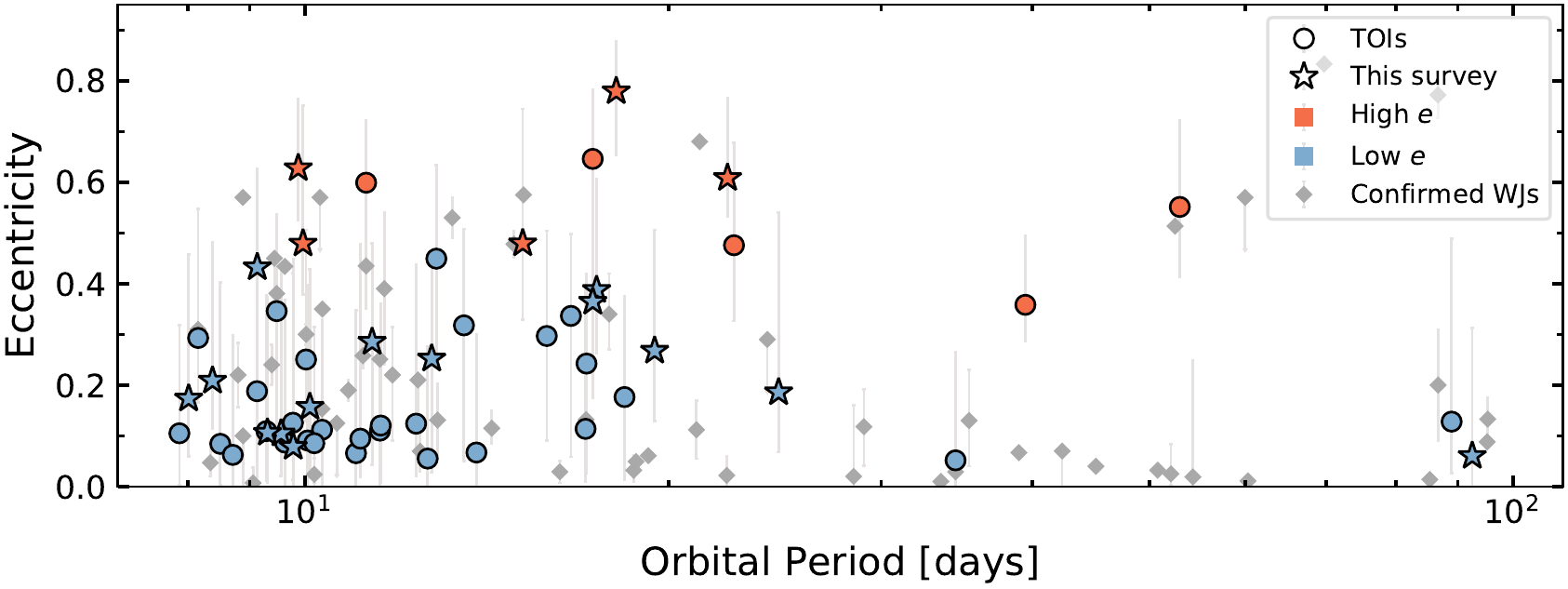}
    \caption{Eccentricity versus orbital period of the \changes{55} Warm Jupiter candidates discovered in Year 1 \TESS FFIs around host stars brighter than Tmag of 12. The eccentricities are inferred from the candidates' stellar densities and presented by their posterior modes and the 68\% highest posterior density (HPD) intervals. TOIs are labeled with circles and unique targets yield from our survey are labeled with stars. If the lower bound of the 95\% HPD interval of the eccentricity is greater than 0.2, we identify the planet as a high-$e$ planet and have it colored in orange; otherwise, we have it colored in blue. We have 10 candidates identified as high-$e$ planets. \changes{All 79 confirmed Warm Jupiters (defined here as planets larger than 6 Earth radii with orbital periods of 8--200 days) are plotted in grey dots, as of March 2021 from NASA Exoplanet Archive.}}
    \label{fig:ecc1}

    \bigskip
    
    \hspace*{-0.5cm}
    
    \includegraphics{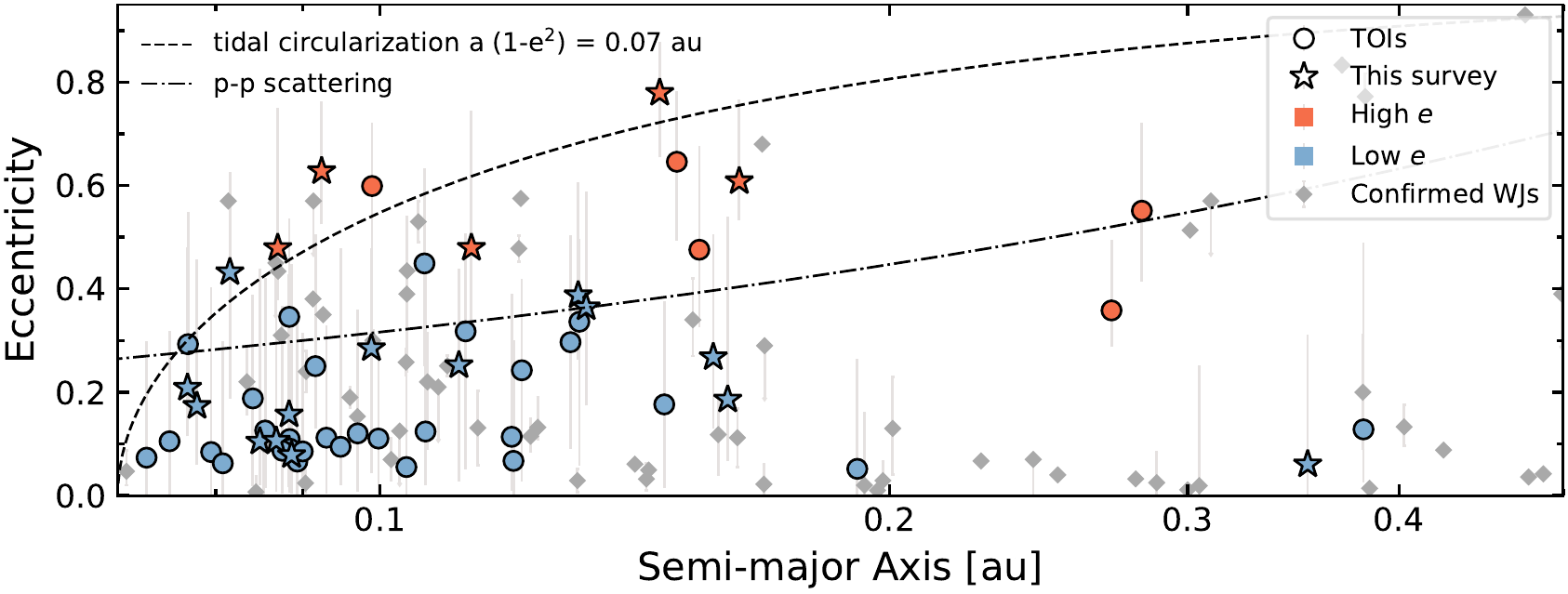}
    \caption{Eccentricity versus semi-major axis of the \changes{55} Warm Jupiter candidates discovered in Year 1 \TESS FFIs around host stars brighter than Tmag of 12. The eccentricities are inferred from the candidates' stellar densities and presented by their posterior modes and the 68\% highest posterior density (HPD) intervals. We color a collection of candidates that are possibly on highly elliptical orbits in orange, according to their eccentricities and HPD intervals (similar to Figure~\ref{fig:ecc1}). We present two reference curves to demonstrate possible Warm Jupiter origin channels. The tidal circularization line (i.e., the dashed line; $a_{\mathrm{final}} = a (1-e^2) = 0.07$ au) illustrates one possible formation pathway of a Warm Jupiter under high-$e$ migration. Above the line, the planet's orbit will shrink to a semi-major axis of 0.07 au or smaller. The planet-planet scattering line (i.e., the dot-dash line) demonstrates the approximate maximum eccentricity to which a planet could be excited by nearby companions without undergoing collisions and having its eccentricity damped. The line can be understood as the maximum eccentricity one would expect for multiple Warm Jupiters formed in situ.}
    \label{fig:ecc2}
\end{figure*}

We characterize the eccentricities of \changes{55} Warm Jupiter candidates using the ``photoeccentric" effect discussed in Section \ref{subsec:post_analysis}. Given the asymmetrical distribution of the eccentricity posteriors and the bimodal distribution of the argument-of-periapse posteriors, we report their modes and the 68\% highest posterior density (HPD) intervals instead of the medians and 68\% quantiles. In Figure \ref{fig:ecc1}, we present the eccentricity-versus-orbital period of these candidates. Each candidate is labeled by the mode of its eccentricity posterior and the grey errorbar indicates the 68\% HPD. As shown in Figure~\ref{fig:ecc1}, we identify a collection of Warm Jupiter candidates that are possibly on highly elliptical orbits and have them colored in orange. To identify these high-$e$ candidates, we use the criterion of whether the lower bound of the 95\% HPD of the eccentricity posterior is greater than 0.2. The criterion here is not driven by theoretical models but to prioritize a list of targets for ground-based follow-up observations. The 10 targets are
\begin{itemize}
    \item TIC-350618622 (TOI-201.01; Hobson et al. submitted)
    \item TIC-130415266 (TOI-588.01)
    \item TIC-280206394 \citep[TOI-677b;][]{jord20}
    \item TIC-147660886 (TOI-2005.01)
    \item TIC-24358417 (TOI-2338.01)
    \item TIC-290403522
    \item TIC-395113305
    \item TIC-180989820
    \item TIC-464300749
    \item TIC-343936388.
\end{itemize}

In Figure \ref{fig:ecc2}, we plot the eccentricity versus the semi-major axis of the candidates to demonstrate possible Warm Jupiter formation pathways. 
If Warm Jupiters are formed at large semi-major axis and migrate inwards via high-eccentricity tidal migration, they will follow a tidal circularization track of constant angular momentum, such as the dashed line shown in Figure~\ref{fig:ecc2}. During the process of the high-$e$ migration, the planet loses orbital energy ($E_{\textrm{p}}=-G M_\star m_{\textrm{p}}/2a$) due to the tidal dissipation in the planet and thus shrinks its orbit. The angular momentum ($L_{\textrm{p}} = m_{\textrm{p}}\sqrt{G M_\star a (1-e^2)}$), however, stays roughly the same. As a result, the planet reduces its eccentricity as its semi-major axis gets smaller, following an evolution track of $a_{\mathrm{final}} = a(1-e^2)$, where $a_{\mathrm{final}}$ is a constant and the semi-major axis the planet ends up with by the end of the migration (considering tidal dissipation in the planet only). The tidal circularization timescale has a strong dependence on $a_{\mathrm{final}}$ (i.e., $\tau \propto a_{\mathrm{final}}^8$; \citealt{eggl89}). Generally, planets with $a_{\mathrm{final}} < 0.05$ au are likely to get circularized in a star's lifetime, whereas planets with $a_{\mathrm{final}} > 0.1$ au are unlikely to do so. Because of the uncertainty in the tidal dissipation efficiency and thus on the tidal circularization timescale, some of our shortest period, circular-orbit warm Jupiters could have have undergone complete tidal circularization. The dashed line in Figure \ref{fig:ecc2} has a final semi-major axis of 0.07 au, an illustrative value for the critical final semi-major axis. Warm Jupiter on and above the dashed line are experiencing high-$e$ migration and will have their final orbits shrunk to semi-major axes of 0.07 au or smaller. Planets below the dashed line could also experience high-$e$ migration if they are coupled to outer companions and undergo eccentricity oscillations (\citealt{socr12}; see \citealt{jack19} for a case study and \citealt{kane14} for an investigation of secular oscillations in radial-velocity discovered giant planet systems). 
If Warm Jupiter are instead formed in situ or arrived via disk migration, we expect to observe them at low eccentricities. The dot-dashed line in Figure~\ref{fig:ecc2} presents the approximate maximum eccentricity for which a planet could be excited via in situ planet-planet scattering \citep[Equation 10 in][]{daws18, petr14, ande20}. We assume a mass of 0.5 $M_{\textrm{Jup}}$ and a radius of 2 $R_{\textrm{Jup}}$ as illustrative values for young planets. Planets below the dot-dashed line are consistent with in situ formation and disk migration. However, \cite{ande17} pointed out Warm Jupiters formed with low eccentricities could have their eccentricities excited by outer companions via secular interactions. In such scenarios, Warm Jupiters formed in situ or via disk migration can be observed above the dot-dashed line.

\begin{figure}
  \centering
  \hspace*{-0.4cm}
  \includegraphics[scale=0.95]{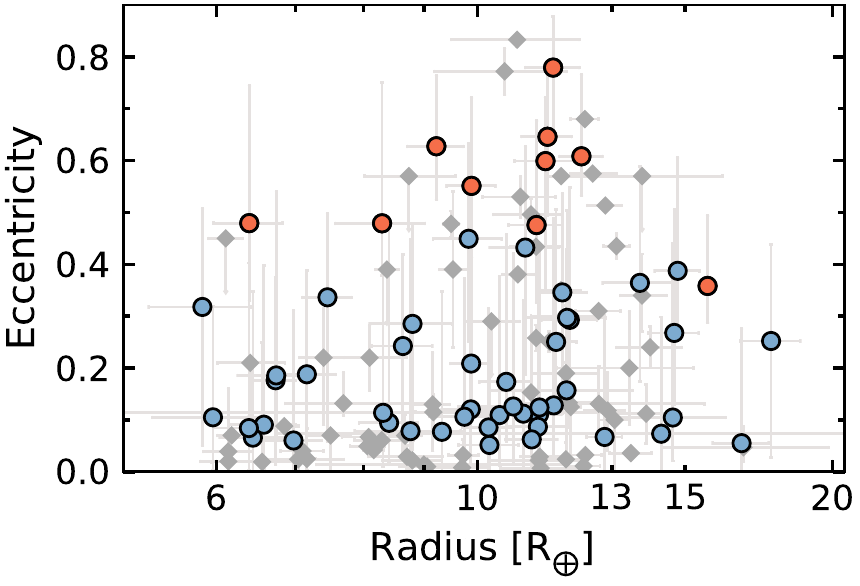}
  \caption{Eccentricity versus planet radius in Earth radii for the \changes{55} Warm Jupiter candidates. The eccentricities are inferred from the candidates' stellar densities and presented by their posterior modes and the 68\% highest posterior density (HPD) intervals. The candidates are categorized into a high-$e$ population colored in orange and a low-$e$ population colored in blue, given their eccentricities and HPD intervals (similar to Figure~\ref{fig:ecc1} and \ref{fig:ecc2}). The major of the high-$e$ candidates (colored in orange) have a planetary size of 8--13 Earth radii.}
  \label{fig:e_vs_r}
\end{figure}

We investigate possible correlations between planets' radii and eccentricities, as shown in Figure~\ref{fig:e_vs_r}. We find the majority of Warm Jupiter candidates on highly elliptical orbits have a size between 8 and 13 Earth radii (for reference, Jupiter is $\sim$11 $R_\earth$). For both small (i.e, $R_p < 8 R_\earth$) and large ($R_p > 12 R_\earth$) planets, we identify a lack of high-$e$ planets. One possible explanation of the lack of high-$e$ planets for the large planets ($R_p > 12 R_\earth$) is that they are inflated by some heating sources, such as tidal heating from the planetary tidal dissipation \citep[e.g.,][]{bode01}, stellar heating from the stellar irradiation \citep[e.g.,][]{guil02}, and thermal tides caused by asynchronous rotation and orbital eccentricity \citep[e.g.,][]{arra10}, and then tidally circularized by their host stars. However, all these theories require a small periapse of the planet, whereas many of our large Warm Jupiters have long orbital periods ($P>8$ days; some even have $P>15$ days) and low eccentricities.
Other possibilities are that the stellar radii could have errors unaccounted for in the uncertainties or the large planets could instead be low mass stars. which might have a different dynamical history and/or circularization distance. For the small planets (i.e, $R_p < 8 R_\earth$), without a proper mass measurement, we cannot tell whether or not they are giant planets or super puff planets. If they are super puff planets instead, the lack of high-$e$ planets could be explained by a different origin channel(s) for low mass planets. Our observations could suffer from small number statistics since only 6/\changes{55} candidates have a mode eccentricity greater than 0.5. Moreover, astrophysical false-positives, e.g., eclipse binaries and brown dwarfs, could contaminate our sample of warm Jupiters and compromise our interpretations. More investigation, especially mass measurements of small and large planets, is required to draw any firm conclusions about the apparent correlation with planet size.

\subsection{Additional candidates}
\label{subsec:catalog_more}
In Table \ref{tbl:pc_missing}, we introduce 11 additional Warm Jupiter candidates that are not included (i.e., Table \ref{tbl:pc}) due to unconstrained ephemerides and eccentricities. The missing information is itemized in Table \ref{tbl:pc_missing}. Eight of these candidates have unconstrained orbital periods due to observation gaps in the \TESS data. Ground-based follow up or the \TESS Extended Mission is likely to resolve their orbital period degeneracy. Four of them have missing stellar parameters and/or \emph{Gaia} parallax according to the TIC-v8 catalog. The missing stellar parameters lead to unconstrained planet eccentricity.

In addition, we list 19 more \textit{possible} Warm Jupiter candidates that are not included due to poorly constrained impact parameters and thus the planet-star radius ratios. The transit duration of these candidates is short and their light curves are usually \emph{v}-shaped. The posterior distributions of the impact parameter show a flat distribution between 0 and 1 with a long tail above 1. Because of the existence of the $b>1$ solution, planet radii could approach the substellar object regime for these candidates. We do not include these candidates in our catalog to minimize false positives. Still, we list their TIC IDs here:
\begin{itemize}
    \item 4588737
    \item 38815574
    \item 76761591
    \item 117817010
    \item 304418238 
    \item 305345992 
    \item 359271114
    \item 359732062
    \item 379717270
    \item 381982417 
    \item 384164973
    \item 410395660 
    \item 412484721
    \item 460205581 \citep[TOI-837.01;][]{boum20}
    \item 461968320
    \item 462162948 (TOI-684.01)
    \item 464124454 
    \item 467807015
    \item 467807116.
\end{itemize}
Finer cadence photometric observations (e.g., from the \TESS Extended Mission) may improve the constraints on the impact parameters and make some of the candidates qualified for the catalog.

\section{The eccentricity distribution}
\label{sec:ecc_dist}
Different origin channels (i.e., in situ formation, disk/high-$e$ migration) make different predictions for Warm Jupiters' eccentricities. To shed light on which one or more origin channels predominantly contribute to the Warm Jupiter population, we characterize the eccentricity distribution of our catalog using hierarchical Bayesian modeling (HBM). The philosophy of HBM is that each Warm Jupiter is a member of a specific population, and that members of each population share properties in common. Consequently, individual members reflect the properties of the population and the population helps to make better inferences of the properties of individual members. Studying warm Jupiters as a catalog both provides information on the eccentricity distribution and improves the inference of the eccentricity of a single planet.

Transit durations can serve a proxies for the eccentricities of transiting planets and can be used to infer a population-wide eccentricity distribution under the assumption of a uniform distribution of impact parameters (e.g., as performed by studies such as \citealt{moor11,kane12} for {\it Kepler} planet candidates). For planets with large SNR transits, transit shapes are well resolved and we may constrain the impact parameters and their $\rho_{\textrm{circ}}$ assuming circular orbits. With both well-constrained $\rho_{\textrm{circ}}$ and $\rho_\star$ (the true stellar density), we can constrain the planets' individual eccentricities using the ``photoeccentric" effect \citep{kipp14a}. For example, \cite{daws15} applied the approach to search for super-eccentric Warm Jupiters in the {\it Kepler} sample and \cite{vinc19} used the approach to constrain the eccentricity distribution of small {\it Kepler} planets. Here we apply the method to a large sample of \TESS Warm Jupiters. We note one caveat of this study is that although we have removed several targets that have been dispositioned as FPs by TFOP groups, many targets in our sample are still planet candidates that have not yet been confirmed. The contamination of astrophysical FPs could compromise our interpretation of Warm Jupiters' eccentricity distribution.

We examine three functional forms of the eccentricity distribution, including a Beta distribution, a Rayleigh distribution, and a mixture distribution with two Gaussian components. Both the Beta distribution and the Rayleigh distribution have been broadly used in exoplanet eccentricity distribution studies (e.g., \citealt{exoplanet:kipping13} on all radial-velocity discovered planets; \citealt{fabr14} and \citealt{he19} on {\it Kepler} systems; \citealt{shab16} on short-period {\it Kepler} planet candidates). The Beta distribution is known for its flexibility in shape and is bounded in $[0, 1]$, which can be conveniently adopted for the eccentricity distribution. The Rayleigh distribution is motivated by the planet-planet scattering origin of Warm Jupiters. \cite{ida92} find the Rayleigh distribution is a good descriptor for the eccentricity distribution generated by planet-planet scattering. We also introduce the two-component mixture distribution, inspired by the low-$e$ and high-$e$ populations predicted by Warm Jupiter's origin theories. The disk migration and in situ origin channels predict Warm Jupiters on circular or moderately elliptical orbits, whereas the high-$e$ tidal migration origin channel predicts a group of Warm Jupiters on highly elliptical orbits. The two-component model is flexible enough to learn the fractions and expected eccentricities of the low-$e$ and high-$e$ populations of the sample.

\begin{figure}
  \centering
  \resizebox{3in}{!}{%
    \begin{tikzpicture}[thick]
          \node[input]                                         (ocirc) {$\hat{\rho}_{\mathrm{circ},i}$, $\sigma_{\hat{\rho}_{\mathrm{circ},i}}$};
          \factor[above=0.6 of ocirc] {ocirc-f} {left:$\mathcal{N}$} {} {}; %
        
          \node[det, above=1.2 of ocirc, xshift=0cm]           (rhocirc) {$\rho_{\mathrm{circ},i}$};
          
          \node[input, above=0.5 of ocirc, xshift=3cm] (trans) {obs$_i$};
          
          \node[latent, above=0.9 of rhocirc, xshift=0cm]      (e) {$e_i$};
          \node[latent, above=0.9 of rhocirc, xshift=2cm]      (w) {$\omega_i$};
          \node[latent, above=0.9 of rhocirc, xshift=-2cm]     (rhostar) {$\rho_{\star,i}$};
          \node[input, above=0.9 of rhocirc, xshift=4cm]  (P) {$P_i$};
        
          \node[latent, above=1.5 of e, xshift=-0.5cm] (ae) {$\alpha_e$} ; %
          \node[latent, above=1.5 of e, xshift=0.5cm]  (be) {$\beta_e$} ; %
          
          \node[input, above=-2.8 of rhostar, xshift=0cm] (star) {$\hat{\rho}_{\star,i}$, $\sigma_{\hat{\rho}_{\star,i}}$} ; %
          \factor[above=-1.4 of rhostar] {rhostar-f} {left:$\mathcal{N}$} {} {} ; %
        
          \edge {rhocirc} {ocirc} ;
          \edge {e, w, rhostar} {rhocirc};
          
          \factor[above=0.5 of e] {e-f} {left:Beta} {ae,be} {e} ; %
          \factoredge {rhostar} {rhostar-f} {star} ; %
          \factor[above=0.5 of trans]{trans-f} {right:$P_{\textrm{transit}}$}{e, w, P, rhostar} {trans};
        
          \node[const, left=1.9cm of ocirc] (a) {$\textcolor{white}{.}$} ;
          \node[const, right=3.5cm of ocirc] (b) {$\textcolor{white}{.}$} ;
          \node[const, above=5.3cm of ocirc, xshift=2.7cm] (c) {$\textcolor{white}{.}$} ;
        
          \plate {} {(ocirc)(rhocirc)(e)(w)(rhostar)(a)(b)(c)} {$i$ of $N$} ;
    \end{tikzpicture}
}
  \caption{Graphic model of the hierarchical Bayesian model using the Beta distribution as the functional form of the eccentricity distribution. Outside the plate, we have hyperparameters, $\alpha_e$ and $\beta_e$, describing the eccentricity distribution of the Warm Jupiter population. Inside the plate, we have individual parameters for each of the $N$ planets. Each planet $i$ has parameters, $\rho_{\star,i}$, $e_i$ and $\omega_i$, to be inferred from the model, labeled in circles. $\rho_{\textrm{circ},i}$ is a deterministic parameter that can be directly calculated from $\rho_{\star,i}$, $e_i$ and $\omega_i$ and is labeled in diamond. The observed parameters, $\hat{\rho}_{\textrm{circ},i}$, $\sigma_{\hat{\rho}_{\textrm{circ},i}}$, $\hat{\rho}_{\star,i}$, $\sigma_{\hat{\rho}_{\star,i}}$, and $\textrm{obs}_i$, are labeled in lavender boxes. The planet orbital period, $P_i$, is an input parameter and taken as a constant in the model, similar to the other observed parameters.
  }\label{fig:hbm}
\end{figure}
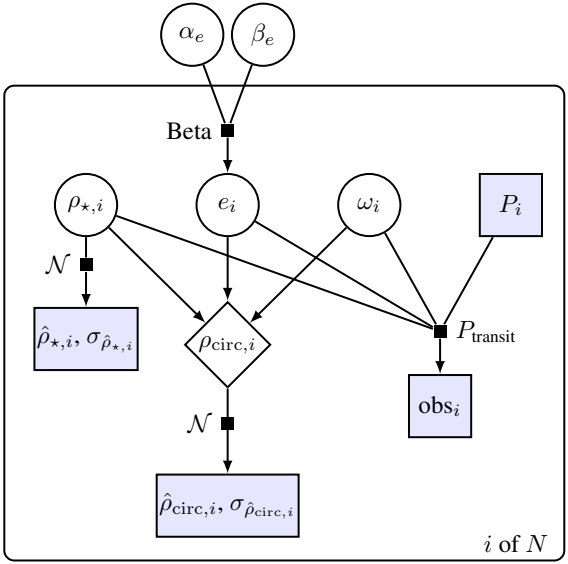

We first introduce our HBM framework using the Beta distribution. The frameworks for the Rayleigh and mixture distributions are similar except with minor changes on hyperparameters. Using the Beta distribution as the functional form of the eccentricity distribution, we build the hierarchical model by introducing two hyperparameters, $\alpha_e$, and $\beta_e$. The probability distribution follows
\begin{equation}
    p(e|\alpha_e,\beta_e)=e^{\alpha_e-1}(1-e)^{\beta_e-1}/B(\alpha_e,\beta_e),
\end{equation}
where $B(\alpha_e,\beta_e)=\Gamma(\alpha_e)\Gamma(\beta_e)/\Gamma(\alpha_e+\beta_e)$ and $\Gamma$ is the Gamma function. In Figure \ref{fig:hbm}, we display the Bayesian model using a directed factor graph. As shown in Figure \ref{fig:hbm}, the hyperparameters $\alpha_e$ and $\beta_e$ are outside the plate. Inside the plate, we have $N$ planets and each planet has one set of the parameters listed in the graph. For each Warm Jupiter, we have two sets of observed parameters from previous analysis, the posterior distributions of stellar densities from isochrone fitting $\rho_{\star}$ and the posterior distributions of stellar densities from light curve fitting $\rho_{\mathrm{circ}}$, assuming all planets have circular orbits. The stellar density posteriors and the transit probabilities are observed parameters, colored in lavender and labeled in boxes in Figure \ref{fig:hbm}. We approximate the stellar density posterior distributions as normal distributions to simplify the model and reduce the computational effort. To relieve the concern that some stellar density posteriors are not nearly Gaussian, we compare the inferred eccentricities (and argument of periapsis) of individual targets using Gaussian-approximated posteriors to the inferred eccentricities using the ``true" posteriors (e.g., \citealt{ford08, kipp12}; see \citealt{daws12} for a detailed description of the method).
The values inferred using two types of posteriors are in good consistency with differences well below 1$\sigma$ uncertainties. The planet's orbital period $P_i$ is also taken as an observed parameter in the model to calculate the transit probability. The relationship between $\rho_{\textrm{circ},i}$ and $\rho_{\star,i}, e_i, \omega_i$ is deterministic, so we label it using a diamond.
We write down the posterior distribution of our hierarchical model as
\begin{align}
p(\bm{\theta},\bm{\beta}|\bm{X}) &\propto p(\bm{X}|\bm{\theta},\bm{\beta})p(\bm{\theta}|\bm{\beta})p(\bm{\beta}) \nonumber\\
&\propto \prod_{i=1}^N\Big\{p(\hat{\rho}_{\textrm{circ},i}, \sigma_{\hat{\rho}_{\textrm{circ},i}}|\rho_{\textrm{circ},i})
p(\hat{\rho}_{\star,i},\sigma_{\hat{\rho}_{\star,i}}|\rho_{\star,i}) \nonumber\\
&\qquad \quad p(\textrm{obs}_i|\rho_{\star,i},e_i,\omega_i,P_i)\Big\} \nonumber\\
&\quad \times \prod_{i=1}^N\Big\{p(e_i|\alpha_e,\beta_e)p(\rho_{\star,i})p(\omega_i)\Big\} \nonumber\\ 
&\qquad \times p(\alpha_e)p(\beta_e),
\end{align}
where $\bm{X}=\{\hat{\rho}_{\textrm{circ},i}, \sigma_{\hat{\rho}_{\textrm{circ},i}}, \hat{\rho}_{\star,i}, \sigma_{\hat{\rho}_{\star,i}}, \textrm{obs}_i\}$ for the observed parameters, $\bm{\theta}=\{\rho_{\textrm{circ},i}, \rho_{\star,i}, e_i, \omega_i\}$ for the individual parameters of each planet, and $\bm{\beta}=\{\alpha_e, \beta_e\}$ for the hyperparameters of the population. $P_\textrm{transit}$ in the model describes the transit probability of a planet given its $\rho_{\star,i},e_i,\omega_i,P_i$. The planet's orbital period $P_i$ is taken from the light curve modeling as a fixed value.
Here we list the probability distributions that we assume for the hierarchical Bayesian model.
\begin{align} \label{statmod}
p(\hat{\rho}_{\textrm{circ},i}, \sigma_{\hat{\rho}_{\textrm{circ},i}}|\rho_{\textrm{circ},i}) & \sim \mathcal{N}(\rho_{\textrm{circ},i}|\hat{\rho}_{\textrm{circ},i}, \sigma^2_{\hat{\rho}_{\textrm{circ},i}}) \nonumber\\
p(\hat{\rho}_{\star,i},\sigma_{\hat{\rho}_{\star,i}}|\rho_{\star,i}) & \sim \mathcal{N}(\rho_{\star,i}|\hat{\rho}_{\star,i},\sigma^2_{\hat{\rho}_{\star,i}})\nonumber\\
p(\textrm{obs}_i|\rho_{\star,i},e_i,\omega_i,P_i) & \sim \begin{cases} 
\frac{1}{\rho_{\star,i}^{1/3} P_i^{2/3}} \frac{1+e_i \sin{\omega_i}}{1-e_i^2} & \mbox{if } e_i < e_{\textrm{max},i} \\
0 &\mbox{if } e_i \ge e_{\textrm{max},i} \end{cases} \nonumber\\
p(e_i|\alpha_e,\beta_e) & \sim \mathcal{B}\emph{eta}(e_i|\alpha_e,\beta_e)\nonumber\\
p(\rho_{\star,i}) & \sim \mathcal{N}(\hat{\rho}_{\star,i},\sigma^2_{\hat{\rho}_{\star,i}}) \nonumber\\
p(\omega_i) & \sim \mathrm{Uniform}(0,2\pi) \nonumber\\
p(\alpha_e) & \sim \mathrm{Uniform}(0,10)\nonumber\\
p(\beta_e) & \sim \mathrm{Uniform}(0,10)
\end{align}
$\mathcal{N}$ indicates a Normal distribution. The $e_{\textrm{max},i}$ is the maximum eccentricity a planet can reach without getting tidally disrupted (i.e., $e_{\textrm{max},i} = 1-R_{\star,i}/a_i$). The $p(\textrm{obs}_i|\rho_{\star,i},e_i,\omega_i,P_i)$ is the transit probability to correct the observation biases in the eccentricities of transiting planets \citep[Equation 9 in][]{winn10, burk08, kipp14b}.

Two more functional forms of the eccentricity distribution are examined: the Rayleigh distribution and the mixture distribution. The Rayleigh distribution can be written as
\begin{equation}
    p(e|\sigma_e)=\frac{e}{\sigma_e} \exp{(-e^2/2\sigma^2_e)},
\end{equation}
where $\sigma_e$ is a free hyperparameter.\footnote{The Rayleigh distribution is a special form of the Weibull distribution, $f(e|\alpha, \beta) = \alpha e^{\alpha-1} \exp{(-(e/\beta)^\alpha)/\beta^\alpha}$, where $\alpha=2$ and $\beta=\sqrt{2}\sigma_e$.} We use $p(\sigma_e) \sim \mathrm{Uniform}(0,10/\sqrt{2})$ as the hyperprior for $\sigma_e$.
For the mixture distribution, we use two Gaussian distributions to describe the low-$e$ and high-$e$ populations of Warm Jupiters, respectively. The probability distribution can be written as
\begin{align}
    p(e|f_1, f_2, & \mu_1, \mu_2, \sigma_1, \sigma_2) = \nonumber\\ 
    & f_1\mathcal{N}(e|\mu_1,\sigma_1^2)+f_2\mathcal{N}(e|\mu_2,\sigma_2^2),
\end{align}
where $f_1$ and $f_2$ present the fractions of two distributions and have the sum of 1. For both the Rayleigh distribution and the mixture distribution, we bound the eccentricity distribution between 0 to 1. In the mixture model, we require both $f_1$ and $f_2$ to be greater than 0.05 to avoid divergence of the model. We restrict $\mu_1$ and $\mu_2$ to be ordered to avoid swaps between modes (i.e., label switching). A full description of the hyperpriors in the mixture model is shown below:
\begin{align}
p(f_1) & \sim \mathrm{Uniform}(0,1) \nonumber\\
p(f_2) & = 1-f_1 \nonumber\\
p(\mu_1), p(\mu_2) & \sim \mathrm{Uniform}(0,1) \nonumber\\
p(\sigma_1), p(\sigma_2) & \sim \mathrm{HalfCauchy}(1) \nonumber\\
p(\mbox{prior1}|f_1, f_2) & =
\begin{cases}
    0 &\mbox{if  } f_1, f_2 < 0.05 \\
    1 &\mbox{otherwise} 
\end{cases} \nonumber\\
p(\mbox{prior2}|\mu_1, \mu_2) & =
\begin{cases}
    0 &\mbox{if  } \mu_1 > \mu_2 \\
    1 &\mbox{otherwise} 
\end{cases} \nonumber\\
p(\mbox{prior3}|\sigma_1, \sigma_2) & =
\begin{cases}
    0 &\mbox{if  } \sigma_1, \sigma_2 \notin [0,1] \\
    1 &\mbox{otherwise} 
\end{cases} \nonumber\\
p(\mbox{prior4}|\mu_1, \sigma_1) & = \mbox{CDF}[\mathcal{N}(\mu_1,\sigma_1^2)=1, \mathcal{N}(\mu_1,\sigma_1^2)=0] \nonumber\\
p(\mbox{prior5}|\mu_2, \sigma_2) & = \mbox{CDF}[\mathcal{N}(\mu_2,\sigma_2^2)=1, \mathcal{N}(\mu_2,\sigma_2^2)=0], \nonumber\\
\end{align}
where $\mbox{CDF}[a, b] = \mbox{CDF}[a] - \mbox{CDF}[b]$. Prior 1 avoids the fraction of either Gaussian components to be zero to cause divergence. Prior 2 sets $\mu_1$ and $\mu_2$ to be ordered. Prior 3 bounds $\sigma_1$ and $\sigma_2$ between 0 and 1. Lastly, prior 4 and 5 bound the eccentricity distribution between 0 to 1.

\begin{figure}[htb!]
    \centering
    \hspace{-0.2cm}
    \includegraphics{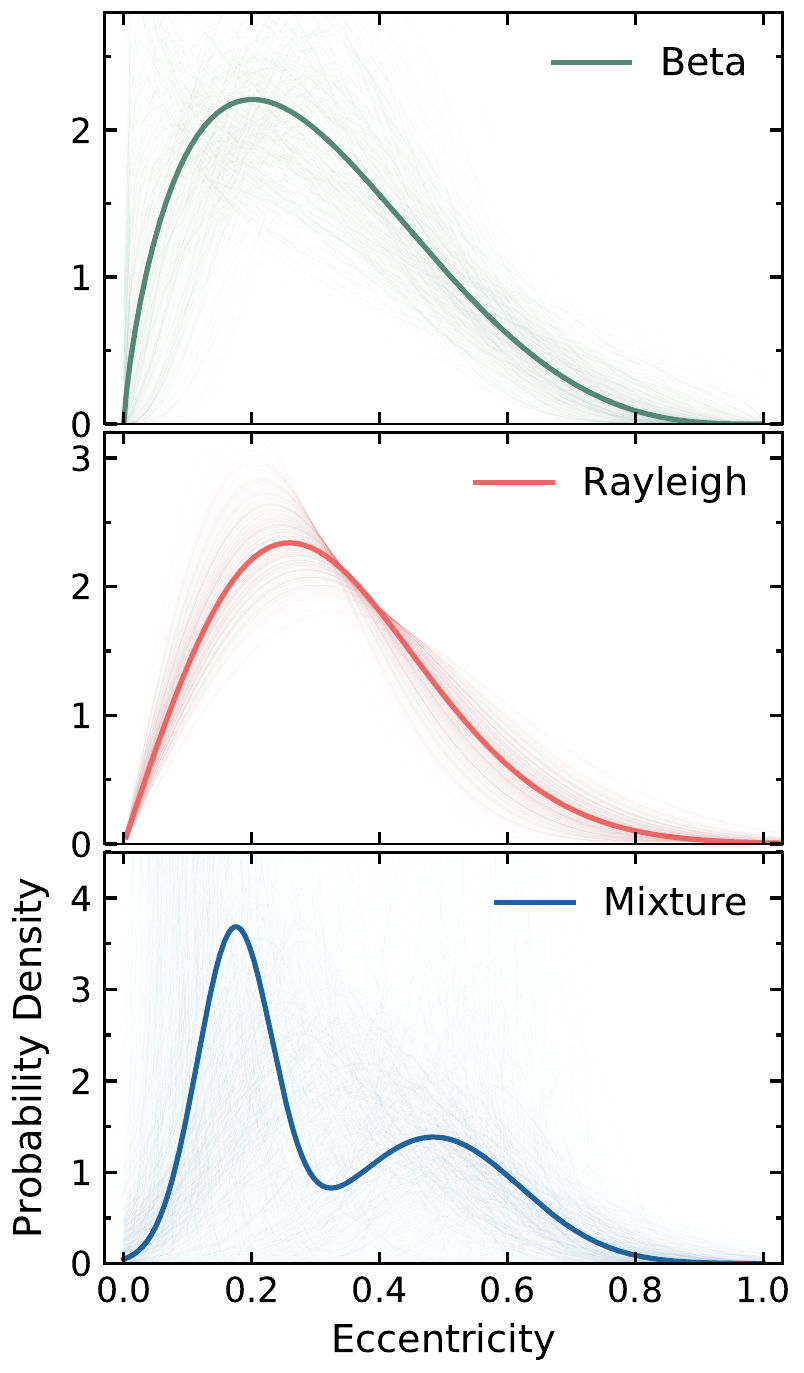}
    \caption{The eccentricity distributions of the catalog of \changes{55} Warm Jupiter candidates discovered in Year 1 \TESS FFIs (Table~\ref{tbl:pc}). The eccentricity distributions are inferred using hierarchical Bayesian modeling with the Beta distribution (upper panel), the Rayleigh distribution (middle panel), and the two-component mixture distribution (lower panel) as the functional forms. Planets' eccentricities inferred from the Beta distribution are generally lower than the eccentricities inferred from the Rayleigh distribution. The mixture model splits roughly $55\%$ of the candidates into a low-$e$ population centered at 0.16 and the rest $45\%$ into a high-$e$ population centered at 0.49.}
    \label{fig:hbm_all}
\end{figure}

\begin{deluxetable}{clll}
\tablecaption{Summary of the posteriors of the hyperparameters for hierarchical Bayesian models. \label{tbl:hbm_post}}
\tabletypesize{\small}
\tablehead{\colhead{Distribution} & \multicolumn{3}{l}{Hyperparameter posteriors}} 
\startdata
Beta & $\alpha_e$ & $\beta_e$ \\
& $1.776^{+1.385}_{-0.771}$ & $4.082^{+2.647}_{-1.634}$ \vspace{0.5em} \\
\hline
Rayleigh & $\sigma_e$ \\
& $0.259^{+0.032}_{-0.029}$ \vspace{0.5em}\\
\hline
Mixture & $f_1$ & $\mu_1$ & $\sigma_1$ \\
& $0.530^{+0.227}_{-0.209}$ & $0.174^{+0.114}_{-0.088}$ & $0.059^{+0.072}_{-0.043}$\\
& $f_2$ & $\mu_2$ & $\sigma_2$ \\
& $0.470^{+0.209}_{-0.227}$ & $0.484^{+0.107}_{-0.095}$ & $0.135^{+0.063}_{-0.071}$\\
\enddata
\tablecomments{We report the medians and 68\% credible intervals of the posteriors.}
\end{deluxetable}

We build the hierarchical Bayesian models and sample posteriors using $\mathtt{PyMC3}$ \citep{exoplanet:pymc3}.\footnote{All codes used in this project are available from J.D. upon request.} We sample 4 parallel chains, each chain with 40,000 tuning steps and 10,000 draws. A target accept rate of 0.99 is used to avoid divergences. The MCMC convergence is evaluated by the Gelman-Rubin diagnostic (i.e., $\hat{\mathcal{R}} < 1.1$ for convergence), trace plots, and corner plots \citep{corner} of the marginal joint distributions. The posterior distributions and covariances of the hyperparameters for the Beta-distribution and the mixture model can be found in Figure~\ref{fig:corner_beta} and \ref{fig:corner_mixture}. A summary of best fit models is shown in Table~\ref{tbl:hbm_post}.

The eccentricity distributions assuming the three functional forms discussed above are shown in Figure~\ref{fig:hbm_all}. In each panel, we plot the best-fitting model in dark line by calculating the distribution using the medians of the posteriors of the hyperparameters, along with 500 draws from the posteriors. Comparing to the Rayleigh distribution, the Beta distribution infers lower eccentricities for the population. A similar distinction was also found by \cite{shab16} when the author did a comparison of the Beta and Rayleigh distribution for their sample of Kepler hot Jupiters. The reason might be the Beta distribution has more flexible shapes compared to the Rayleigh distribution. For our sample of \TESS warm Jupiters, the eccentricity distribution spreads between 0 and 0.8 and peaks at \changes{$\sim$0.19} for the Beta distribution, whereas it peaks at \changes{$\sim$0.26} for the Rayleigh distribution. For the two-component mixture model, the eccentricity distribution is well constrained by a low-$e$ population centered at \changes{$\bar{e} = 0.174^{+0.114}_{-0.088}$ and a high-$e$ population centered at $\bar{e} = 0.484^{+0.107}_{-0.095}$}. About $53 \pm 20$\% of the systems fall into the low-$e$ population and $47 \pm 20$\% of the systems fall into the high-$e$ population. The corresponding widths of two distributions are $\sigma_1 = 0.059^{+0.072}_{-0.043}$ for the low-$e$ population and $\sigma_2 = 0.135^{+0.063}_{-0.071}$ for the high-$e$ population. Naively speaking, the fraction of the systems categorized into the low-$e$ population ($\sim$53\%) could indicate the fraction of Warm Jupiters originated from the disk migration and in situ formation; the fraction of the systems categorized into the high-$e$ population ($\sim$47\%) could indicate the fraction of Warm Jupiters originated from the high-eccentricity tidal migration. However, such a conclusion overlooks the evolution of Warm Jupiters' eccentricities, as discussed in Section~\ref{sec:results}. To correctly interpret the eccentricity distribution of a two-component mixture model, we will have to compare it with population predictions from different theorized origin channels of Warm Jupiters.

While we tested three functional forms for this distribution, we do not make any strong quantitative claims about the relative performance of each. Each model is qualitatively consistent with the others within the uncertainties and, given the size of the dataset and the dimensionality of the model, a formal model comparison (e.g., using the nested sampling algorithm employed by \citealt{kipp14b} to characterize the eccentricity distribution of radial velocity planets) would be computationally expensive and may not provide strong evidence for one model. We do find that the weights for each component of the mixture model are significantly inconsistent with zero, providing weak evidence that the two-component model is required to capture the distribution. Lastly, as a robustness test for the eccentricity distribution, we tried randomly dropping 30\% of the targets from our sample and conduct the same analysis to the new sample. We found the results are consistence within 1-$\sigma$ uncertainties.

\section{Discussion} \label{sec:discussion}
Here we discuss our Warm-Jupiter-candidate catalog in the context of the \TESS Extended Mission and ground-based follow-up observations. We also discuss the implication of our study on Warm Jupiter origins.

\subsection{TESS Extended Mission}
We identified \changes{55} Warm Jupiters candidates in the Year 1 \TESS FFIs. Many of these candidates will be revisited (at the time of this writing) during the \TESS Extended Mission (e.g., in Year 3 data). The longer observing baseline will help with the TTV analysis and identify any additional planets in the system. About 20 out of the \changes{55} candidates only have two transits observed during the \TESS Prime Mission. The \TESS Extended Mission will likely catch two more transits on these targets to allow some preliminary TTV analysis. There is also a group of candidates with possible TTV signals (Figure~\ref{fig:ttv1} and \ref{fig:ttv2}). The extended mission data will allow us to examine the robustness of these signals. Besides, the finer observing cadence of the extended mission data (i.e., from 30-minute to 10-minute) will improve the impact parameter characterization and the eccentricity analysis. Particularly, the 19 \emph{possible} Warm Jupiter candidates that are not selected in our catalog could be included if we can rule out the $b>1$ solution with the fine-cadence data. We also identify 8 candidates with degenerate orbital periods listed in Table~\ref{tbl:pc_missing}. The \TESS extended mission data is likely to break the degeneracy of the orbital periods.

Most of our Warm Jupiter candidates have orbital periods less than 20 days due to their short observing baselines (i.e., 27 or 54 days). Combing the \TESS Prime and Extended Mission data will allow the discovery of Warm Jupiters in the longer orbital period range, especially the targets identified as single-transit events in the Year 1 data. 

\subsection{Ground-based follow-up observations} \label{subsec:ground}
Ground-based follow-up observations are essential to validate planet candidates in our catalog. Due to the coarse angular resolution of \TESS, ground-based photometric follow-up will be required to confirm on-target transits. If the target has no aperture contamination, radial velocity follow-up will further rule out brown dwarfs or low-mass stars. As discussed in Section~\ref{subsec:tfop}, the \TESS community has already made great progress on validating planet candidates and ruling out FPs in our catalog.

Here we propose a follow-up strategy for the Warm Jupiter candidates. First, we propose to prioritize follow-up of candidates showing evidence of high eccentricities. The high-$e$ planets play an important role in constructing the eccentricity distribution. Validation will ensure that their contribution to the population-wide eccentricity distribution is real and, moreover, confirm or rule out the high-eccentricity tidal migration scenario as one of the predominant origin channels contributing to the Warm Jupiter population. Radial velocity follow-up will further break the degeneracy between eccentricity and argument of periapse. A list of high-$e$ candidates that have not yet been confirmed (130415266, 147660886, 24358417, 290403522, 395113305, 180989820, 464300749, 343936388) can be found in Section~\ref{sec:results} and Table~\ref{tbl:pc}.
Second, we propose to prioritize candidates showing possible TTV signals. The large transits depths of our candidates (i.e., several parts per thousand) make them feasible for ground-based photometric follow-up observations. For example, in the case of TRAPPIST telescopes, transit events with depths greater than 2.5 parts per thousand will readily be detected. In general, candidates with $> 2\sigma$ TTV detections are worthwhile to follow up (see Figure~\ref{fig:ttv1}). We also listed \changes{four} targets (150299840, 382200986, 371188886, 343936388) showing possible sinusoidal TTV signals in Section~\ref{sec:results}. Extensive transit follow-up will help to detect any non-transiting, nearby planets, and shed light on the dynamical history of the system.
Lastly, we recommend to prioritize the remaining targets given their transmission spectroscopy metric \citep[TSM;][]{kemp18} for follow-up atmospheric characterization observations. In Table~\ref{tbl:pc} Column 13, we calculate a TSM score using the empirical mass–radius relationship of \cite{chen17} for each target on a scale of 1 to 100, with 100 being most favorable. The confirmation of these targets will help to select ideal candidates for Warm Jupiter atmospheric characterization for future missions (e.g., \emph{JWST}).

Follow-up observations on candidates with missing information listed in Table~\ref{tbl:pc_missing} are also important. Photometric and radial velocity follow-up observations will help to break the orbital period degeneracy due to \TESS observation gaps. Characterization on stellar densities will allow the photoeccentric analysis.  

\subsection{Implication of the eccentricity distribution}
We identified a large sample of Warm Jupiter candidates to conduct a preliminary eccentricity distribution study of the population using the ``photoeccentric" approach to constrain the eccentricity from the transit light curve. There are several caveats to our results. Although we have incorporated uncertainties in stellar density, systematic errors in stellar parameters could impact our inference of eccentricities. Since many of the candidates have not yet been statistically validated, our interpretation of the eccentricity distribution could also be compromised by astrophysical false-positives, e.g., binaries and brown dwarfs, which may have different dynamical histories and thus different eccentricity distributions from Warm Jupiters'. Due to the short \TESS-sector observing baseline, most of our candidates have orbital periods less than 20 days. The eccentricity distribution of longer period Warm Jupiters is not well addressed in this work.

In our preliminary study, we found both single-component models (i.e., the Beta distribution and the Rayleigh distribution) and a two-component model could be used as the functional forms of the eccentricity distribution. Our two-component model is flexible enough to describe both single-component and two-component distributions, benefiting from its five free parameters. The two-component eccentricity distribution showed that slightly more than half of the Warm Jupiters have nearly circular orbits in support of the disk migration and in situ origin scenarios; slightly less than half of the Warm Jupiters have moderately to highly elliptical orbits in support of high-eccentricity tidal migration. However, as discussed in Section~\ref{sec:ecc_dist}, evolution of planetary eccentricities could modify the shape of the distribution. A statistical study on a \emph{clean} Warm Jupiter catalog will require extensive ground-based follow-up observations and is deferred to future work. In future studies, the eccentricity distribution can be compared more directly with predictions from different origin theories \citep[e.g.,][]{ande20} to shed light on one or more origin channels that predominantly contribute to the Warm Jupiter population.

\changes{Benefiting from the extensive follow-up observations, a dozen of Warm Jupiter candidates in the catalog have been confirmed. In Appendix~\ref{appendix:hbm}, we demonstrate one approach to incorporate eccentricity measurements from different sources However, as to be discussed in Appendix~\ref{appendix:hbm}, further experiments would be necessary to robustly account for variations in selection effects.} 

We note the eccentricity distributions we found here are the \emph{observed} eccentricity distributions of a sample of \emph{transiting} Warm Jupiter candidates. While we have taken account of the transit probability to correct the observation biases for the eccentricity inferences, the detection efficiency of the transit search will also be a function of the eccentricity \citep[e.g., targets with high eccentricities will have a better chance to transit;][]{kipp14b}.
To find the \emph{intrinsic} eccentricity distribution, we will need to weight planets differently in the hierarchical Bayesian model according to their detection efficiency which will depend on its $\rho_\star, e, \omega$, and $P$. To more directly compare a model to the observations, in future work we can forward model the detection efficiency to characterize the eccentricity distribution of simulated detected transiting planets. The vetting efficiency is less of a concern since the large transit SNRs of giant planets make them readily vetted, and they are readily detected around our bright targets even if a high eccentricity shortens their transit duration.

\changes{It will also be interesting to see how the eccentricity distribution varies as a function of semi-major axis. In a preliminary study, we separated our candidates into two groups, $a < 0.1$ au (29 candidates) and $a > 0.1$ au (26 candidates), and inferred their eccentricity distributions. We found planets with greater semi-major axes have a higher median eccentricity compared to that of planets with smaller semi-major axes. This finding is consistent with theories since both the in-situ formation scenario that leads to planet-planet scattering and the high-$e$ migration scenario predict an increase of eccentricity as the semi-major axis increases. However, the tidal circularization effect also needs to be taken into account here since with $a < 0.1$ au, some planets could have been circularized in their systems' lifetime. The issue can be solved by increasing the number of semi-major axis bins if more Warm Jupiters at large semi-major axes are detected.}

\section{Summary}
\label{sec:summary}
We systematically searched for Warm Jupiter candidates, i.e., planets greater than 6 Earth-radii with orbital periods of 8--200 days, around host stars brighter than Tmag of 12 in Year 1 \TESS Full-Frame Images (Figure~\ref{fig:WaLEs_south}). We characterized each candidate's \TESS light curve with a quadratic limb darkening transit model along with Gaussian processes. For candidates with more than 2 transits, we analyzed their transit-timing variations (Figure~\ref{fig:ttv1} and \ref{fig:ttv2}). We inferred each planet's eccentricity using the ``photoeccentric" effect (Figure~\ref{fig:ecc1} and \ref{fig:ecc2}). In Table~\ref{tbl:pc}, we tabulate the catalog of Warm Jupiter candidates with their host star and planet properties. 
Furthermore, we  derived the preliminary eccentricity distribution of the Warm-Jupiter-candidate catalog using hierarchical Bayesian modeling (Figure~\ref{fig:hbm}). We investigated three functional forms for the eccentricity distribution, the Beta distribution, the Rayleigh distribution, and the mixture distribution, and found a set of well-constrained hyperparameters for each functional form (Figure~\ref{fig:hbm_all} and Table~\ref{tbl:hbm_post}). Extensive ground-based follow-up observations will be required to identify FPs in the sample and to construct a clean Warm Jupiter catalog. We proposed a follow-up strategy in Section~\ref{subsec:ground}. In future studies, the eccentricity distribution can be directly compared with predictions from different origin theories, with detection effects accounted for, to shed light on origin channels that predominantly contribute to the Warm Jupiter population.

\acknowledgments
This research made use of $\mathtt{exoplanet}$ \citep{exoplanet:exoplanet} and its dependencies \citep{exoplanet:agol20, exoplanet:astropy13, exoplanet:astropy18, exoplanet:exoplanet, exoplanet:foremanmackey17, exoplanet:foremanmackey18, exoplanet:kipping13, exoplanet:luger18, exoplanet:pymc3, exoplanet:theano}. This research made use of $\mathtt{Lightkurve}$, a Python package for {\it Kepler} and \TESS data analysis \citep{lightkurve18}. This research made use of $\mathtt{astroquery}$ \citep{astroquery19}. Computations for this research were performed on the Pennsylvania State University’s Institute for CyberScience Advanced CyberInfrastructure (ICS-ACI). This content is solely the responsibility of the authors and does not necessarily represent the views of the Institute for CyberScience. 

This work has made use of data from the European Space Agency (ESA) mission {\it Gaia} (\url{https://www.cosmos.esa.int/gaia}), processed by the {\it Gaia} Data Processing and Analysis Consortium (DPAC, \url{https://www.cosmos.esa.int/web/gaia/dpac/consortium}). Funding for the DPAC has been provided by national institutions, in particular the institutions participating in the {\it Gaia} Multilateral Agreement. 

We acknowledge the use of TESS High Level Science Products (HLSP) produced by the Quick-Look Pipeline (QLP) at the TESS Science Office at MIT, which are publicly available from the Mikulski Archive for Space Telescopes (MAST). Funding for the TESS mission is provided by NASA's Science Mission directorate.

This paper includes data collected by the \TESS mission, which are publicly available from the Mikulski Archive for Space Telescopes (MAST). Resources supporting this work were provided by the NASA High-End Computing (HEC) Program through the NASA Advanced Supercomputing (NAS) Division at Ames Research Center for the production of the SPOC data products.

This research has made use of the NASA Exoplanet Archive, which is operated by the California Institute of Technology, under contract with the National Aeronautics and Space Administration under the Exoplanet Exploration Program.

This work makes use of observations from the LCOGT network (proposal IDs: 2020B-0189, 2020B-0202). 

The research leading to these results has received funding from  the ARC grant for Concerted Research Actions, financed by the Wallonia-Brussels Federation. TRAPPIST is funded by the Belgian Fund for Scientific Research (Fond National de la Recherche Scientifique, FNRS) under the grant FRFC 2.5.594.09.F, with the participation of the Swiss National Science Fundation (SNF). TRAPPIST-North is a project funded by the University of Liege (Belgium), in collaboration with Cadi Ayyad University of Marrakech (Morocco) MG and EJ are F.R.S.-FNRS Senior Research Associate. This publication benefits from the support of the French Community of Belgium in the context of the FRIA Doctoral Grant awarded to MT.

This research received funding from the European Research Council (ERC) under the European Union's Horizon 2020 research and innovation programme (grant agreement n$^\circ$ 803193/BEBOP), and from the Science and Technology Facilities Council (STFC; grant n$^\circ$ ST/S00193X/1).

RB\ acknowledges support from FONDECYT Project 11200751, CORFO project N$^\circ$14ENI2-26865. 
RB and AJ acknowledge support from ANID -- Millennium Science Initiative -- ICN12\_009. AJ acknowledges additional support from FONDECYT project 1210718.

This work has been carried out within the framework of the NCCR PlanetS supported by the Swiss National Science Foundation. We made use of PlanetS’ Data \& Analysis Center for Exoplanets (DACE), which is available at \url{https://dace.unige.ch}.

This research has made use of the Exoplanet Follow-up Observation Program website, which is operated by the California Institute of Technology, under contract with the National Aeronautics and Space Administration under the Exoplanet Exploration Program.

We gratefully acknowledge support by NASA XRP NNX16AB50G,  NASA \TESS GO 80NSSC18K1695, and the Alfred P. Sloan Foundation's Sloan Research Fellowship. The Center for Exoplanets and Habitable Worlds is supported by the Pennsylvania State University, the Eberly College of Science, and the Pennsylvania Space Grant Consortium. JD gratefully acknowledges support and hospitality from the pre-doctoral program at the Center for Computational Astrophysics, Flatiron Institute. Research at the Flatiron Institute is supported by the Simons Foundation. J.D. thanks the Flatiron Astronomical Data Group and Planet Formation Group for helpful discussions. J.D. gratefully acknowledges Angie Wolfgang and Tom Loredo for inspiring lectures on Bayesian statistics. J.D. thanks Eric Ford, Kassandra Anderson, Juliette Becker, Vincent Van Eylen, Matthias He, Sarah Millholland, Ruth Murray-Clay, Cristobal Petrovich, and Keivan Stassun for helpful discussions.

\software{$\mathtt{ArviZ}$ \citep{Kumar2019}, $\mathtt{AstroImageJ}$ \citep{Collins:2017}, $\mathtt{astropy}$ \citep{exoplanet:astropy13, exoplanet:astropy18}, $\mathtt{astroquery}$ \citep{astroquery19}, $\mathtt{celerite}$ \citep{exoplanet:foremanmackey17, exoplanet:foremanmackey18}, $\mathtt{exoplanet}$ \citep{exoplanet:exoplanet}, $\mathtt{Jupyter}$ \citep{kluy16}, $\mathtt{lightkurve}$ \citep{lightkurve18}, $\mathtt{Matplotlib}$ \citep{hunt07, droe16}, $\mathtt{NumPy}$ \citep{vand11, harr20}, $\mathtt{pandas}$ \citep{mckinney-proc-scipy-2010}, $\mathtt{PyMC3}$ \citep{exoplanet:pymc3}, $\mathtt{SciPy}$ \citep{2020SciPy-NMeth}, $\mathtt{starry}$ \citep{exoplanet:luger18}, $\mathtt{TAPIR}$ \citep{Jensen:2013}, $\mathtt{Theano}$ \citep{exoplanet:theano}}

\facilities{\TESS, \emph{Gaia}, LCOGT, Exoplanet Archive}

\clearpage
\pagebreak[4]
\global\pdfpageattr\expandafter{\the\pdfpageattr/Rotate 90}
\begin{longrotatetable}
\begin{deluxetable*}{llccccccccccccccl}
\centerwidetable
\tabletypesize{\notsotiny}
\tablecaption{Warm Jupiter Candidates Discovered in Year 1 \TESS Full-Frame Images. A brief description of the columns: (1) \TESS Input Catalog ID, TIC ID \changes{(2) \TESS Objects of Interest name, TOI Name} (3) stellar radius in solar radius, $R_\star$ (4) stellar density in the mean solar density, $\rho_\star$ (5) stellar density inferred from light curves assuming a circular-orbit planet in the mean solar density, $\rho_{\textrm{circ}}$ (6) transit depth in parts-per-thousand (ppt), $\delta$ (7) planet radius in Earth radii, $R_{\textrm{p}}$ (8) impact parameter, $b$ (9) orbital period in days, $P$ (10) conjunction time, $T_c$ (11) the mode and 68\% highest posterior density (HPD) interval of the eccentricity inferred from light curves, $e$ (12) the 68\% HPD intervals of the argument of periapse in degrees inferred from light curves, $\omega$ \changes{(13) the median and 68\% credible interval of the eccentricity from literature (see Column 17) (14) the median and 68\% credible interval of the argument of periapse from literature (see Column 17)} (15) the significance level of the TTV signal using the metric discussed in Section~\ref{subsec:post_analysis} for candidates with 3$+$ transits (16) scores the candidate gets from the Transmission Spectroscopy Metric \citep[TSM;][]{kemp18}, scaled between 1--100 (17) other names the candidates have and their references. \label{tbl:pc}}
\tablehead{
\colhead{TIC ID} & \colhead{TOI name} & \colhead{$R_\star$} & \colhead{$\rho_\star$} & \colhead{$\rho_{\textrm{circ}}$} & \colhead{$\delta$} & \colhead{$R_{\textrm{p}}$} & \colhead{$b$} & \colhead{$P$} & \colhead{$T_c$} & \colhead{$e$} & \colhead{$\omega$} & \colhead{$e_{\mathrm{RV}}$} & \colhead{$\omega_{\mathrm{RV}}$} & \colhead{$\sigma_{\textrm{TTV}}$} & \colhead{TSM} & \colhead{Other Name/Reference}\\ 
\colhead{} & \colhead{} & \colhead{[R$_\sun$]} & \colhead{[$\rho_\sun$]} & \colhead{[$\rho_\sun$]} & \colhead{[ppt]} & \colhead{[$R_\earth$]} & \colhead{} & \colhead{[day]} & \colhead{[BJD-2457000]} & \colhead{} & \colhead{[$^\circ$]} & \colhead{} & \colhead{} & \colhead{} & \colhead{} & \colhead{}
}
\decimalcolnumbers
\startdata
394137592 & 120b & 2.292$^{+0.093}_{-0.093}$ & 0.083$^{+0.007}_{-0.007}$ & 0.114$^{+0.009}_{-0.022}$ & 2.025$^{+0.076}_{-0.057}$ & 11.274$^{+0.497}_{-0.497}$ & 0.234$^{+0.200}_{-0.160}$ & 11.537$^{+0.001}_{-0.001}$ & 1320.543$^{+0.002}_{-0.002}$ & 0.11$^{+0.19}_{-0.11}$ & [-29,40], [133,210] & 0.22$^{+0.027}_{-0.026}$ & 35$^{+9}_{-10}$ & 2.1 & 43 & HD1397b; \citetalias{brah19,niel19} \\
29857954 & 172b & 1.666$^{+0.066}_{-0.066}$ & 0.222$^{+0.032}_{-0.032}$ & 0.574$^{+0.096}_{-0.190}$ & 4.220$^{+0.251}_{-0.242}$ & 11.798$^{+0.596}_{-0.573}$ & 0.307$^{+0.257}_{-0.213}$ & 9.476$^{+0.001}_{-0.001}$ & 1317.444$^{+0.002}_{-0.002}$ & 0.35$^{+0.19}_{-0.23}$ & [-12,66], [113,189] & 0.38$^{+0.0093}_{-0.009}$ & 57$^{+2}_{-2}$ & 0.5 & 12 & \citetalias{rodr19} \\
166739520 & 190b & 1.219$^{+0.043}_{-0.043}$ & 0.563$^{+0.087}_{-0.087}$ & 0.301$^{+0.066}_{-0.068}$ & 7.677$^{+0.362}_{-0.339}$ & 11.656$^{+0.481}_{-0.472}$ & 0.410$^{+0.148}_{-0.241}$ & 10.021$^{+0.000}_{-0.000}$ & 1347.549$^{+0.001}_{-0.001}$ & 0.25$^{+0.26}_{-0.12}$ & [-68,-26], [205,250] & 0.3$^{+0.023}_{-0.023}$ & 242$^{+3}_{-2}$ & 0.9 & 38 & WASP-117b; \citetalias{lend14} \\
183532609 & 191b & 0.957$^{+0.047}_{-0.047}$ & 1.046$^{+0.129}_{-0.129}$ & 0.466$^{+0.085}_{-0.055}$ & 13.215$^{+0.596}_{-0.709}$ & 11.973$^{+0.653}_{-0.651}$ & 0.606$^{+0.052}_{-0.083}$ & 8.159$^{+0.000}_{-0.000}$ & 1350.762$^{+0.000}_{-0.000}$ & 0.29$^{+0.26}_{-0.13}$ & [-71,-27], [209,250] & 0.31$^{+0.0029}_{-0.0024}$ & 274$^{+0}_{-0}$ & 0.5 & 81 & WASP-8b; \citetalias{quel10} \\
410214986 & 200b & 0.913$^{+0.055}_{-0.055}$ & 1.197$^{+1.541}_{-1.541}$ & 1.576$^{+0.163}_{-0.342}$ & 3.591$^{+0.723}_{-0.658}$ & 5.959$^{+0.679}_{-0.659}$ & 0.261$^{+0.209}_{-0.183}$ & 7.872$^{+0.001}_{-0.001}$ & 1332.576$^{+0.001}_{-0.001}$ & 0.11$^{+0.21}_{-0.1}$ & [-32,35], [136,215] & - & - & - & 64 & DS Tuc Ab; \citetalias{newt19} \\
350618622 & 201b & 1.229$^{+0.043}_{-0.043}$ & 0.573$^{+0.107}_{-0.107}$ & 3.078$^{+0.833}_{-1.056}$ & 5.390$^{+0.429}_{-0.279}$ & 9.879$^{+0.490}_{-0.445}$ & 0.419$^{+0.204}_{-0.263}$ & 52.978$^{+0.000}_{-0.000}$ & 1323.075$^{+0.001}_{-0.001}$ & 0.55$^{+0.17}_{-0.14}$ & [30,159] & - & - & 1.1 & 29 & \citetalias{hobs21} \\
55652896 & 216b & 0.759$^{+0.044}_{-0.044}$ & 1.793$^{+0.220}_{-0.220}$ & 1.178$^{+0.562}_{-0.279}$ & 9.990$^{+24.072}_{-4.694}$ & 8.311$^{+6.917}_{-2.275}$ & 0.987$^{+0.109}_{-0.055}$ & 17.072$^{+0.001}_{-0.001}$ & 1308.338$^{+0.017}_{-0.017}$ & 0.11$^{+0.28}_{-0.1}$ & [-65,-1], [179,246] & 0.16$^{+0.003}_{-0.002}$ & 292$^{+1}_{-1}$ & 27.1 & 15 & \citetalias{daws21, daws19, kipp19} \\
55652896 & 216c & 0.759$^{+0.044}_{-0.044}$ & 1.793$^{+0.220}_{-0.220}$ & 1.714$^{+0.057}_{-0.096}$ & 15.261$^{+0.281}_{-0.249}$ & 10.232$^{+0.590}_{-0.608}$ & 0.136$^{+0.108}_{-0.092}$ & 34.554$^{+0.001}_{-0.001}$ & 1296.714$^{+0.006}_{-0.006}$ & 0.052$^{+0.21}_{-0.052}$ & [-43,9], [171,222] & 0.0046$^{+0.0027}_{-0.0012}$ & 190$^{+30}_{-50}$ & 24.5 & 15 & \citetalias{daws21, daws19, kipp19} \\
200723869 & 257b & 1.719$^{+0.048}_{-0.048}$ & 0.248$^{+0.035}_{-0.035}$ & 0.380$^{+0.093}_{-0.158}$ & 1.284$^{+0.086}_{-0.061}$ & 6.736$^{+0.289}_{-0.258}$ & 0.389$^{+0.251}_{-0.264}$ & 18.387$^{+0.001}_{-0.001}$ & 1367.377$^{+0.003}_{-0.003}$ & 0.18$^{+0.2}_{-0.16}$ & [-41,49], [137,219] & 0.24$^{+0.04}_{-0.065}$ & 96$^{+22}_{-22}$ & 0.8 & 30 & HD19916b; \citetalias{addi20} \\
339672028 & 481b & 1.605$^{+0.066}_{-0.066}$ & 0.243$^{+0.033}_{-0.033}$ & 0.350$^{+0.056}_{-0.091}$ & 3.878$^{+0.161}_{-0.088}$ & 10.931$^{+0.481}_{-0.468}$ & 0.330$^{+0.199}_{-0.220}$ & 10.331$^{+0.000}_{-0.000}$ & 1377.336$^{+0.001}_{-0.001}$ & 0.11$^{+0.22}_{-0.11}$ & [-30,48], [138,210] & 0.15$^{+0.006}_{-0.007}$ & 65$^{+2}_{-2}$ & 3.3 & 24 & \citetalias{brah20} \\
130415266 & 588.01 & 1.694$^{+0.016}_{-0.016}$ & 0.348$^{+0.009}_{-0.009}$ & 1.018$^{+0.080}_{-0.117}$ & 7.179$^{+0.146}_{-0.101}$ & 15.673$^{+0.202}_{-0.188}$ & 0.250$^{+0.126}_{-0.159}$ & 39.471$^{+0.000}_{-0.000}$ & 1481.793$^{+0.000}_{-0.000}$ & 0.36$^{+0.14}_{-0.071}$ & [27,146] & - & - & - & 87 & - \\
53189332 & 660b & 1.293$^{+0.050}_{-0.050}$ & 0.494$^{+0.077}_{-0.077}$ & 0.393$^{+0.050}_{-0.094}$ & 5.480$^{+0.219}_{-0.220}$ & 10.437$^{+0.458}_{-0.447}$ & 0.277$^{+0.216}_{-0.189}$ & 9.293$^{+0.001}_{-0.001}$ & 1535.348$^{+0.001}_{-0.001}$ & 0.11$^{+0.27}_{-0.098}$ & [-59,-8], [187,240] & - & - & 0.4 & 15 & WASP-106b; \citetalias{smit14} \\
280206394 & 677b & 1.193$^{+0.051}_{-0.051}$ & 0.602$^{+0.093}_{-0.093}$ & 2.241$^{+2.156}_{-0.708}$ & 7.810$^{+0.444}_{-0.833}$ & 11.416$^{+0.651}_{-0.696}$ & 0.705$^{+0.084}_{-0.280}$ & 11.237$^{+0.000}_{-0.000}$ & 1536.233$^{+0.001}_{-0.001}$ & 0.6$^{+0.12}_{-0.25}$ & [26,162] & 0.43$^{+0.024}_{-0.024}$ & 70$^{+4}_{-4}$ & 0.9 & 46 & \citetalias{jord20} \\
286864983 & 772.01 & 0.766$^{+0.042}_{-0.042}$ & 1.746$^{+0.187}_{-0.187}$ & 1.634$^{+0.371}_{-0.628}$ & 5.957$^{+0.398}_{-0.355}$ & 6.443$^{+0.411}_{-0.396}$ & 0.367$^{+0.258}_{-0.249}$ & 11.022$^{+0.003}_{-0.003}$ & 1575.939$^{+0.002}_{-0.002}$ & 0.066$^{+0.28}_{-0.066}$ & [-61,11], [167,238] & - & - & - & 17 & - \\
334305570 & 777.01 & 1.474$^{+0.056}_{-0.056}$ & 0.341$^{+0.062}_{-0.062}$ & 0.871$^{+0.239}_{-0.397}$ & 2.144$^{+0.141}_{-0.116}$ & 7.454$^{+0.367}_{-0.346}$ & 0.400$^{+0.269}_{-0.269}$ & 16.603$^{+0.003}_{-0.003}$ & 1575.780$^{+0.002}_{-0.002}$ & 0.34$^{+0.16}_{-0.28}$ & [-18,51], [104,201] & - & - & - & 13 & - \\
306472057 & 791.01 & 1.279$^{+0.048}_{-0.048}$ & 0.505$^{+0.102}_{-0.102}$ & 0.445$^{+0.097}_{-0.148}$ & 4.461$^{+0.227}_{-0.202}$ & 9.324$^{+0.417}_{-0.411}$ & 0.388$^{+0.215}_{-0.265}$ & 139.309$^{+0.003}_{-0.003}$ & 1427.624$^{+0.002}_{-0.002}$ & 0.077$^{+0.27}_{-0.077}$ & [-60,2], [176,240] & - & - & - & 4 & - \\
363914762 & 812.01 & 1.438$^{+0.048}_{-0.048}$ & 0.403$^{+0.056}_{-0.056}$ & 0.372$^{+0.016}_{-0.039}$ & 6.683$^{+0.112}_{-0.109}$ & 12.816$^{+0.444}_{-0.445}$ & 0.165$^{+0.150}_{-0.113}$ & 13.866$^{+0.000}_{-0.000}$ & 1431.463$^{+0.001}_{-0.001}$ & 0.067$^{+0.23}_{-0.067}$ & [-52,4], [178,228] & - & - & 1.9 & 15 & - \\
158978373 & 823.01 & 1.194$^{+0.045}_{-0.045}$ & 0.606$^{+0.091}_{-0.091}$ & 1.120$^{+0.666}_{-0.787}$ & 1.993$^{+0.506}_{-0.246}$ & 5.837$^{+0.711}_{-0.451}$ & 0.529$^{+0.306}_{-0.357}$ & 13.540$^{+0.005}_{-0.005}$ & 1609.005$^{+0.003}_{-0.003}$ & 0.32$^{+0.19}_{-0.27}$ & [-45,58], [124,220] & - & - & - & 10 & - \\
243200602 & 826.01 & 0.975$^{+0.045}_{-0.045}$ & 0.983$^{+0.118}_{-0.118}$ & 1.347$^{+0.412}_{-0.497}$ & 8.556$^{+0.665}_{-0.533}$ & 9.866$^{+0.551}_{-0.536}$ & 0.447$^{+0.212}_{-0.286}$ & 11.554$^{+0.001}_{-0.001}$ & 1607.581$^{+0.001}_{-0.001}$ & 0.12$^{+0.24}_{-0.11}$ & [-42,41], [141,220] & - & - & - & 26 & WASP-130b; \citetalias{hell17} \\
123846039 & 883.01 & 0.967$^{+0.043}_{-0.043}$ & 1.011$^{+0.117}_{-0.117}$ & 0.944$^{+0.401}_{-0.545}$ & 3.863$^{+0.483}_{-0.276}$ & 6.583$^{+0.453}_{-0.387}$ & 0.486$^{+0.279}_{-0.326}$ & 10.051$^{+0.002}_{-0.002}$ & 1476.531$^{+0.001}_{-0.001}$ & 0.091$^{+0.31}_{-0.086}$ & [-63,17], [161,245] & - & - & - & 28 & - \\
177162886 & 899.01 & 1.051$^{+0.049}_{-0.049}$ & 0.834$^{+1.541}_{-1.541}$ & 2.265$^{+1.237}_{-0.961}$ & 7.264$^{+0.935}_{-0.615}$ & 9.823$^{+0.695}_{-0.624}$ & 0.563$^{+0.194}_{-0.303}$ & 12.846$^{+0.000}_{-0.000}$ & 1313.638$^{+0.001}_{-0.001}$ & 0.45$^{+0.18}_{-0.2}$ & [10,65], [89,174] & - & - & 2.7 & 12 & - \\
260130483 & 933.01 & 1.644$^{+0.070}_{-0.070}$ & 0.210$^{+0.038}_{-0.038}$ & 0.313$^{+0.173}_{-0.067}$ & 4.210$^{+0.220}_{-0.237}$ & 11.603$^{+0.596}_{-0.652}$ & 0.817$^{+0.033}_{-0.089}$ & 88.932$^{+0.001}_{-0.001}$ & 1256.009$^{+0.002}_{-0.002}$ & 0.13$^{+0.36}_{-0.1}$ & [-22,67], [133,198] & - & - & 0.3 & 4 & - \\
408137826 & 941.01 & 1.070$^{+0.047}_{-0.047}$ & 0.755$^{+0.107}_{-0.107}$ & 0.698$^{+0.276}_{-0.394}$ & 2.997$^{+0.340}_{-0.271}$ & 6.393$^{+0.455}_{-0.412}$ & 0.441$^{+0.299}_{-0.303}$ & 8.511$^{+0.001}_{-0.001}$ & 1431.971$^{+0.001}_{-0.001}$ & 0.084$^{+0.32}_{-0.082}$ & [-65,12], [162,245] & - & - & 0.2 & 10 & - \\
399871011 & 943.01 & 1.284$^{+0.050}_{-0.050}$ & 0.496$^{+0.099}_{-0.099}$ & 0.362$^{+0.078}_{-0.144}$ & 6.490$^{+0.365}_{-0.351}$ & 11.283$^{+0.551}_{-0.533}$ & 0.375$^{+0.264}_{-0.260}$ & 12.360$^{+0.002}_{-0.002}$ & 1444.268$^{+0.002}_{-0.002}$ & 0.12$^{+0.31}_{-0.1}$ & [-66,-10], [193,244] & - & - & - & 16 & - \\
25799609 & 963.01 & 1.015$^{+0.041}_{-0.041}$ & 0.879$^{+0.113}_{-0.113}$ & 0.341$^{+0.617}_{-0.131}$ & 5.802$^{+0.403}_{-0.528}$ & 8.409$^{+0.466}_{-0.494}$ & 0.761$^{+0.073}_{-0.392}$ & 11.114$^{+0.003}_{-0.003}$ & 1501.309$^{+0.002}_{-0.002}$ & 0.094$^{+0.38}_{-0.075}$ & [-77,3], [179,255] & - & - & - & 18 & - \\
73038411 & 978.01 & 1.434$^{+0.054}_{-0.054}$ & 0.392$^{+0.060}_{-0.060}$ & 0.847$^{+0.180}_{-0.232}$ & 5.801$^{+0.833}_{-0.785}$ & 11.910$^{+0.937}_{-0.945}$ & 0.289$^{+0.227}_{-0.200}$ & 15.854$^{+1.437}_{-1.437}$ & 1502.778$^{+0.002}_{-0.002}$ & 0.3$^{+0.21}_{-0.21}$ & [-13,62], [113,195] & - & - & - & 17 & - \\
254113311 & 1130c & 0.695$^{+0.036}_{-0.036}$ & 2.195$^{+0.265}_{-0.265}$ & 1.920$^{+0.478}_{-0.250}$ & 35.686$^{+76.608}_{-16.610}$ & 14.292$^{+11.245}_{-3.810}$ & 0.986$^{+0.182}_{-0.091}$ & 8.351$^{+0.000}_{-0.000}$ & 1649.547$^{+0.001}_{-0.001}$ & 0.073$^{+0.22}_{-0.073}$ & [-51,12], [168,230] & 0.047$^{+0.04}_{-0.027}$ & 332$^{+24}_{-55}$ & 1.1 & 100 & \citetalias{huan20} \\
409794137 & 1478b & 0.988$^{+0.044}_{-0.044}$ & 0.974$^{+0.126}_{-0.126}$ & 1.116$^{+0.200}_{-0.333}$ & 8.944$^{+0.502}_{-0.358}$ & 10.213$^{+0.525}_{-0.509}$ & 0.339$^{+0.218}_{-0.225}$ & 10.181$^{+0.000}_{-0.000}$ & 1485.738$^{+0.001}_{-0.001}$ & 0.085$^{+0.23}_{-0.085}$ & [-44,30], [153,224] & 0.024$^{+0.032}_{-0.017}$ & 250$^{+120}_{-130}$ & 1.0 & 33 & \citetalias{rodr21} \\
308098254 & 1906.01 & 1.113$^{+0.048}_{-0.048}$ & 0.701$^{+0.090}_{-0.090}$ & 0.619$^{+0.260}_{-0.269}$ & 8.517$^{+0.903}_{-0.613}$ & 11.249$^{+0.710}_{-0.648}$ & 0.494$^{+0.214}_{-0.318}$ & 9.627$^{+0.000}_{-0.000}$ & 1538.718$^{+0.000}_{-0.000}$ & 0.087$^{+0.28}_{-0.083}$ & [-62,12], [168,241] & 0.43$^{+0.005}_{-0.005}$ & 358$^{+2}_{-2}$ & 0.1 & 15 & WASP-162b; \citetalias{hell19} \\
190271688 & 1963.01 & 1.062$^{+0.054}_{-0.054}$ & 0.776$^{+0.107}_{-0.107}$ & 0.814$^{+0.097}_{-0.073}$ & 21.059$^{+0.777}_{-1.058}$ & 16.751$^{+0.922}_{-0.912}$ & 0.659$^{+0.036}_{-0.051}$ & 12.636$^{+0.001}_{-0.001}$ & 1607.245$^{+0.000}_{-0.000}$ & 0.055$^{+0.22}_{-0.055}$ & [-35,26], [152,213] & - & - & - & 57 & - \\
371188886 & 2000.01 & 1.074$^{+0.053}_{-0.053}$ & 0.769$^{+0.104}_{-0.104}$ & 1.221$^{+0.298}_{-0.544}$ & 3.730$^{+0.258}_{-0.200}$ & 7.160$^{+0.433}_{-0.405}$ & 0.389$^{+0.273}_{-0.261}$ & 9.127$^{+0.001}_{-0.001}$ & 1535.058$^{+0.003}_{-0.003}$ & 0.19$^{+0.2}_{-0.17}$ & [-41,51], [137,221] & - & - & 2.3 & 16 & - \\
147660886 & 2005.01 & 2.103$^{+0.077}_{-0.077}$ & 0.161$^{+0.021}_{-0.021}$ & 1.394$^{+0.314}_{-0.559}$ & 2.498$^{+0.164}_{-0.149}$ & 11.461$^{+0.578}_{-0.546}$ & 0.366$^{+0.257}_{-0.250}$ & 17.306$^{+0.002}_{-0.002}$ & 1549.835$^{+0.002}_{-0.002}$ & 0.65$^{+0.14}_{-0.15}$ & [29,148] & - & - & - & 13 & - \\
394287035 & 2328.01 & 0.829$^{+0.042}_{-0.042}$ & 1.429$^{+0.167}_{-0.167}$ & 2.693$^{+0.508}_{-0.937}$ & 9.130$^{+0.598}_{-0.538}$ & 8.639$^{+0.519}_{-0.504}$ & 0.327$^{+0.254}_{-0.219}$ & 17.102$^{+0.000}_{-0.000}$ & 1313.386$^{+0.000}_{-0.000}$ & 0.24$^{+0.18}_{-0.22}$ & [-27,52], [123,209] & - & - & 0.1 & 14 & CTOI \\
24358417 & 2338.01 & 1.028$^{+0.047}_{-0.047}$ & 0.881$^{+0.125}_{-0.125}$ & 3.631$^{+0.592}_{-1.020}$ & 10.028$^{+0.508}_{-0.547}$ & 11.218$^{+0.600}_{-0.595}$ & 0.278$^{+0.230}_{-0.191}$ & 22.652$^{+0.003}_{-0.003}$ & 1481.407$^{+0.002}_{-0.002}$ & 0.48$^{+0.2}_{-0.15}$ & [20,55], [68,170] & - & - & - & 11 & - \\
279514271 & 2339.01 & 2.166$^{+0.096}_{-0.096}$ & 0.086$^{+0.015}_{-0.015}$ & 0.085$^{+0.050}_{-0.059}$ & 2.032$^{+0.526}_{-0.239}$ & 10.720$^{+1.334}_{-0.862}$ & 0.528$^{+0.297}_{-0.355}$ & 9.774$^{+0.000}_{-0.000}$ & 1351.802$^{+0.008}_{-0.008}$ & 0.13$^{+0.29}_{-0.11}$ & [-64,29], [152,243] & - & - & 1.2 & 1 & - \\
204671232 & 2361.01 & 0.984$^{+0.045}_{-0.045}$ & 0.971$^{+0.125}_{-0.125}$ & 1.001$^{+0.134}_{-0.230}$ & 10.749$^{+0.520}_{-0.539}$ & 11.114$^{+0.572}_{-0.565}$ & 0.259$^{+0.212}_{-0.177}$ & 8.716$^{+0.003}_{-0.003}$ & 1569.062$^{+0.007}_{-0.007}$ & 0.062$^{+0.24}_{-0.062}$ & [-47,19], [160,228] & - & - & 2.3 & 15 & CTOI; \citetalias{mont20} \\
238542895 & - & 0.896$^{+0.030}_{-0.030}$ & 1.204$^{+0.137}_{-0.137}$ & 0.730$^{+0.076}_{-0.136}$ & 10.233$^{+0.431}_{-0.432}$ & 9.872$^{+0.398}_{-0.391}$ & 0.229$^{+0.203}_{-0.157}$ & 8.385$^{+0.000}_{-0.000}$ & 1485.880$^{+0.001}_{-0.001}$ & 0.21$^{+0.27}_{-0.094}$ & [-66,-24], [203,247] & - & - & 1.6 & 18 & CTOI \\
282498590 & - & 1.051$^{+0.145}_{-0.145}$ & 0.778$^{+0.279}_{-0.279}$ & 0.556$^{+0.101}_{-0.172}$ & 10.780$^{+0.656}_{-0.543}$ & 11.896$^{+1.674}_{-1.603}$ & 0.343$^{+0.231}_{-0.232}$ & 10.095$^{+0.002}_{-0.002}$ & 1474.825$^{+0.001}_{-0.001}$ & 0.16$^{+0.27}_{-0.14}$ & [-66,-8], [189,246] & - & - & - & 22 & CTOI; \citetalias{mont20}  \\
290403522 & - & 1.384$^{+0.052}_{-0.052}$ & 0.435$^{+0.067}_{-0.067}$ & 0.059$^{+0.010}_{-0.007}$ & 6.596$^{+0.231}_{-0.246}$ & 12.247$^{+0.512}_{-0.499}$ & 0.538$^{+0.057}_{-0.092}$ & 22.375$^{+0.000}_{-0.000}$ & 1324.188$^{+0.003}_{-0.003}$ & 0.61$^{+0.16}_{-0.076}$ & [-86,-52], [232,266] & - & - & 1.5 & 18 & CTOI \\
380836882 & - & 1.197$^{+0.053}_{-0.053}$ & 0.574$^{+0.087}_{-0.087}$ & 0.392$^{+0.052}_{-0.102}$ & 6.541$^{+0.367}_{-0.302}$ & 10.573$^{+0.553}_{-0.534}$ & 0.282$^{+0.219}_{-0.195}$ & 8.012$^{+0.000}_{-0.000}$ & 1652.826$^{+0.001}_{-0.001}$ & 0.17$^{+0.28}_{-0.11}$ & [-65,-17], [197,243] & - & - & 0.3 & 18 & CTOI \\
382200986 & - & 0.680$^{+0.036}_{-0.036}$ & 2.182$^{+0.232}_{-0.232}$ & 2.002$^{+0.251}_{-0.449}$ & 8.810$^{+0.417}_{-0.346}$ & 6.975$^{+0.399}_{-0.394}$ & 0.279$^{+0.202}_{-0.191}$ & 92.498$^{+0.005}_{-0.005}$ & 1243.912$^{+0.010}_{-0.010}$ & 0.06$^{+0.25}_{-0.06}$ & [-51,1], [175,233] & - & - & 3.7 & 5 & CTOI; \citetalias{mont20} \\
357202877 & - & 1.116$^{+0.044}_{-0.044}$ & 0.645$^{+0.084}_{-0.084}$ & 1.143$^{+0.693}_{-0.549}$ & 14.388$^{+3.440}_{-2.518}$ & 14.643$^{+1.728}_{-1.468}$ & 0.570$^{+0.223}_{-0.372}$ & 9.555$^{+0.001}_{-0.001}$ & 1534.787$^{+0.003}_{-0.003}$ & 0.1$^{+0.33}_{-0.084}$ & [-35,56], [132,213] & - & - & 1.0 & 20 & - \\
254142310 & - & 1.633$^{+0.074}_{-0.074}$ & 0.231$^{+0.045}_{-0.045}$ & 0.264$^{+0.080}_{-0.123}$ & 2.986$^{+0.322}_{-0.252}$ & 9.743$^{+0.675}_{-0.616}$ & 0.414$^{+0.267}_{-0.279}$ & 9.311$^{+0.002}_{-0.002}$ & 1652.052$^{+0.004}_{-0.004}$ & 0.11$^{+0.27}_{-0.099}$ & [-57,27], [152,236] & - & - & 0.6 & 8 & - \\
395113305 & - & 0.886$^{+0.043}_{-0.043}$ & 1.272$^{+0.162}_{-0.162}$ & 0.284$^{+0.150}_{-0.159}$ & 7.345$^{+1.252}_{-0.917}$ & 8.294$^{+0.787}_{-0.667}$ & 0.506$^{+0.259}_{-0.338}$ & 9.963$^{+0.008}_{-0.008}$ & 1607.420$^{+0.006}_{-0.006}$ & 0.48$^{+0.27}_{-0.1}$ & [-82,-45], [224,262] & - & - & - & 11 & - \\
180989820 & - & 1.256$^{+0.058}_{-0.058}$ & 0.544$^{+0.088}_{-0.088}$ & 0.084$^{+0.017}_{-0.026}$ & 4.508$^{+0.338}_{-0.236}$ & 9.223$^{+0.524}_{-0.484}$ & 0.356$^{+0.216}_{-0.243}$ & 9.873$^{+0.001}_{-0.001}$ & 1511.287$^{+0.004}_{-0.004}$ & 0.63$^{+0.14}_{-0.1}$ & [-86,-52], [233,265] & - & - & 1.4 & 12 & - \\
4598935 & - & 1.042$^{+0.040}_{-0.040}$ & 0.801$^{+0.108}_{-0.108}$ & 0.460$^{+0.159}_{-0.231}$ & 3.515$^{+0.475}_{-0.367}$ & 6.744$^{+0.525}_{-0.455}$ & 0.438$^{+0.271}_{-0.299}$ & 24.663$^{+0.007}_{-0.007}$ & 1539.451$^{+0.006}_{-0.006}$ & 0.19$^{+0.35}_{-0.12}$ & [-71,-19], [202,248] & - & - & - & 5 & - \\
464300749 & - & 1.678$^{+0.056}_{-0.056}$ & 0.270$^{+0.044}_{-0.044}$ & 4.896$^{+1.439}_{-2.229}$ & 4.007$^{+0.334}_{-0.316}$ & 11.589$^{+0.620}_{-0.601}$ & 0.385$^{+0.281}_{-0.265}$ & 18.098$^{+0.003}_{-0.003}$ & 1529.321$^{+0.006}_{-0.006}$ & 0.78$^{+0.098}_{-0.13}$ & [34,143] & - & - & 1.1 & 11 & - \\
7536985 & - & 1.908$^{+0.086}_{-0.086}$ & 0.163$^{+0.036}_{-0.036}$ & 0.315$^{+0.091}_{-0.133}$ & 7.271$^{+0.557}_{-0.503}$ & 17.745$^{+1.074}_{-1.015}$ & 0.397$^{+0.253}_{-0.268}$ & 12.732$^{+0.005}_{-0.005}$ & 1502.907$^{+0.004}_{-0.004}$ & 0.25$^{+0.19}_{-0.22}$ & [-32,54], [128,210] & - & - & - & 8 & - \\
76228620 & - & 1.194$^{+0.051}_{-0.051}$ & 0.571$^{+0.090}_{-0.090}$ & 0.343$^{+0.227}_{-0.217}$ & 4.492$^{+1.118}_{-0.626}$ & 8.768$^{+1.045}_{-0.768}$ & 0.567$^{+0.254}_{-0.381}$ & 9.779$^{+0.002}_{-0.002}$ & 1645.400$^{+0.004}_{-0.004}$ & 0.078$^{+0.37}_{-0.075}$ & [-71,-4], [186,253] & - & - & 0.5 & 11 & - \\
87422071 & - & 1.342$^{+0.055}_{-0.055}$ & 0.413$^{+0.073}_{-0.073}$ & 0.794$^{+0.358}_{-0.447}$ & 3.622$^{+0.524}_{-0.433}$ & 8.803$^{+0.753}_{-0.660}$ & 0.444$^{+0.295}_{-0.304}$ & 11.366$^{+0.004}_{-0.004}$ & 1646.564$^{+0.008}_{-0.008}$ & 0.29$^{+0.19}_{-0.24}$ & [-35,58], [121,216] & - & - & 0.7 & 8 & - \\
253126207 & - & 1.249$^{+0.046}_{-0.046}$ & 0.506$^{+0.085}_{-0.085}$ & 1.547$^{+0.208}_{-0.325}$ & 11.792$^{+0.551}_{-0.588}$ & 14.784$^{+0.648}_{-0.646}$ & 0.255$^{+0.207}_{-0.174}$ & 17.434$^{+0.002}_{-0.002}$ & 1604.089$^{+0.001}_{-0.001}$ & 0.39$^{+0.22}_{-0.12}$ & [8,77], [101,172] & - & - & - & 22 & - \\
342167886 & - & 1.437$^{+0.095}_{-0.095}$ & 0.348$^{+0.080}_{-0.080}$ & 0.983$^{+0.153}_{-0.318}$ & 7.701$^{+0.314}_{-0.353}$ & 13.731$^{+0.963}_{-0.967}$ & 0.313$^{+0.248}_{-0.218}$ & 17.308$^{+0.001}_{-0.001}$ & 1510.187$^{+0.003}_{-0.003}$ & 0.36$^{+0.22}_{-0.19}$ & [-4,60], [105,188] & - & - & 1.5 & 11 & - \\
343936388 & - & 0.855$^{+0.033}_{-0.033}$ & 1.350$^{+0.155}_{-0.155}$ & 0.303$^{+0.228}_{-0.184}$ & 4.702$^{+0.584}_{-0.489}$ & 6.397$^{+0.454}_{-0.409}$ & 0.583$^{+0.227}_{-0.389}$ & 15.144$^{+0.009}_{-0.009}$ & 1469.040$^{+0.019}_{-0.019}$ & 0.48$^{+0.26}_{-0.15}$ & [-81,-42], [222,261] & - & - & 1.9 & 6 & - \\
394492336 & - & 0.934$^{+0.044}_{-0.044}$ & 1.063$^{+0.113}_{-0.113}$ & 3.489$^{+1.140}_{-1.470}$ & 11.543$^{+1.345}_{-0.961}$ & 10.975$^{+0.776}_{-0.699}$ & 0.436$^{+0.243}_{-0.286}$ & 9.131$^{+0.000}_{-0.000}$ & 1597.926$^{+0.002}_{-0.002}$ & 0.43$^{+0.19}_{-0.24}$ & [4,63], [95,183] & - & - & 1.4 & 22 & - \\
150299840 & - & 1.839$^{+0.061}_{-0.061}$ & 0.220$^{+0.031}_{-0.031}$ & 0.443$^{+0.046}_{-0.096}$ & 5.362$^{+0.422}_{-0.369}$ & 14.685$^{+0.740}_{-0.699}$ & 0.259$^{+0.202}_{-0.175}$ & 19.473$^{+0.001}_{-0.001}$ & 1308.841$^{+0.005}_{-0.005}$ & 0.27$^{+0.24}_{-0.14}$ & [-6,65], [116,190] & - & - & 3.7 & 13 & - \\
\enddata
\tablecomments{Stellar parameters $R_\star$ and $\rho_\star$ are derived from the isochrone fitting (Section~\ref{sec:selection}). Planet parameters $\rho_{\textrm{circ}}$, $\delta$, $R_{\textrm{p}}$, $b$, $P$, $T_c$, $e$, and $\omega$ are derived from the light curve fitting (Section~\ref{sec:lc}). We report the medians and 68\% credible intervals of posteriors for all parameters except for $e$ and $\omega$. For $e$, we report the modes and 68\% highest posterior density (HPD) intervals. For $\omega$, we report the 68\% HPD intervals of which have taken account of multimodal distributions.}
\tablerefs{Reference abbreviations used in this able are listed. Br19: \cite{brah19}; Ni9: \cite{niel19}; Ro19: \cite{rodr19}; Le14: \cite{lend14}; Qu10: \cite{quel10}; Ne19: \cite{newt19}; Ho21: \cite{hobs21}; Da21: \cite{daws21}; Da19: \cite{daws19}; Ki19: \cite{kipp19}; Ad20: \cite{addi20}; Br20: \cite{brah20}; Sm14: \cite{smit14}; Jo20: \cite{jord20}; He17: \cite{hell17}; Hu20: \cite{huan20}; He19: \cite{hell19}.}
\end{deluxetable*}
\end{longrotatetable}
\global\pdfpageattr\expandafter{\the\pdfpageattr/Rotate 0}
\begin{deluxetable*}{lcc}
\tablecaption{Warm Jupiter candidates with missing information. \label{tbl:pc_missing}}
\tabletypesize{\small}
\tablehead{
\colhead{TIC ID} & \colhead{TOI name} & \colhead{Missing Information}
}
\startdata
149601557 & TOI-1033.01 & $\sigma_{T_{\textrm{eff}}}$\\
296863792 & - & $T_{\textrm{eff}}$, $\sigma_{T_{\textrm{eff}}}$, Gmag, $\sigma_{\textrm{Gmag}}$ \\
306919690 & - & $T_{\textrm{eff}}$, $\sigma_{T_{\textrm{eff}}}$, plx, $\sigma_{\textrm{plx}}$\\
270341214 & TOI-173.01 & 14 possible orbital periods\tablenotemark{\scriptsize $*$}\\
262746281 & TOI-603.01 & $P$: 8.09 or 16.18 days\tablenotemark{\scriptsize $*$}\\
308994098 & TOI-790.01 & $P$: 99.77 or 199.55 days\\
437329044 & TOI-1982.01 & $P$: 8.58 or 17.16 days\\
39218269  & TOI-2366.01 & $P$: 8.60 or 17.19 days\\
99133239  & - & $P$: 9.17 or 18.34 days\\
398466662 & - & $P$: 8.77 or 17.54 days\\
412635642 & - & $\sigma_{T_{\textrm{eff}}}$, plx, $\sigma_{\textrm{plx}}$, $P$: 8.71 or 17.42 days \\
\enddata
\tablenotetext{$*$}{Unique orbital period is later determined by ground-based follow-up observations.}
\tablecomments{Gmag and plx stand for \emph{Gaia} DR2 apparent \emph{G} magnitude and parallax. $P$ stands for orbital period, which is unconstrained due to observation gaps.}
\end{deluxetable*}
\begin{deluxetable*}{llcc}
\centerwidetable
\tablecaption{False-Positives (FPs) and False-Alarms (FAs) identified by TFOP WG SG1 and SG2 \label{tbl:fp}}
\tablewidth{0pt}
\tabletypesize{\small}
\tablehead{
\colhead{TIC ID} & \colhead{TOI name} & \colhead{SG1/2 Disposition} & \colhead{Comments/Reference}
}
\decimalcolnumbers
\startdata
207081058 & TOI-121.01 & SB1 & - \\
281710229 & TOI-308.01 & BEB & - \\
167418898 & TOI-383.01 & NPC, BD/SB1 & - \\
271900960 & TOI-389.01 & NEB & - \\
123482865 & TOI-569b & BD & \citet{carm20a} \\
196286587 & TOI-592.01 & SB1 & - \\
101395259 & TOI-623.01 & SEB1 & - \\
293853437 & TOI-629.01 & BD & Carmichael et al. 2021 in prep \\
151681127 & TOI-671.01 & SB1 & - \\
151959065 & TOI-673.01 & SEB1 & - \\
308050066 & TOI-679.01 & SB1 & - \\
55383975 & TOI-694b & SB1 & \citet{mire20} \\
309402106 & TOI-710.01 & SEB1 & - \\
131081852 & TOI-758.01 & SB2 & - \\
294780517 & TOI-792.01 & NEB & - \\
143526444 & TOI-803.01 & NEB & - \\
100757807 & TOI-811b & BD & \citet{carm20b} \\
461271719 & TOI-838.01 & SB1 & - \\
216935214 & TOI-902.01 & SB2 & - \\
399144800 & TOI-1213.01 & SEB1 & - \\
231736113 & TOI-1406b & BD & \citet{carm20a} \\
235067594 & - & NEB & CTOI; \citet{mont20}; actual star TIC-235067595 \\
23733479 & - & NEB & actual star TIC-23733473 \\
92833442 & - & NEB & actual star TIC-92833424 \\
140344868 & - & NEB & actual star TIC-140344846 \\
177350401 & - & NEB & actual star TIC-177350397 \\
308885493 & - & BEB & actual star TIC-308885490 \\
394662124 & - & NEB & actual star TIC-394662125 \\
418883593 & - & NEB & actual star TIC-414477523 \\
425170378 & - & FA & - \\
\enddata
\tablecomments{{\bf Photometric Dispositions} EB: Eclipsing Binary; BEB: Blended EB; NEB: Nearby EB; NPC: Nearby planet candidate; FA: False Alarm;
{\bf Spectroscopic Dispositions} SB: Spectroscopy Binary; SB1: Single-lined spectra showing RV variation too large to be caused by a planet; SEB1: SB1 with orbital solution; SB2: Double-lined SB moving in phase with the photometric orbit; BD: Brown Dwarf}
\end{deluxetable*}

\appendix

\restartappendixnumbering
\section{Light Curves of non-TOI Candidates} \label{appendix:lc}
\begin{figure}[hb!]
\hspace*{-0.5cm}
\plotone{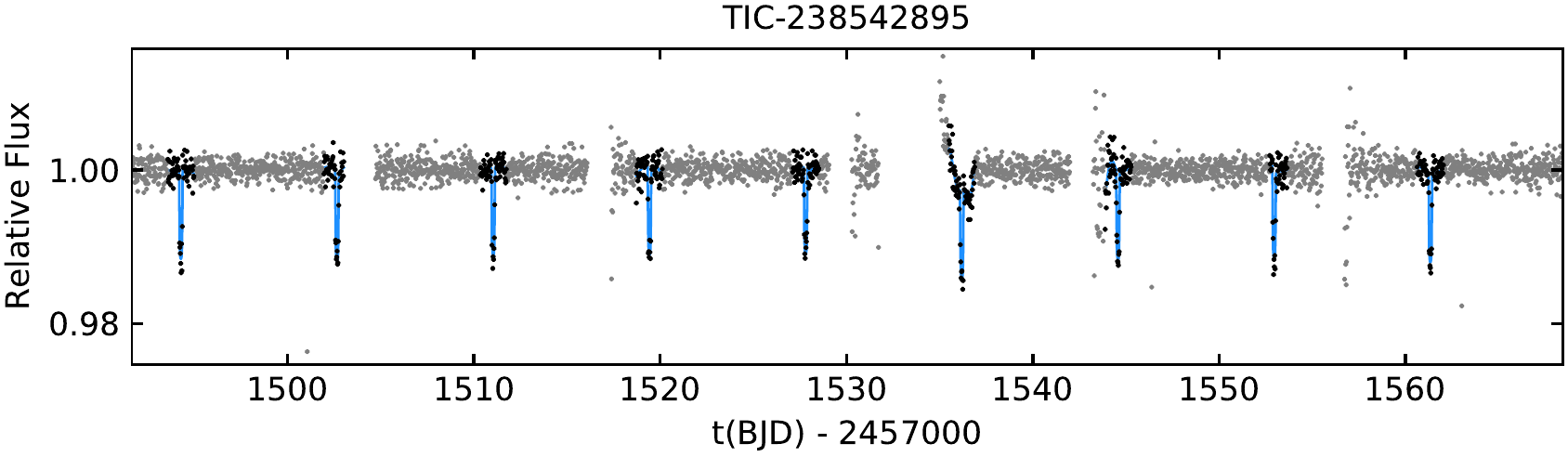}
\caption{\TESS Full-Frame-Image light curves of TIC-238542895 processed by the Quick Look Pipeline. A full light curve is plotted as grey dots, the trimmed light curves used for modeling are colored in black, and the best-fitted light curve models are shown as blue lines. The complete figure set (21 figures) is available in the online journal.\label{fig:lc_demo}}
\end{figure}

We present the FFI light curves of non-TOI candidates obtained from the QLP. In each figure, we plot the target's full light curve in grey dots, the trimmed light curves in black dots, and the best-fitted light curve models in blue lines. See Figure~\ref{fig:lc_demo} for demonstration. The complete figure set for all non-TOI candidates (\changes{19} figures) is available in the online journal.

\figsetstart
\figsetnum{1}
\figsettitle{Light curves of non-TOI candidates}

\figsetgrpstart
\figsetgrpnum{1.1}
\figsetgrptitle{TIC-238542895}
\figsetplot{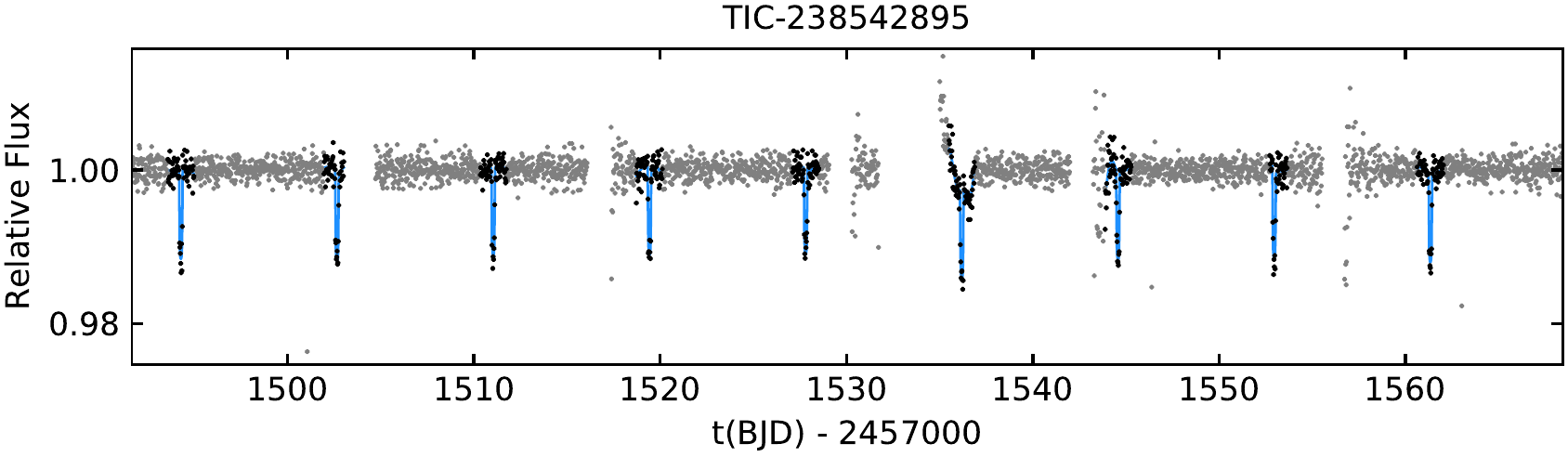}
\figsetgrpnote{\TESS FFI light curves of TIC-238542895}
\figsetgrpend

\figsetgrpstart
\figsetgrpnum{1.2}
\figsetgrptitle{TIC-282498590}
\figsetplot{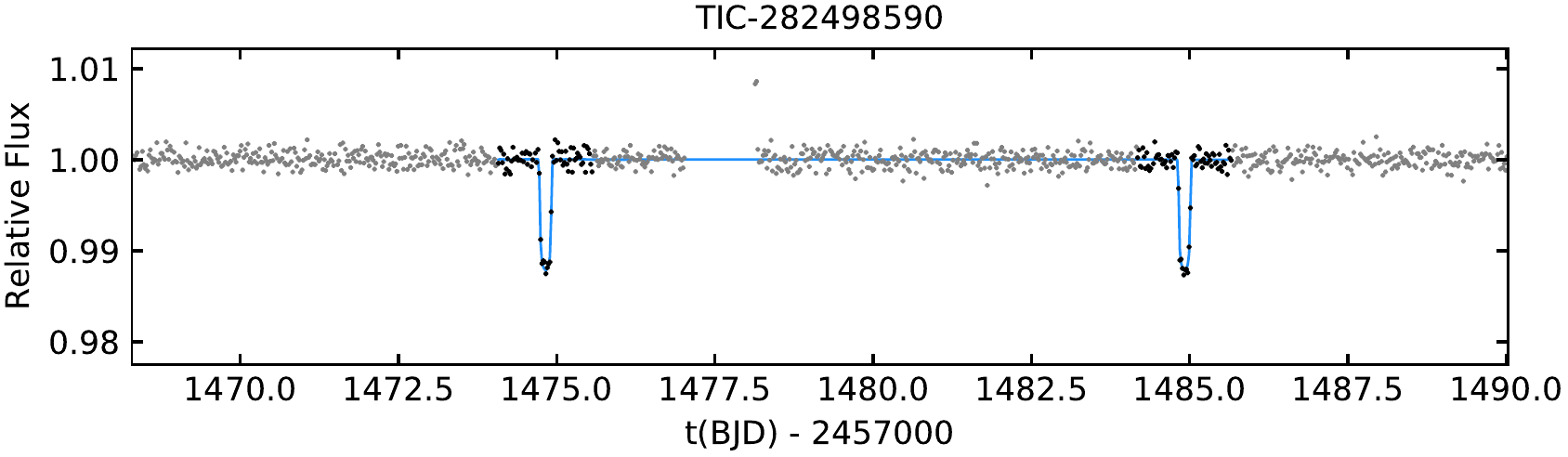}
\figsetgrpnote{\TESS FFI light curves of TIC-282498590}
\figsetgrpend

\figsetgrpstart
\figsetgrpnum{1.3}
\figsetgrptitle{TIC-290403522}
\figsetplot{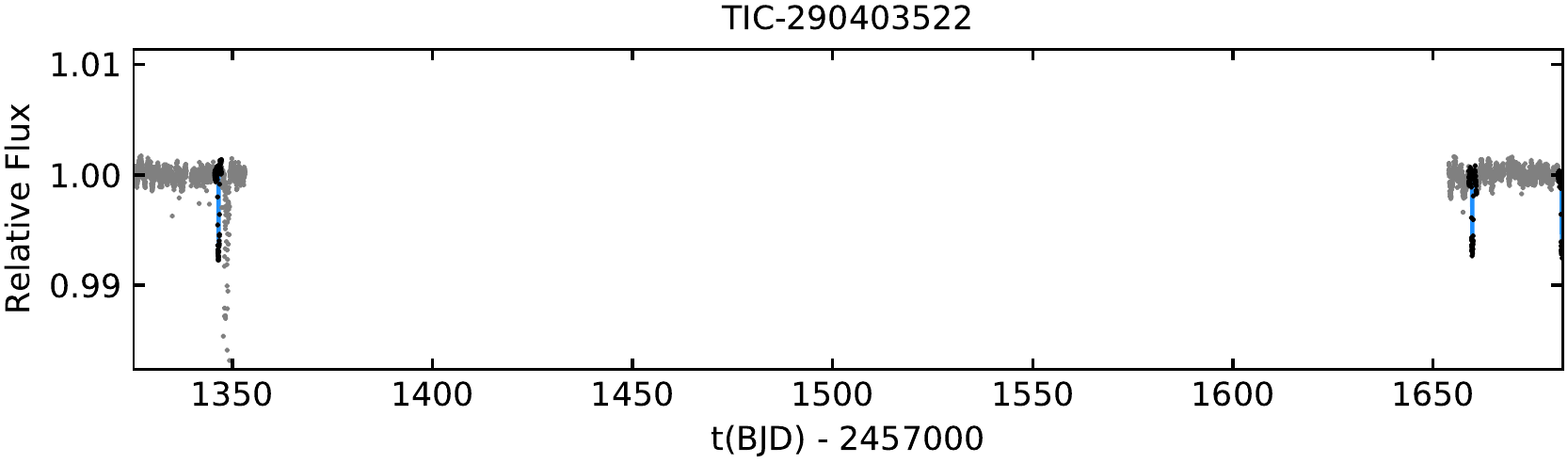}
\figsetgrpnote{\TESS FFI light curves of TIC-290403522}
\figsetgrpend

\figsetgrpstart
\figsetgrpnum{1.4}
\figsetgrptitle{TIC-380836882}
\figsetplot{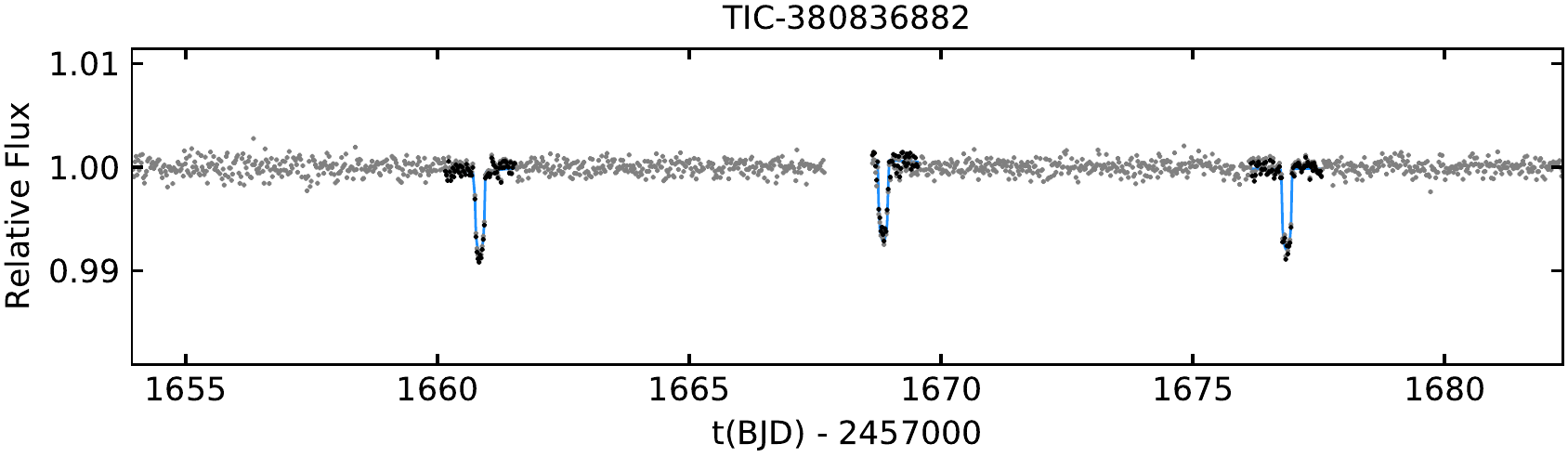}
\figsetgrpnote{\TESS FFI light curves of TIC-380836882}
\figsetgrpend

\figsetgrpstart
\figsetgrpnum{1.5}
\figsetgrptitle{TIC-382200986}
\figsetplot{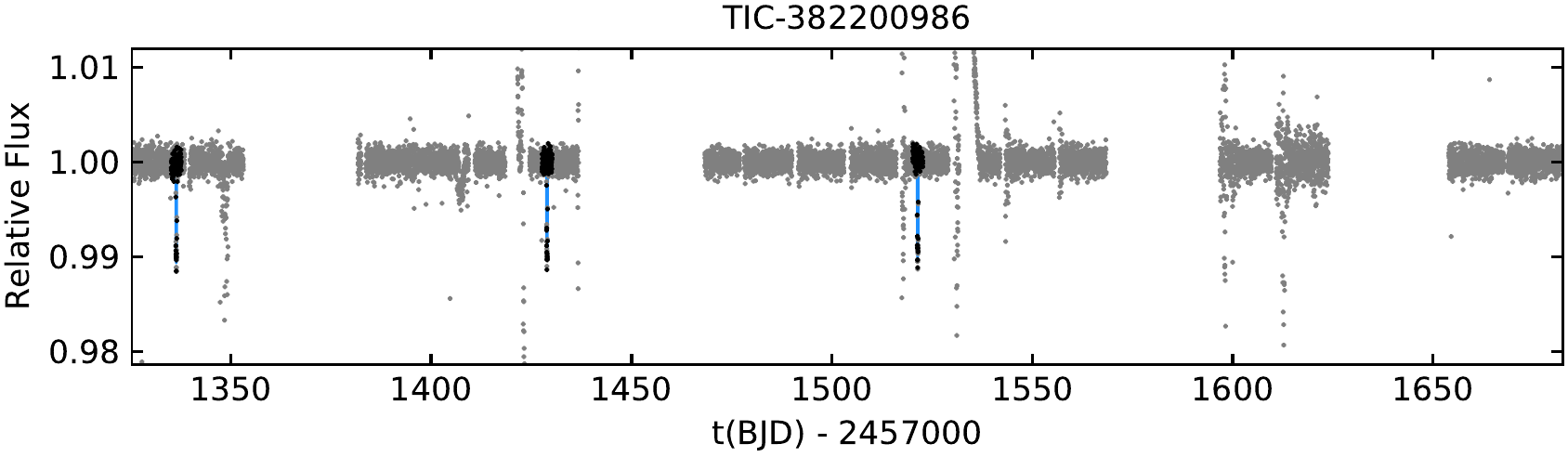}
\figsetgrpnote{\TESS FFI light curves of TIC-382200986}
\figsetgrpend

\figsetgrpstart
\figsetgrpnum{1.6}
\figsetgrptitle{TIC-357202877}
\figsetplot{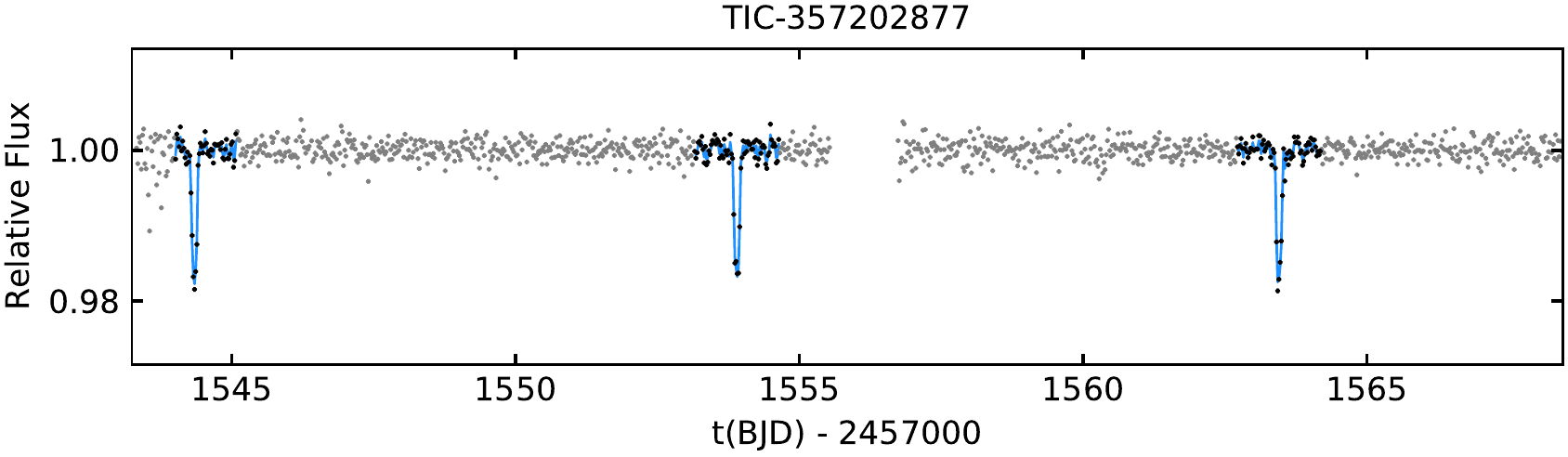}
\figsetgrpnote{\TESS FFI light curves of TIC-357202877}
\figsetgrpend

\figsetgrpstart
\figsetgrpnum{1.7}
\figsetgrptitle{TIC-254142310}
\figsetplot{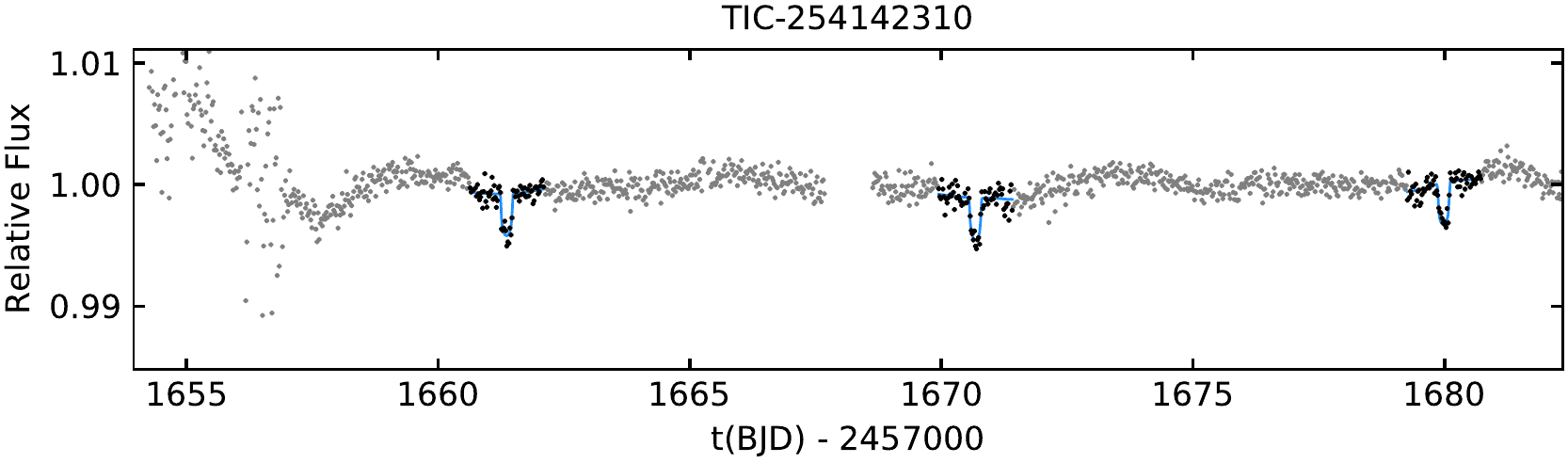}
\figsetgrpnote{\TESS FFI light curves of TIC-254142310}
\figsetgrpend

\figsetgrpstart
\figsetgrpnum{1.8}
\figsetgrptitle{TIC-395113305}
\figsetplot{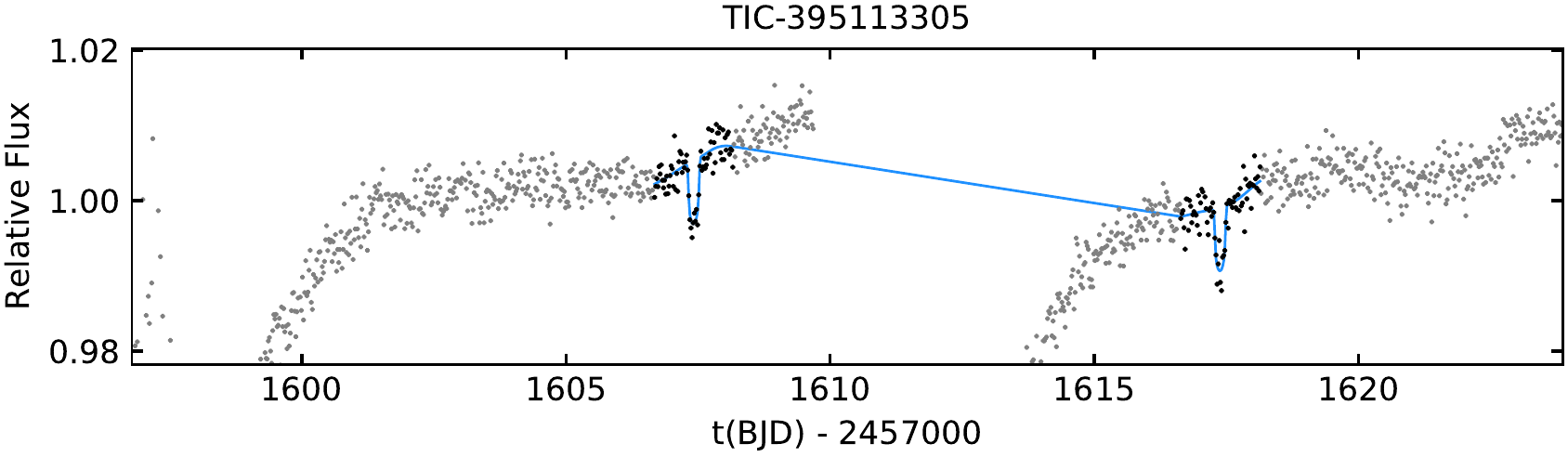}
\figsetgrpnote{\TESS FFI light curves of TIC-395113305}
\figsetgrpend

\figsetgrpstart
\figsetgrpnum{1.9}
\figsetgrptitle{TIC-180989820}
\figsetplot{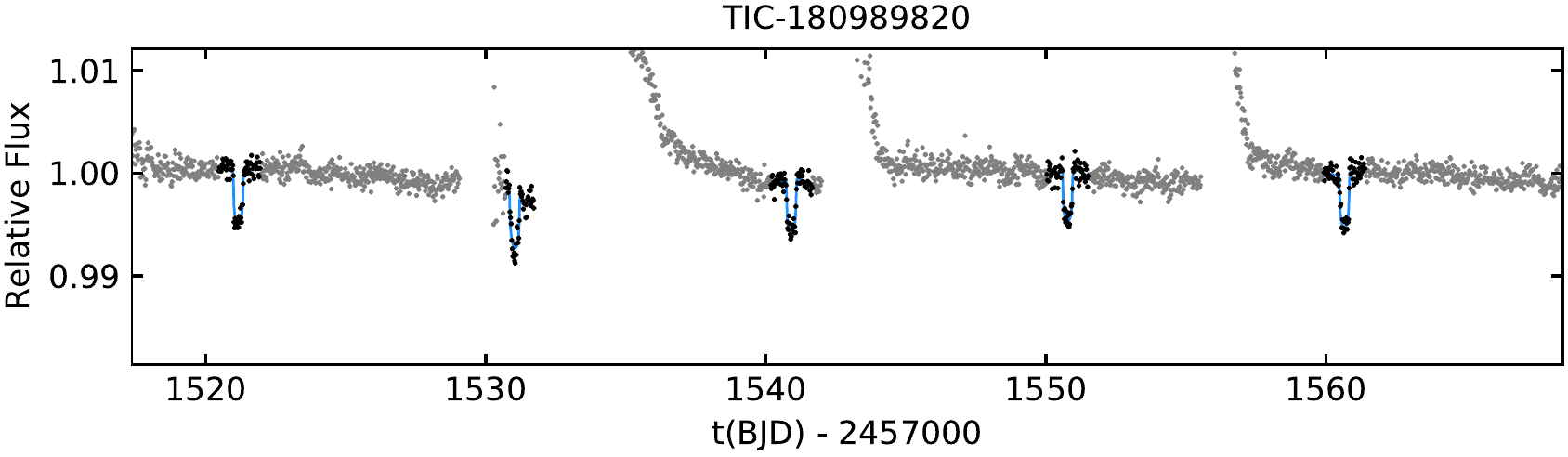}
\figsetgrpnote{\TESS FFI light curves of TIC-180989820}
\figsetgrpend

\figsetgrpstart
\figsetgrpnum{1.10}
\figsetgrptitle{TIC-4598935}
\figsetplot{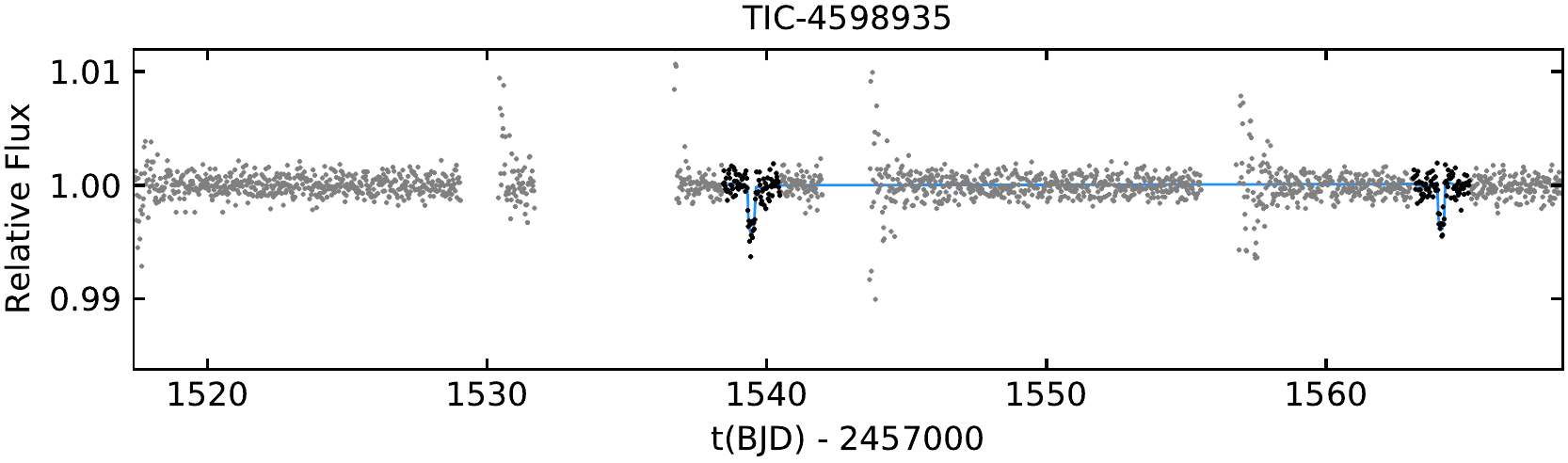}
\figsetgrpnote{\TESS FFI light curves of TIC-4598935}
\figsetgrpend

\figsetgrpstart
\figsetgrpnum{1.11}
\figsetgrptitle{TIC-464300749}
\figsetplot{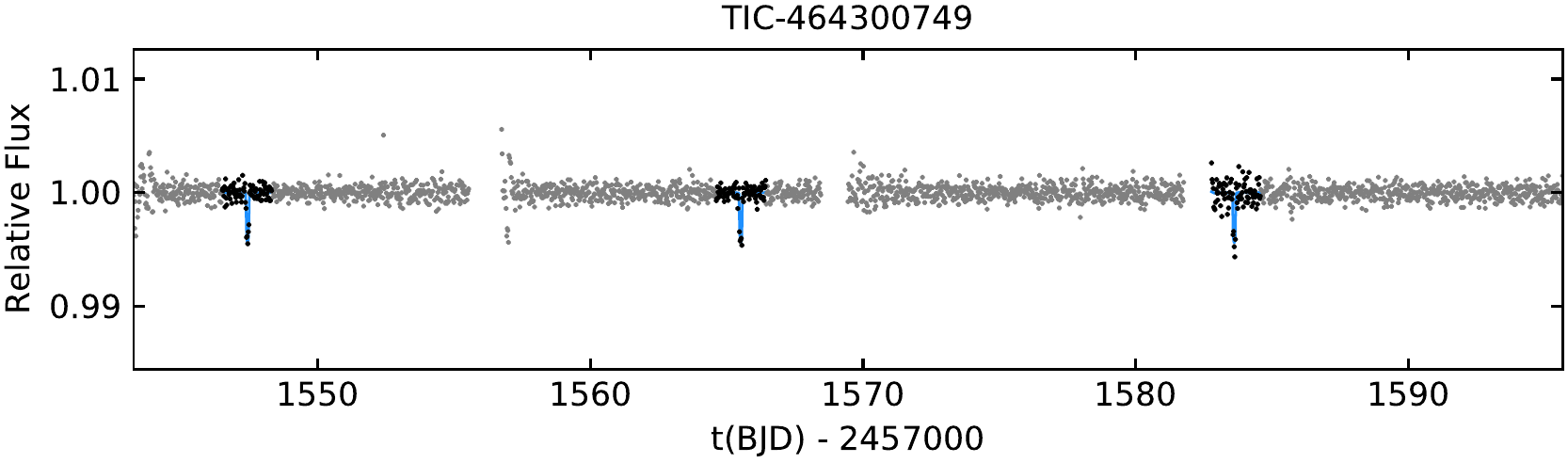}
\figsetgrpnote{\TESS FFI light curves of TIC-464300749}
\figsetgrpend

\figsetgrpstart
\figsetgrpnum{1.12}
\figsetgrptitle{TIC-7536985}
\figsetplot{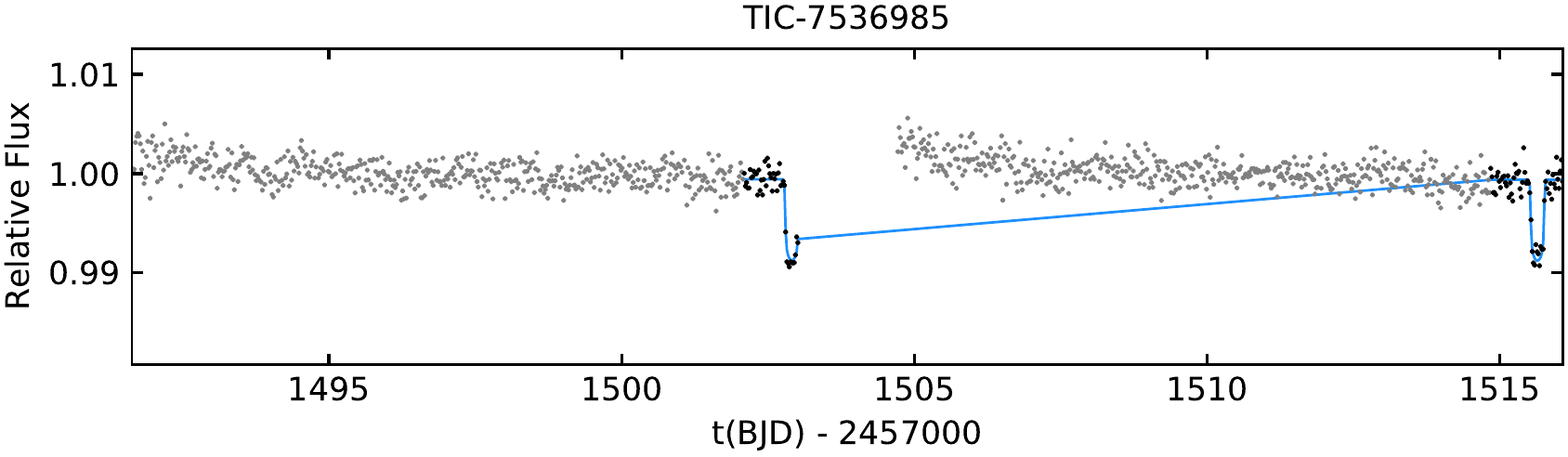}
\figsetgrpnote{\TESS FFI light curves of TIC-7536985}
\figsetgrpend

\figsetgrpstart
\figsetgrpnum{1.13}
\figsetgrptitle{TIC-76228620}
\figsetplot{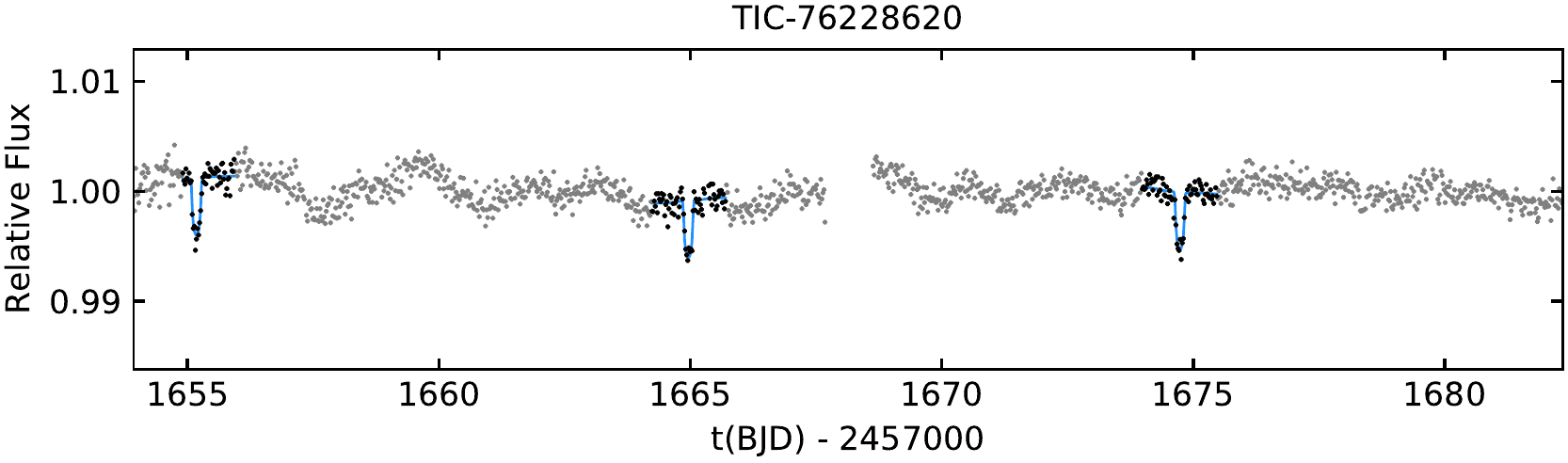}
\figsetgrpnote{\TESS FFI light curves of TIC-76228620}
\figsetgrpend

\figsetgrpstart
\figsetgrpnum{1.14}
\figsetgrptitle{TIC-87422071}
\figsetplot{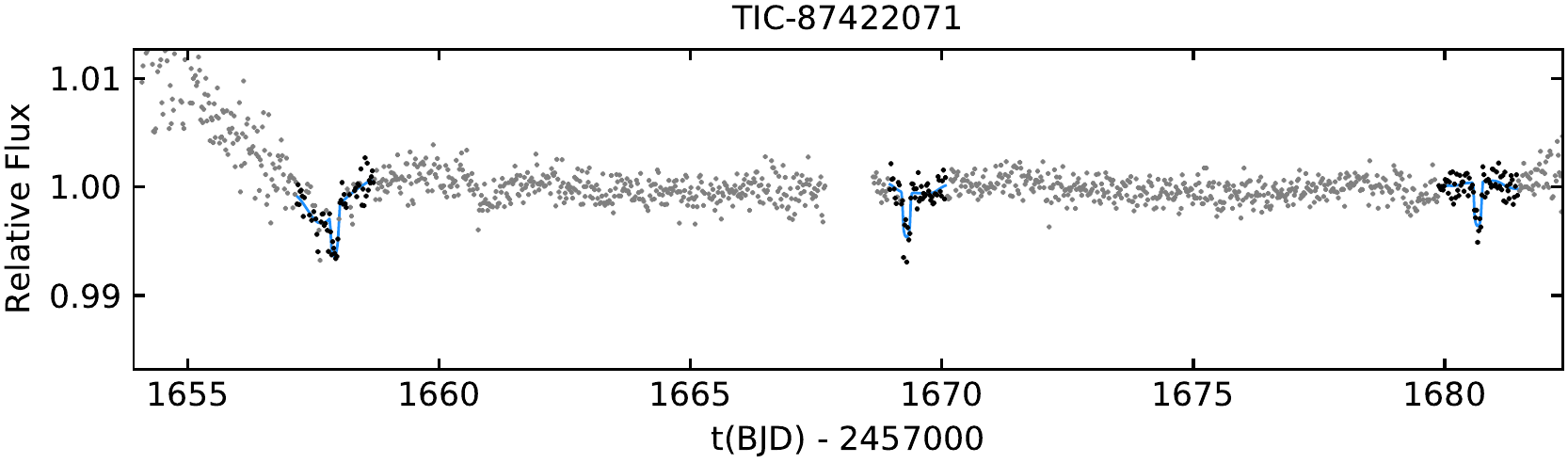}
\figsetgrpnote{\TESS FFI light curves of TIC-87422071}
\figsetgrpend

\figsetgrpstart
\figsetgrpnum{1.15}
\figsetgrptitle{TIC-253126207}
\figsetplot{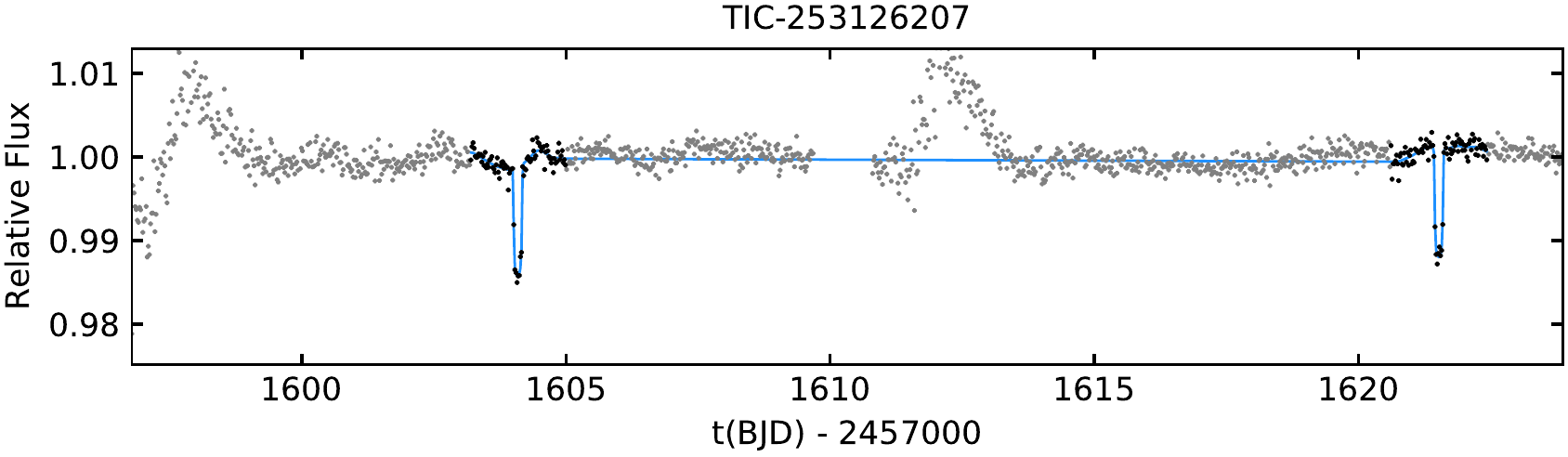}
\figsetgrpnote{\TESS FFI light curves of TIC-253126207}
\figsetgrpend

\figsetgrpstart
\figsetgrpnum{1.16}
\figsetgrptitle{TIC-342167886}
\figsetplot{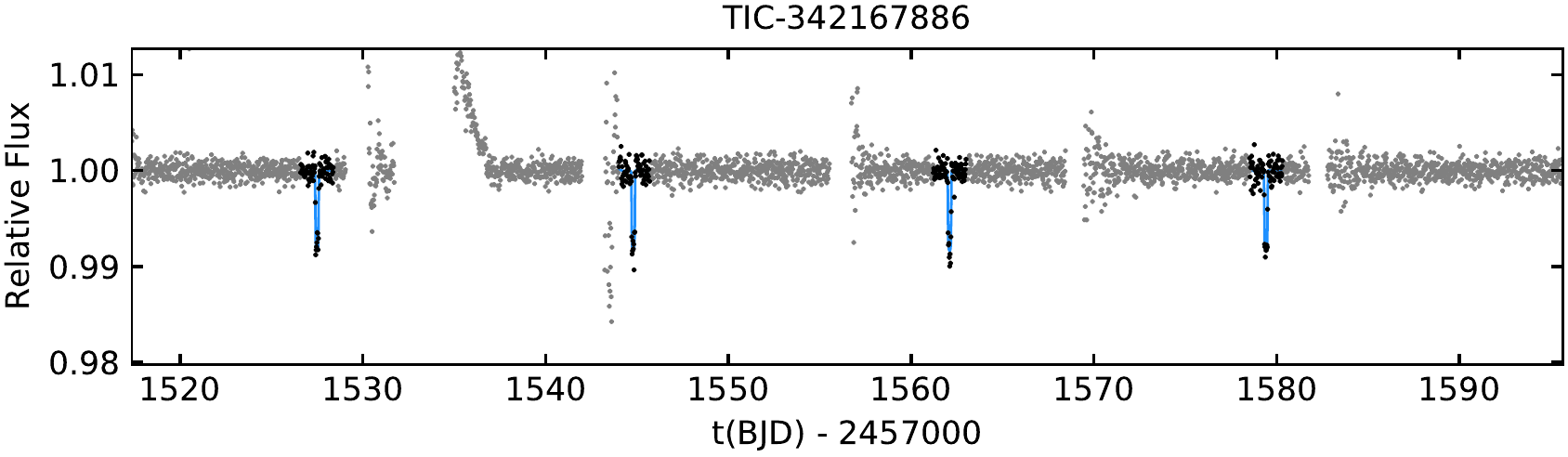}
\figsetgrpnote{\TESS FFI light curves of TIC-342167886}
\figsetgrpend

\figsetgrpstart
\figsetgrpnum{1.17}
\figsetgrptitle{TIC-343936388}
\figsetplot{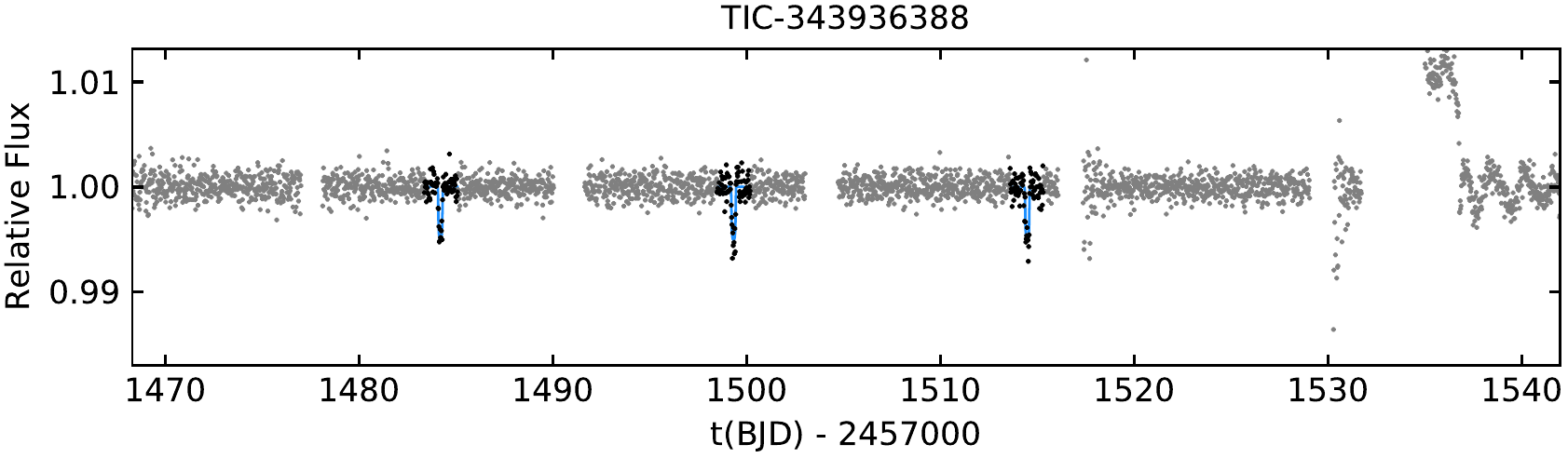}
\figsetgrpnote{\TESS FFI light curves of TIC-343936388}
\figsetgrpend

\figsetgrpstart
\figsetgrpnum{1.18}
\figsetgrptitle{TIC-394492336}
\figsetplot{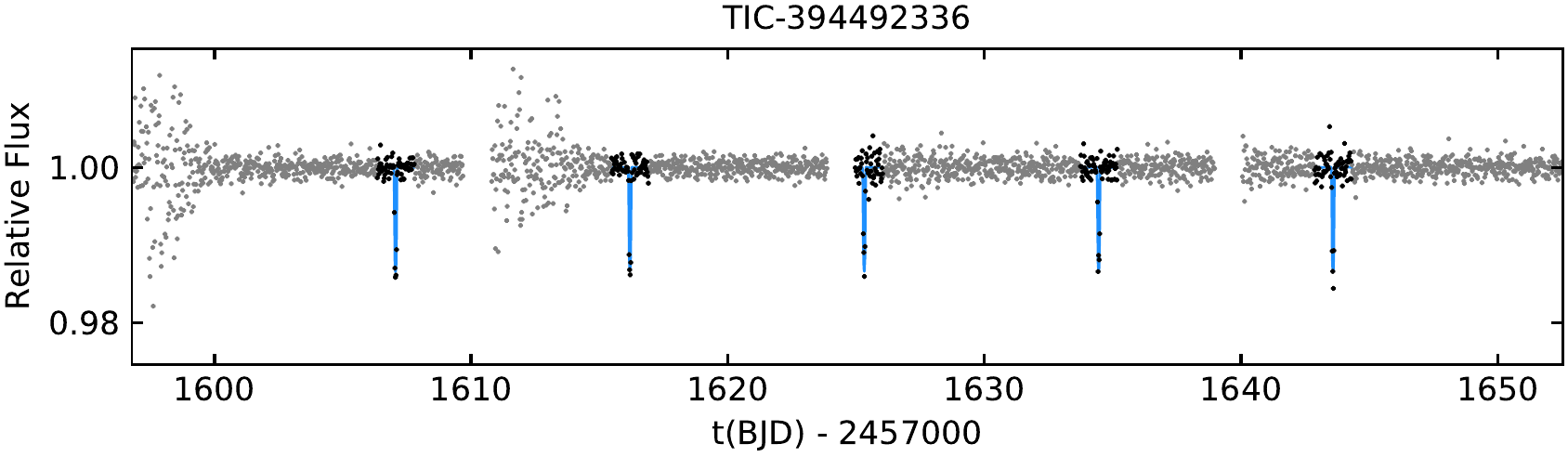}
\figsetgrpnote{\TESS FFI light curves of TIC-394492336}
\figsetgrpend

\figsetgrpstart
\figsetgrpnum{1.19}
\figsetgrptitle{TIC-150299840}
\figsetplot{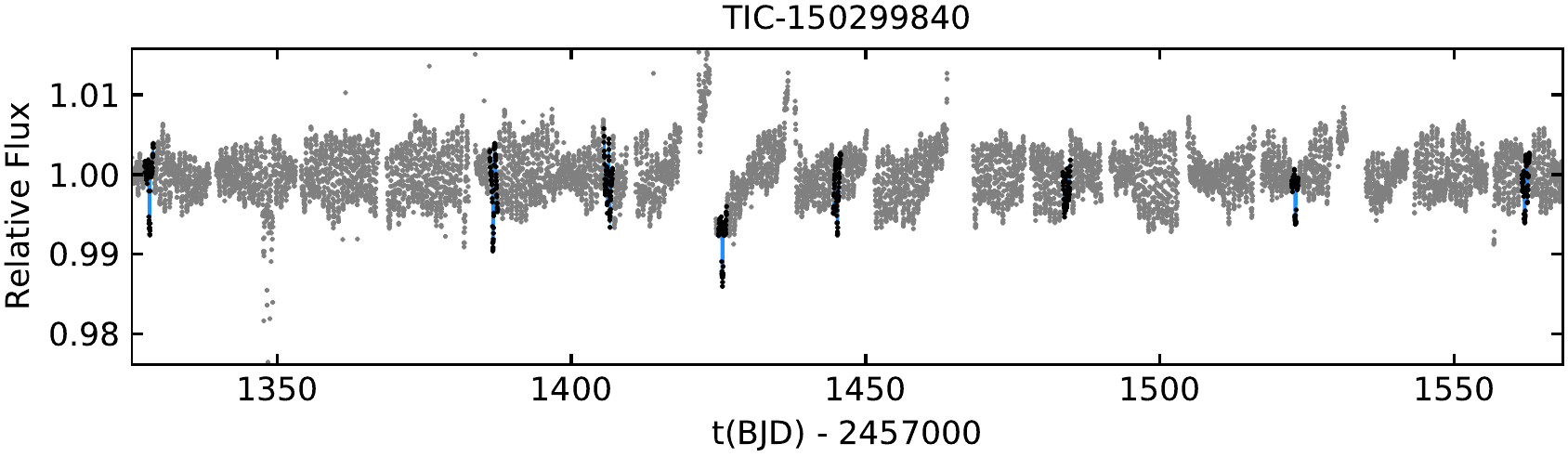}
\figsetgrpnote{\TESS FFI light curves of TIC-150299840}
\figsetgrpend

\figsetend

\restartappendixnumbering

\section{Hyperparameter posterior distributions}
\begin{figure}[hb!]
    \centering
    \includegraphics{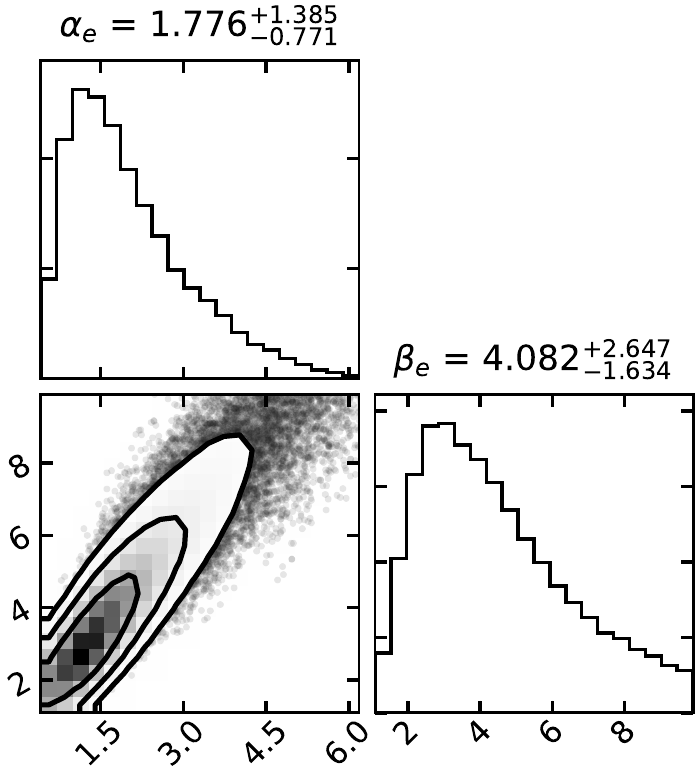}
    \caption{Posterior distributions of hyperparameters using the Beta distribution as the functional form of the eccentricity distribution. The probability distribution follows $p(e|\alpha_e,\beta_e)=e^{\alpha-1}(1-e)^{\beta-1}/B(\alpha,\beta)$ where $B(\alpha,\beta)=\Gamma(\alpha)\Gamma(\beta)/\Gamma(\alpha+\beta)$ and $\Gamma$ is the Gamma function.}
    \label{fig:corner_beta}
\end{figure}

\begin{figure}
    \centering
    \includegraphics[scale=0.9]{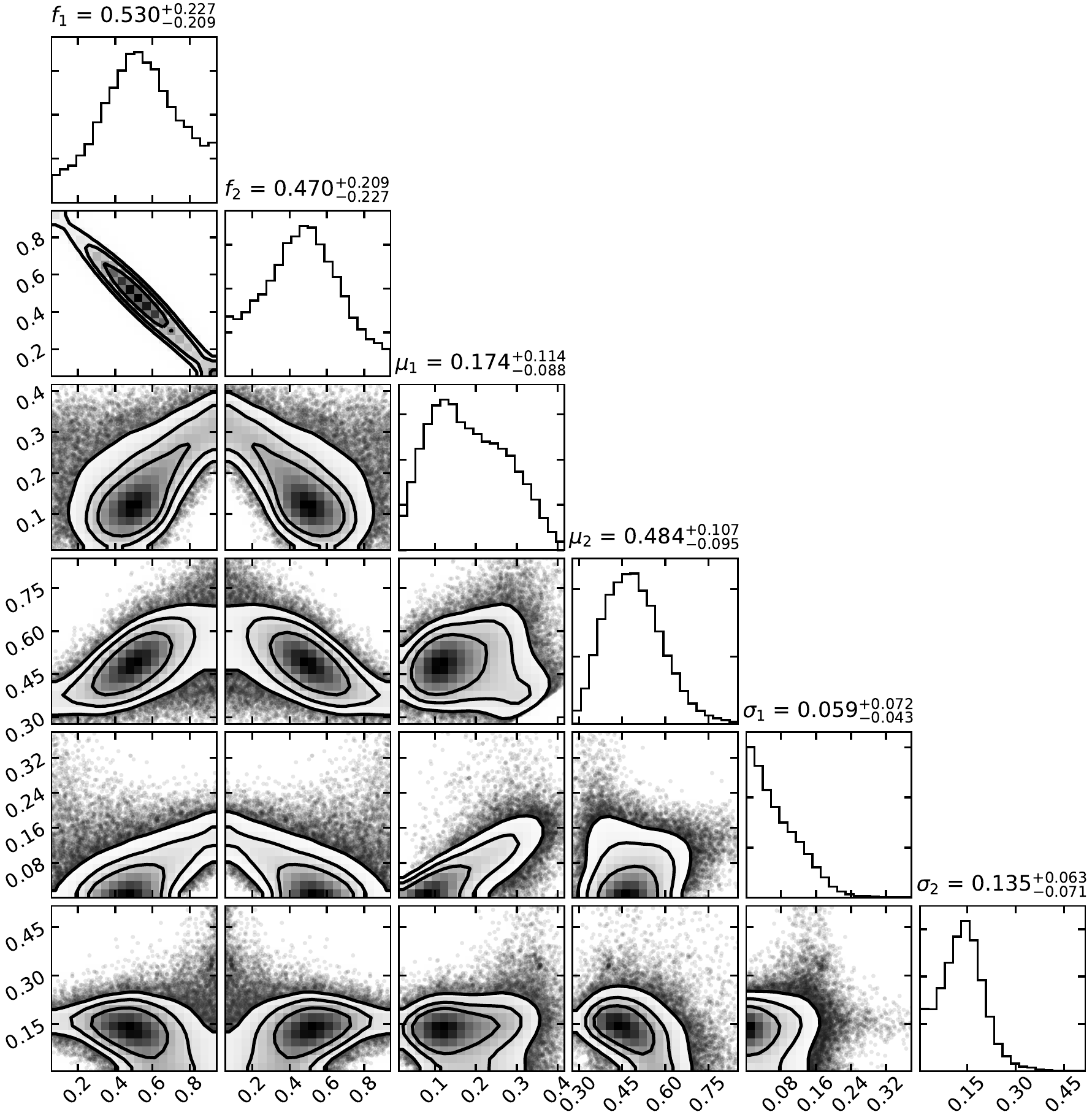}
    \caption{Posterior distributions of hyperparameters using the two-component mixture distribution as the functional form of the eccentricity distribution. The probability distribution follows $p(e|f_1, f_2, \mu_1, \mu_2, \sigma_1, \sigma_2) = f_1\mathcal{N}(e|\mu_1,\sigma_1^2)+f_2\mathcal{N}(e|\mu_2,\sigma_2^2)$ where $\mathcal{N}$ is the Normal distribution.}
    \label{fig:corner_mixture}
\end{figure}

Corner plots for the posterior distributions of the hyperparameters assuming the Beta distribution as the functional form of the eccentricity distribution (Figure \ref{fig:corner_beta}) and the two-component mixture distribution as the functional form (Figure \ref{fig:corner_mixture}). See Section~\ref{sec:ecc_dist} for more details.

\restartappendixnumbering

\section{Incorporating eccentricity measurements from different sources} \label{appendix:hbm}
\begin{figure}
  \centering
  \resizebox{3in}{!}{%
    \begin{tikzpicture}[thick]
          \node[input]                                         (ocirc) {$\hat{\rho}_{\mathrm{circ},i}$, $\sigma_{\rho_{\mathrm{circ},i}}$};
          \factor[above=0.6 of ocirc] {ocirc-f} {left:$\mathcal{N}$} {} {}; %
        
          \node[det, above=1.2 of ocirc, xshift=0cm]           (rhocirc) {$\rho_{\mathrm{circ},i}$};
          
          \node[input, above=0.5 of ocirc, xshift=3cm] (trans) {obs$_i$};
          
          \node[latent, above=0.9 of rhocirc, xshift=0cm]      (e) {$e_i$};
          \node[latent, above=0.9 of rhocirc, xshift=2cm]      (w) {$\omega_i$};
          \node[latent, above=0.9 of rhocirc, xshift=-2cm]     (rhostar) {$\rho_{\star,i}$};
          \node[input, above=0.9 of rhocirc, xshift=4cm]  (P) {$P_i$};
    
          \node[input, above=-2.8 of rhostar, xshift=0cm] (star) {$\hat{\rho}_{\star,i}$, $\sigma_{\rho_{\star,i}}$} ; %
          \factor[above=-1.4 of rhostar] {rhostar-f} {left:$\mathcal{N}$} {} {} ; %
        
          \edge {rhocirc} {ocirc} ;
          \edge {e, w, rhostar} {rhocirc};
          
          \factoredge {rhostar} {rhostar-f} {star} ; %
          \factor[above=0.5 of trans]{trans-f} {right:$P_{\textrm{transit}}$}{e, w, P, rhostar} {trans};
        
          \node[const, left=1.9cm of ocirc] (a) {$\textcolor{white}{.}$} ;
          \node[const, right=3.5cm of ocirc] (b) {$\textcolor{white}{.}$} ;
          \node[const, above=4.5cm of ocirc, xshift=2.7cm] (c) {$\textcolor{white}{.}$} ;
          
          \node[latent, above=0.9 of rhocirc, xshift= 6cm]      (ej) {$e_j$} ;
          \node[input, xshift= 6cm, above=-0.5cm of ocirc] (obs_ej) {$\hat{e}_j$, $\sigma_{\hat{e}_{j}}$};
          \factor[above=0.5 of obs_ej] {ej-f} {left:$\mathcal{N}$} {} {} ; %
          \edge{ej} {obs_ej} ;
          
          \node[const, xshift= 6.8cm, above=-0.88cm of ocirc] (m) {$\textcolor{white}{.}$} ;
          \node[const, xshift= 5.25cm, above=4.5cm of ocirc] (n) {$\textcolor{white}{.}$} ;
          
          \plate {} {(ocirc)(rhocirc)(e)(w)(rhostar)(a)(b)(c)} {$i$ of $N$} ;
          \plate {} {(ej)(obs_ej)(m)(n)} {$j$ of $M$} ;
          
          \node[latent, above=1.5 of e, xshift= -0.5cm] (ae) {$\alpha_e$} ; %
          \node[latent, above=1.5 of e, xshift= 0.5cm] (be) {$\beta_e$} ; %
          \factor[above=1. of e, xshift= 0cm] {e-f} {left:Beta} {ae, be} {e, ej} ; %
          
    \end{tikzpicture}
}
  \caption{Graphic model of the extended hierarchical Bayesian model. The model is an extension of the model shown in Figure~\ref{fig:hbm}. Since a few planets in our catalog have further constrained eccentricities from ground-based follow-up observations, we adopt their eccentricity posteriors to evaluate the eccentricity distribution directly. For planets without further constrained eccentricities, we continue to adopt the photoeccentric approach.
  }\label{fig:hbm_ecc}
\end{figure}
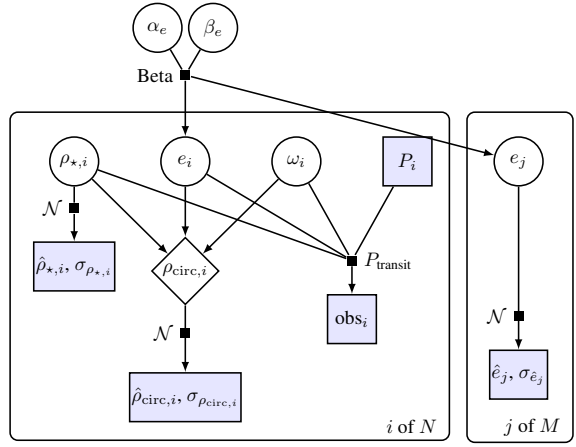

\changes{
Benefiting from the extensive follow-up observations of \TESS planet candidates, 12 targets in our catalog have been confirmed. These confirmed planets have radial-velocity or transit-timing-variation constrained eccentricities that can be used for the eccentricity distribution inference. Here we conduct the hierarchical Bayesian modeling using a similar framework as the one shown in Section~\ref{sec:ecc_dist} but include the information of planets with better constrained eccentricities. In Figure~\ref{fig:hbm_ecc}, we present the graphic model of the extended hierarchical Bayesian model. Planets are separated into two panels: the left panel is for the ones without further constrained eccentricities, so we continue to adopt the photoeccentric approach (similar to Figure~\ref{fig:hbm}); the right panel is for the ones with further constrained eccentricities from ground-based follow-up observations. The median and 1$\sigma$ uncertainty of eccentricities are extracted from literature (see Table~\ref{tbl:pc} Column 17 for references) to construct the observed parameters $\hat{e}_j$ and $\sigma_{\hat{e}_{j}}$. 
Planets that have been followed up have much smaller eccentricity uncertainties ($\sigma_e \sim 0.01$) compared to the ones that have only been analyzed from their light curves ($\sigma_e \sim 0.2$). The inclusion of these precise eccentricity measurements risks biasing the inferred eccentricity distribution, since they are likely to be drawn from a different underlying population than the full sample, and this subset will be censored by different selection effects. Furthermore, there exist more recently discovered likely high-$e$ planets that have yet to be followed up. This model is a demonstration of a method for incorporating eccentricity measurements from different sources, but further experiments would be necessary to robustly account for variations in selection effects.}

\begin{figure}[htb!]
    \centering
    \hspace{-0.2cm}
    \includegraphics{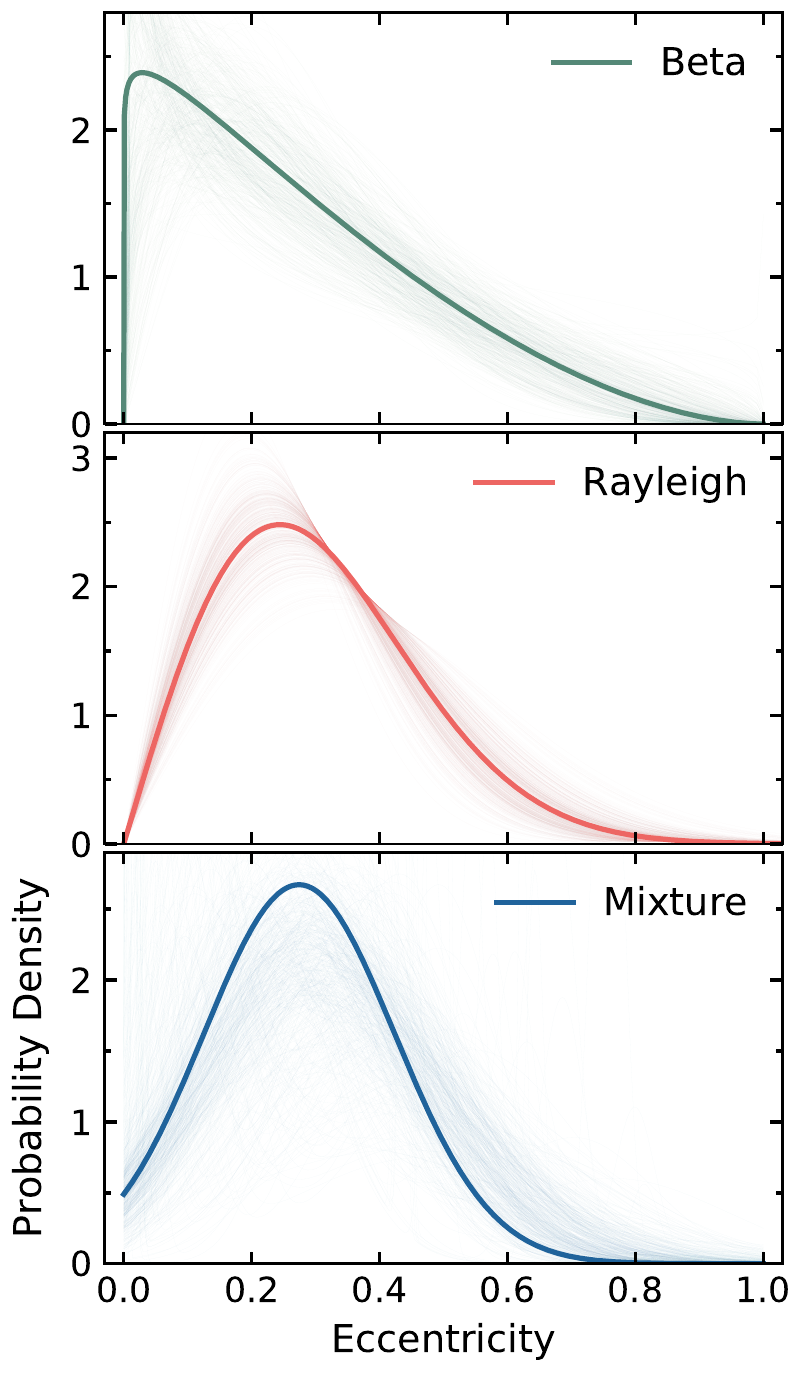}
    \caption{The eccentricity distributions inferred from the extended hierarchical Bayesian model shown in Figure~\ref{fig:hbm_ecc}. The extended model includes the information of confirmed planets with further constrained eccentricities. Comparing to the eccentricity distributions inferred without including these information (i.e., Figure~\ref{fig:hbm_all}), the Rayleigh distribution is consistent with the previous distribution (i.e., the hyperparameter is consistent within 1$\sigma$), whereas both the Beta distribution and the mixture distribution present obvious changes on the distributions. The Beta distribution becomes more right-skewed. The mixture distribution is composed of a single component instead of two and now has a similar shape to the Rayleigh distribution.}
    \label{fig:hbm_all_ecc}
\end{figure}

\begin{figure}
    \centering
    \includegraphics[scale=0.9]{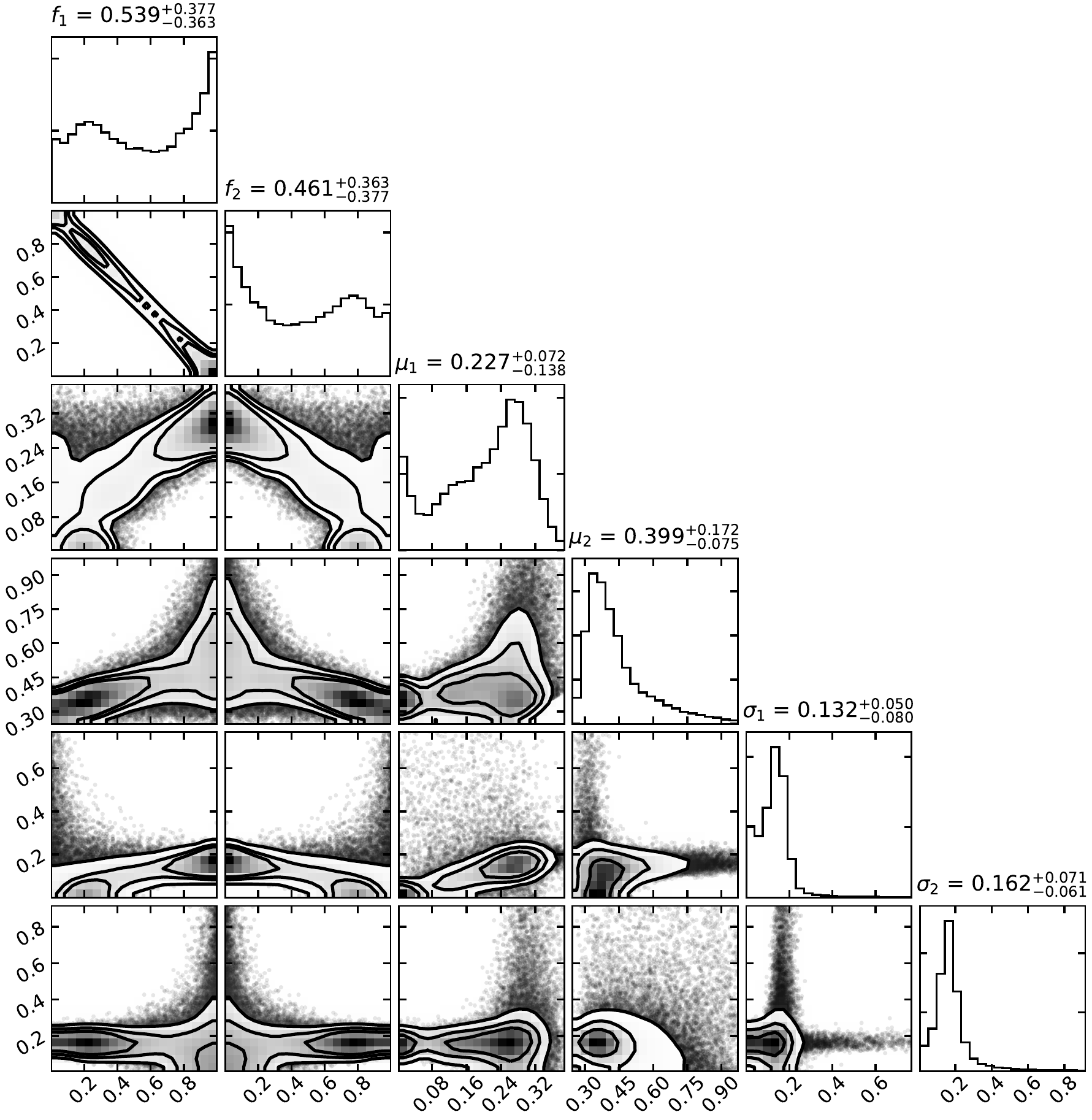}
    \caption{Posterior distributions of hyperparameters using the mixture distribution as the functional form of the eccentricity distribution for the extended hierarchical Bayesian model. The distributions are bimodal with a preference of a single-component Normal distribution.}
    \label{fig:corner_mixture_ecc}
\end{figure}

\changes{
In Figure~\ref{fig:hbm_all_ecc}, we show the eccentricity distributions using three different functional forms inferred from the extended model. Comparing to the eccentricity distributions inferred without including these information (i.e., Figure~\ref{fig:hbm_all}), the Rayleigh distribution is consistent with the previous distribution within 1$\sigma$ uncertainty of the hyperparameter. For the Beta distribution, however, it is more right-skewed and has a mode eccentricity close to zero. For the mixture model, a single-component distribution is favored over the two-component distribution. As shown in Figure~\ref{fig:corner_mixture_ecc}, the joint and marginal posterior distributions of hyperparameters are bimodal but have a strong preference to a single-component model. The best-fitted mixture distribution now has a shape similar to the Rayleigh distribution. The changes on the eccentricity distributions are likely caused by the small eccentricity uncertainties of the confirmed planets, as discussed earlier. To examine the statement, we modify the eccentricity uncertainties of the confirmed planets to the typical eccentricity uncertainties from the photoeccentric analysis (i.e, $\sigma_e \sim 0.2$) and redo the analysis. We find the inferred eccentricity distributions are consistent with the ones shown in Figure~\ref{fig:hbm_all}.}

\begin{deluxetable}{cll}
\tablecaption{Summary of the posteriors of the hyperparameters for the extended hierarchical Bayesian models. \label{tbl:hbm_post_ecc}}
\tabletypesize{\small}
\tablehead{\colhead{Distribution} & \multicolumn{2}{l}{Hyperparameter posteriors}} 
\startdata
Beta & $\alpha_e$ & $\beta_e$ \\
& $1.053^{+0.379}_{-0.285}$ & $2.771^{+1.036}_{-0.778}$ \vspace{0.5em} \\
\hline
Rayleigh & $\sigma_e$ \\
& $0.243^{+0.028}_{-0.025}$ \vspace{0.5em}\\
\hline
Mixture & Mode & 68\% HPD intervals \\
$f_1$ & 0.947, 0.228 & [0.742, 1.000], [0.078, 0.411] \\
$\mu_1$ & 0.271, 0.016 & [0.150, 0.330], [0.013, 0.019] \\
$\sigma_1$ & 0.148, 0.011 & [0.074, 0.193], [0.009, 0.013] \\
$f_2$ & 0.053, 0.772 & [0.000, 0.258], [0.589, 0.922] \\
$\mu_2$ & 0.350 & [0.283, 0.478] \\
$\sigma_2$ & 0.159 & [0.091, 0.224] 
\enddata
\tablecomments{We report the medians and 68\% credible intervals of the posteriors for the Beta and Rayleigh distributions. For the mixture model, the posterior distributions are bimodal, as shown in Figure~\ref{fig:corner_mixture_ecc}. We instead report the modes and the 68\% highest posterior density intervals.}
\end{deluxetable}

\clearpage
\bibliographystyle{aasjournal}
\bibliography{wales}

\begin{thebibliography}{}
\expandafter\ifx\csname natexlab\endcsname\relax\def\natexlab#1{#1}\fi
\providecommand{\url}[1]{\href{#1}{#1}}
\providecommand{\dodoi}[1]{doi:~\href{http://doi.org/#1}{\nolinkurl{#1}}}
\providecommand{\doeprint}[1]{\href{http://ascl.net/#1}{\nolinkurl{http://ascl.net/#1}}}
\providecommand{\doarXiv}[1]{\href{https://arxiv.org/abs/#1}{\nolinkurl{https://arxiv.org/abs/#1}}}

\bibitem[{{Addison} {et~al.}(2021){Addison}, {Wright}, {Nicholson}, {Cale},
  {Mocnik}, {Huber}, {Plavchan}, {Wittenmyer}, {Vanderburg}, {Chaplin},
  {Chontos}, {Clark}, {Eastman}, {Ziegler}, {Brahm}, {Carter}, {Clerte},
  {Espinoza}, {Horner}, {Bentley}, {Jord{\'a}n}, {Kane}, {Kielkopf},
  {Laychock}, {Mengel}, {Okumura}, {Stassun}, {Bedding}, {Bowler}, {Burnelis},
  {Blanco-Cuaresma}, {Collins}, {Crossfield}, {Davis}, {Evensberget},
  {Heitzmann}, {Howell}, {Law}, {Mann}, {Marsden}, {Matson}, {O'Connor},
  {Shporer}, {Stevens}, {Tinney}, {Tylor}, {Wang}, {Zhang}, {Henning},
  {Kossakowski}, {Ricker}, {Sarkis}, {Schlecker}, {Torres}, {Vanderspek},
  {Latham}, {Seager}, {Winn}, {Jenkins}, {Mireles}, {Rowden}, {Pepper},
  {Daylan}, {Schlieder}, {Collins}, {Collins}, {Tan}, {Ball}, {Basu}, {Buzasi},
  {Campante}, {Corsaro}, {Gonz{\'a}lez-Cuesta}, {Davies}, {de Almeida}, {do
  Nascimento}, {Garc{\'\i}a}, {Guo}, {Handberg}, {Hekker}, {Hey}, {Kallinger},
  {Kawaler}, {Kayhan}, {Kuszlewicz}, {Lund}, {Lyttle}, {Mathur}, {Miglio},
  {Mosser}, {Nielsen}, {Serenelli}, {Aguirre}, \& {Theme{\ss}l}}]{addi20}
{Addison}, B.~C., {Wright}, D.~J., {Nicholson}, B.~A., {et~al.} 2021, \mnras,
  502, 3704, \dodoi{10.1093/mnras/staa3960}

\bibitem[{{Agol} {et~al.}(2020){Agol}, {Luger}, \&
  {Foreman-Mackey}}]{exoplanet:agol20}
{Agol}, E., {Luger}, R., \& {Foreman-Mackey}, D. 2020, \aj, 159, 123,
  \dodoi{10.3847/1538-3881/ab4fee}

\bibitem[{{Anderson} \& {Lai}(2017)}]{ande17}
{Anderson}, K.~R., \& {Lai}, D. 2017, \mnras, 472, 3692,
  \dodoi{10.1093/mnras/stx2250}

\bibitem[{{Anderson} {et~al.}(2019){Anderson}, {Lai}, \& {Pu}}]{ande19}
{Anderson}, K.~R., {Lai}, D., \& {Pu}, B. 2019, arXiv e-prints,
  arXiv:1908.04300.
\newblock \doarXiv{1908.04300}

\bibitem[{{Anderson} {et~al.}(2020){Anderson}, {Lai}, \& {Pu}}]{ande20}
---. 2020, \mnras, 491, 1369, \dodoi{10.1093/mnras/stz3119}

\bibitem[{{Arras} \& {Socrates}(2010)}]{arra10}
{Arras}, P., \& {Socrates}, A. 2010, \apj, 714, 1,
  \dodoi{10.1088/0004-637X/714/1/1}

\bibitem[{{Astropy Collaboration} {et~al.}(2013){Astropy Collaboration},
  {Robitaille}, {Tollerud}, {Greenfield}, {Droettboom}, {Bray}, {Aldcroft},
  {Davis}, {Ginsburg}, {Price-Whelan}, {Kerzendorf}, {Conley}, {Crighton},
  {Barbary}, {Muna}, {Ferguson}, {Grollier}, {Parikh}, {Nair}, {Unther},
  {Deil}, {Woillez}, {Conseil}, {Kramer}, {Turner}, {Singer}, {Fox}, {Weaver},
  {Zabalza}, {Edwards}, {Azalee Bostroem}, {Burke}, {Casey}, {Crawford},
  {Dencheva}, {Ely}, {Jenness}, {Labrie}, {Lim}, {Pierfederici}, {Pontzen},
  {Ptak}, {Refsdal}, {Servillat}, \& {Streicher}}]{exoplanet:astropy13}
{Astropy Collaboration}, {Robitaille}, T.~P., {Tollerud}, E.~J., {et~al.} 2013,
  \aap, 558, A33, \dodoi{10.1051/0004-6361/201322068}

\bibitem[{{Astropy Collaboration} {et~al.}(2018){Astropy Collaboration},
  {Price-Whelan}, {Sip{\H o}cz}, {G{\"u}nther}, {Lim}, {Crawford}, {Conseil},
  {Shupe}, {Craig}, {Dencheva}, {Ginsburg}, {VanderPlas}, {Bradley},
  {P{\'e}rez-Su{\'a}rez}, {de Val-Borro}, {Aldcroft}, {Cruz}, {Robitaille},
  {Tollerud}, {Ardelean}, {Babej}, {Bach}, {Bachetti}, {Bakanov}, {Bamford},
  {Barentsen}, {Barmby}, {Baumbach}, {Berry}, {Biscani}, {Boquien}, {Bostroem},
  {Bouma}, {Brammer}, {Bray}, {Breytenbach}, {Buddelmeijer}, {Burke},
  {Calderone}, {Cano Rodr{\'{\i}}guez}, {Cara}, {Cardoso}, {Cheedella},
  {Copin}, {Corrales}, {Crichton}, {D'Avella}, {Deil}, {Depagne}, {Dietrich},
  {Donath}, {Droettboom}, {Earl}, {Erben}, {Fabbro}, {Ferreira}, {Finethy},
  {Fox}, {Garrison}, {Gibbons}, {Goldstein}, {Gommers}, {Greco}, {Greenfield},
  {Groener}, {Grollier}, {Hagen}, {Hirst}, {Homeier}, {Horton}, {Hosseinzadeh},
  {Hu}, {Hunkeler}, {Ivezi{\'c}}, {Jain}, {Jenness}, {Kanarek}, {Kendrew},
  {Kern}, {Kerzendorf}, {Khvalko}, {King}, {Kirkby}, {Kulkarni}, {Kumar},
  {Lee}, {Lenz}, {Littlefair}, {Ma}, {Macleod}, {Mastropietro}, {McCully},
  {Montagnac}, {Morris}, {Mueller}, {Mumford}, {Muna}, {Murphy}, {Nelson},
  {Nguyen}, {Ninan}, {N{\"o}the}, {Ogaz}, {Oh}, {Parejko}, {Parley}, {Pascual},
  {Patil}, {Patil}, {Plunkett}, {Prochaska}, {Rastogi}, {Reddy Janga},
  {Sabater}, {Sakurikar}, {Seifert}, {Sherbert}, {Sherwood-Taylor}, {Shih},
  {Sick}, {Silbiger}, {Singanamalla}, {Singer}, {Sladen}, {Sooley},
  {Sornarajah}, {Streicher}, {Teuben}, {Thomas}, {Tremblay}, {Turner},
  {Terr{\'o}n}, {van Kerkwijk}, {de la Vega}, {Watkins}, {Weaver}, {Whitmore},
  {Woillez}, {Zabalza}, \& {Astropy Contributors}}]{exoplanet:astropy18}
{Astropy Collaboration}, {Price-Whelan}, A.~M., {Sip{\H o}cz}, B.~M., {et~al.}
  2018, \aj, 156, 123, \dodoi{10.3847/1538-3881/aabc4f}

\bibitem[{{Baruteau} {et~al.}(2014){Baruteau}, {Crida}, {Paardekooper},
  {Masset}, {Guilet}, {Bitsch}, {Nelson}, {Kley}, \& {Papaloizou}}]{baru14}
{Baruteau}, C., {Crida}, A., {Paardekooper}, S.~J., {et~al.} 2014, in
  Protostars and Planets VI, ed. H.~{Beuther}, R.~S. {Klessen}, C.~P.
  {Dullemond}, \& T.~{Henning}, 667,
  \dodoi{10.2458/azu_uapress_9780816531240-ch029}

\bibitem[{{Batygin} {et~al.}(2016){Batygin}, {Bodenheimer}, \&
  {Laughlin}}]{baty16}
{Batygin}, K., {Bodenheimer}, P.~H., \& {Laughlin}, G.~P. 2016, \apj, 829, 114,
  \dodoi{10.3847/0004-637X/829/2/114}

\bibitem[{{Betancourt}(2017)}]{beta17}
{Betancourt}, M. 2017, arXiv e-prints, arXiv:1701.02434.
\newblock \doarXiv{1701.02434}

\bibitem[{{Bodenheimer} {et~al.}(2001){Bodenheimer}, {Lin}, \&
  {Mardling}}]{bode01}
{Bodenheimer}, P., {Lin}, D.~N.~C., \& {Mardling}, R.~A. 2001, \apj, 548, 466,
  \dodoi{10.1086/318667}

\bibitem[{{Boley} {et~al.}(2016){Boley}, {Granados Contreras}, \&
  {Gladman}}]{bole16}
{Boley}, A.~C., {Granados Contreras}, A.~P., \& {Gladman}, B. 2016, \apjl, 817,
  L17, \dodoi{10.3847/2041-8205/817/2/L17}

\bibitem[{{Bouma} {et~al.}(2020){Bouma}, {Hartman}, {Brahm}, {Evans},
  {Collins}, {Zhou}, {Sarkis}, {Quinn}, {de Leon}, {Livingston}, {Bergmann},
  {Stassun}, {Bhatti}, {Winn}, {Bakos}, {Abe}, {Crouzet}, {Dransfield},
  {Guillot}, {Marie-Sainte}, {M{\'e}karnia}, {Triaud}, {Tinney}, {Henning},
  {Espinoza}, {Jord{\'a}n}, {Barbieri}, {Nandakumar}, {Trifonov}, {Vines},
  {Vuckovic}, {Ziegler}, {Law}, {Mann}, {Ricker}, {Vanderspek}, {Seager},
  {Jenkins}, {Burke}, {Dragomir}, {Levine}, {Quintana}, {Rodriguez}, {Smith},
  \& {Wohler}}]{boum20}
{Bouma}, L.~G., {Hartman}, J.~D., {Brahm}, R., {et~al.} 2020, \aj, 160, 239,
  \dodoi{10.3847/1538-3881/abb9ab}

\bibitem[{{Brahm} {et~al.}(2019){Brahm}, {Espinoza}, {Jord{\'a}n}, {Henning},
  {Sarkis}, {Jones}, {D{\'\i}az}, {Jenkins}, {Vanzi}, {Zapata}, {Petrovich},
  {Kossakowski}, {Rabus}, {Rojas}, \& {Torres}}]{brah19}
{Brahm}, R., {Espinoza}, N., {Jord{\'a}n}, A., {et~al.} 2019, \aj, 158, 45,
  \dodoi{10.3847/1538-3881/ab279a}

\bibitem[{{Brahm} {et~al.}(2020){Brahm}, {Nielsen}, {Wittenmyer}, {Wang},
  {Rodriguez}, {Espinoza}, {Jones}, {Jord{\'a}n}, {Henning}, {Hobson},
  {Kossakowski}, {Rojas}, {Sarkis}, {Schlecker}, {Trifonov}, {Shahaf},
  {Ricker}, {Vand erspek}, {Latham}, {Seager}, {Winn}, {Jenkins}, {Addison},
  {Bakos}, {Bhatti}, {Bayliss}, {Berlind}, {Bieryla}, {Bouchy}, {Bowler},
  {Brice{\~n}o}, {Brown}, {Bryant}, {Caldwell}, {Charbonneau}, {Collins},
  {Davis}, {Esquerdo}, {Fulton}, {Guerrero}, {Henze}, {Hogan}, {Horner},
  {Huang}, {Irwin}, {Kane}, {Kielkopf}, {Mann}, {Mazeh}, {McCormac}, {McCully},
  {Mengel}, {Mireles}, {Okumura}, {Plavchan}, {Quinn}, {Rabus}, {Saesen},
  {Schlieder}, {Segransan}, {Shiao}, {Shporer}, {Siverd}, {Stassun}, {Suc},
  {Tan}, {Torres}, {Tinney}, {Udry}, {Vanzi}, {Vezie}, {Vines}, {Vuckovic},
  {Wright}, {Yahalomi}, {Zapata}, {Zhang}, \& {Ziegler}}]{brah20}
{Brahm}, R., {Nielsen}, L.~D., {Wittenmyer}, R.~A., {et~al.} 2020, arXiv
  e-prints, arXiv:2009.08881.
\newblock \doarXiv{2009.08881}

\bibitem[{{Burke}(2008)}]{burk08}
{Burke}, C.~J. 2008, \apj, 679, 1566, \dodoi{10.1086/587798}

\bibitem[{{Carmichael} {et~al.}(2020{\natexlab{a}}){Carmichael}, {Quinn},
  {Mustill}, {Huang}, {Zhou}, {Persson}, {Nielsen}, {Collins}, {Ziegler},
  {Collins}, {Rodriguez}, {Shporer}, {Brahm}, {Mann}, {Bouchy}, {Fridlund},
  {Stassun}, {Hellier}, {Seidel}, {Stalport}, {Udry}, {Pepe}, {Ireland},
  {{\v{Z}}erjal}, {Brice{\~n}o}, {Law}, {Jord{\'a}n}, {Espinoza}, {Henning},
  {Sarkis}, \& {Latham}}]{carm20a}
{Carmichael}, T.~W., {Quinn}, S.~N., {Mustill}, A.~J., {et~al.}
  2020{\natexlab{a}}, \aj, 160, 53, \dodoi{10.3847/1538-3881/ab9b84}

\bibitem[{{Carmichael} {et~al.}(2020{\natexlab{b}}){Carmichael}, {Quinn},
  {Zhou}, {Grieves}, {Bouchy}, {Collins}, {Kielkopf}, {Schwarz}, {Vand erburg},
  {Irwin}, {Charbonneau}, {Ziegler}, {Briceno}, {Law}, {Mann}, {Huang},
  {Shporer}, {Rodriguez}, {Stassun}, \& {Latham}}]{carm20b}
{Carmichael}, T.~W., {Quinn}, S.~N., {Zhou}, G., {et~al.} 2020{\natexlab{b}},
  arXiv e-prints, arXiv:2009.13515.
\newblock \doarXiv{2009.13515}

\bibitem[{{Chen} \& {Kipping}(2017)}]{chen17}
{Chen}, J., \& {Kipping}, D. 2017, \apj, 834, 17,
  \dodoi{10.3847/1538-4357/834/1/17}

\bibitem[{{Collins} {et~al.}(2017){Collins}, {Kielkopf}, {Stassun}, \&
  {Hessman}}]{Collins:2017}
{Collins}, K.~A., {Kielkopf}, J.~F., {Stassun}, K.~G., \& {Hessman}, F.~V.
  2017, \aj, 153, 77, \dodoi{10.3847/1538-3881/153/2/77}

\bibitem[{{Collins} {et~al.}(2018){Collins}, {Collins}, {Pepper},
  {Labadie-Bartz}, {Stassun}, {Gaudi}, {Bayliss}, {Bento}, {COL{\'O}N},
  {Feliz}, {James}, {Johnson}, {Kuhn}, {Lund}, {Penny}, {Rodriguez}, {Siverd},
  {Stevens}, {Yao}, {Zhou}, {Akshay}, {Aldi}, {Ashcraft}, {Awiphan},
  {Ba{\textcommabelow s}t{\"u}rk}, {Baker}, {Beatty}, {Benni}, {Berlind},
  {Berriman}, {Berta-Thompson}, {Bieryla}, {Bozza}, {Calchi Novati}, {Calkins},
  {Cann}, {Ciardi}, {Clark}, {Cochran}, {Cohen}, {Conti}, {Crepp}, {Curtis},
  {D'Ago}, {Diazeguigure}, {Dressing}, {Dubois}, {Ellingson}, {Ellis},
  {Esquerdo}, {Evans}, {Friedli}, {Fukui}, {Fulton}, {Gonzales}, {Good},
  {Gregorio}, {Gumusayak}, {Hancock}, {Harada}, {Hart}, {Hintz},
  {Jang-Condell}, {Jeffery}, {Jensen}, {Jofr{\'e}}, {Joner}, {Kar}, {Kasper},
  {Keten}, {Kielkopf}, {Komonjinda}, {Kotnik}, {Latham}, {Leuquire}, {Lewis},
  {Logie}, {Lowther}, {Macqueen}, {Martin}, {Mawet}, {Mcleod}, {Murawski},
  {Narita}, {Nordhausen}, {Oberst}, {Odden}, {Panka}, {Petrucci}, {Plavchan},
  {Quinn}, {Rau}, {Reed}, {Relles}, {Renaud}, {Scarpetta}, {Sorber}, {Spencer},
  {Spencer}, {Stephens}, {Stockdale}, {Tan}, {Trueblood}, {Trueblood},
  {Vanaverbeke}, {Villanueva}, {Warner}, {West}, {Yal{\c{c}}{\i}nkaya},
  {Yeigh}, \& {Zambelli}}]{coll18}
{Collins}, K.~A., {Collins}, K.~I., {Pepper}, J., {et~al.} 2018, \aj, 156, 234,
  \dodoi{10.3847/1538-3881/aae582}

\bibitem[{{Dawson} \& {Johnson}(2012)}]{daws12}
{Dawson}, R.~I., \& {Johnson}, J.~A. 2012, \apj, 756, 122,
  \dodoi{10.1088/0004-637X/756/2/122}

\bibitem[{{Dawson} \& {Johnson}(2018)}]{daws18}
---. 2018, \araa, 56, 175, \dodoi{10.1146/annurev-astro-081817-051853}

\bibitem[{{Dawson} \& {Murray-Clay}(2013)}]{daws13}
{Dawson}, R.~I., \& {Murray-Clay}, R.~A. 2013, \apjl, 767, L24,
  \dodoi{10.1088/2041-8205/767/2/L24}

\bibitem[{{Dawson} {et~al.}(2015){Dawson}, {Murray-Clay}, \&
  {Johnson}}]{daws15}
{Dawson}, R.~I., {Murray-Clay}, R.~A., \& {Johnson}, J.~A. 2015, \apj, 798, 66,
  \dodoi{10.1088/0004-637X/798/2/66}

\bibitem[{{Dawson} {et~al.}(2019){Dawson}, {Huang}, {Lissauer}, {Collins},
  {Sha}, {Armstrong}, {Conti}, {Collins}, {Evans}, {Gan}, {Horne}, {Ireland},
  {Murgas}, {Myers}, {Relles}, {Sefako}, {Shporer}, {Stockdale},
  {{\v{Z}}erjal}, {Zhou}, {Ricker}, {Vand erspek}, {Latham}, {Seager}, {Winn},
  {Jenkins}, {Bouma}, {Caldwell}, {Daylan}, {Doty}, {Dynes}, {Esquerdo},
  {Rose}, {Smith}, \& {Yu}}]{daws19}
{Dawson}, R.~I., {Huang}, C.~X., {Lissauer}, J.~J., {et~al.} 2019, \aj, 158,
  65, \dodoi{10.3847/1538-3881/ab24ba}

\bibitem[{{Dawson} {et~al.}(2021){Dawson}, {Huang}, {Brahm}, {Collins},
  {Hobson}, {Jord{\'a}n}, {Dong}, {Korth}, {Trifonov}, {Abe}, {Agabi}, {Bruni},
  {Butler}, {Barbieri}, {Collins}, {Conti}, {Crane}, {Crouzet}, {Dransfield},
  {Evans}, {Espinoza}, {Gan}, {Guillot}, {Henning}, {Lissauer}, {Jensen},
  {Sainte}, {M{\'e}karnia}, {Myers}, {Nandakumar}, {Relles}, {Sarkis},
  {Torres}, {Shectman}, {Schmider}, {Shporer}, {Stockdale}, {Teske}, {Triaud},
  {Wang}, {Ziegler}, {Ricker}, {Vanderspek}, {Latham}, {Seager}, {Winn},
  {Jenkins}, {Bouma}, {Burt}, {Charbonneau}, {Levine}, {McDermott}, {McLean},
  {Rose}, {Vanderburg}, \& {Wohler}}]{daws21}
{Dawson}, R.~I., {Huang}, C.~X., {Brahm}, R., {et~al.} 2021, \aj, 161, 161,
  \dodoi{10.3847/1538-3881/abd8d0}

\bibitem[{{Dong} {et~al.}(2014){Dong}, {Katz}, \& {Socrates}}]{dong14}
{Dong}, S., {Katz}, B., \& {Socrates}, A. 2014, \apjl, 781, L5,
  \dodoi{10.1088/2041-8205/781/1/L5}

\bibitem[{{Dotter} {et~al.}(2008){Dotter}, {Chaboyer}, {Jevremovi{\'c}},
  {Kostov}, {Baron}, \& {Ferguson}}]{dott08}
{Dotter}, A., {Chaboyer}, B., {Jevremovi{\'c}}, D., {et~al.} 2008, \apjs, 178,
  89, \dodoi{10.1086/589654}

\bibitem[{{Droettboom} {et~al.}(2016){Droettboom}, {Hunter}, {Caswell},
  {Firing}, {Nielsen}, {Elson}, {Root}, {Dale}, {Lee}, {Sepp{\"a}nen},
  {McDougall}, {Straw}, {May}, {Varoquaux}, {Yu}, {Ma}, {Moad}, {Silvester},
  {Gohlke}, {W{\"u}rtz}, {Hisch}, {Ariza}, {Cimarron}, {Thomas}, {Evans},
  {Ivanov}, {Whitaker}, {Hobson}, {mdehoon}, \& {Giuca}}]{droe16}
{Droettboom}, M., {Hunter}, J., {Caswell}, T.~A., {et~al.} 2016, {Matplotlib:
  Matplotlib V1.5.1}, v1.5.1,  Zenodo, \dodoi{10.5281/zenodo.44579}

\bibitem[{{Duffell} \& {Chiang}(2015)}]{duff14}
{Duffell}, P.~C., \& {Chiang}, E. 2015, \apj, 812, 94,
  \dodoi{10.1088/0004-637X/812/2/94}

\bibitem[{{Eggleton} {et~al.}(1998){Eggleton}, {Kiseleva}, \& {Hut}}]{eggl89}
{Eggleton}, P.~P., {Kiseleva}, L.~G., \& {Hut}, P. 1998, \apj, 499, 853,
  \dodoi{10.1086/305670}

\bibitem[{{Fabrycky} {et~al.}(2014){Fabrycky}, {Lissauer}, {Ragozzine}, {Rowe},
  {Steffen}, {Agol}, {Barclay}, {Batalha}, {Borucki}, {Ciardi}, {Ford},
  {Gautier}, {Geary}, {Holman}, {Jenkins}, {Li}, {Morehead}, {Morris},
  {Shporer}, {Smith}, {Still}, \& {Van Cleve}}]{fabr14}
{Fabrycky}, D.~C., {Lissauer}, J.~J., {Ragozzine}, D., {et~al.} 2014, \apj,
  790, 146, \dodoi{10.1088/0004-637X/790/2/146}

\bibitem[{{Ford} \& {Rasio}(2008)}]{ford08}
{Ford}, E.~B., \& {Rasio}, F.~A. 2008, \apj, 686, 621, \dodoi{10.1086/590926}

\bibitem[{Foreman-Mackey(2016)}]{corner}
Foreman-Mackey, D. 2016, The Journal of Open Source Software, 24,
  \dodoi{10.21105/joss.00024}

\bibitem[{{Foreman-Mackey}(2018)}]{exoplanet:foremanmackey18}
{Foreman-Mackey}, D. 2018, Research Notes of the American Astronomical Society,
  2, 31, \dodoi{10.3847/2515-5172/aaaf6c}

\bibitem[{{Foreman-Mackey} {et~al.}(2017){Foreman-Mackey}, {Agol},
  {Ambikasaran}, \& {Angus}}]{exoplanet:foremanmackey17}
{Foreman-Mackey}, D., {Agol}, E., {Ambikasaran}, S., \& {Angus}, R. 2017, \aj,
  154, 220, \dodoi{10.3847/1538-3881/aa9332}

\bibitem[{Foreman-Mackey {et~al.}(2019)Foreman-Mackey, Czekala, Luger, Agol,
  Barentsen, \& Barclay}]{exoplanet:exoplanet}
Foreman-Mackey, D., Czekala, I., Luger, R., {et~al.} 2019, dfm/exoplanet:
  exoplanet v0.2.1, \dodoi{10.5281/zenodo.3462740}

\bibitem[{{Frelikh} {et~al.}(2019){Frelikh}, {Jang}, {Murray-Clay}, \&
  {Petrovich}}]{frel19}
{Frelikh}, R., {Jang}, H., {Murray-Clay}, R.~A., \& {Petrovich}, C. 2019,
  \apjl, 884, L47, \dodoi{10.3847/2041-8213/ab4a7b}

\bibitem[{{Gaia Collaboration} {et~al.}(2016){Gaia Collaboration}, {Prusti},
  {de Bruijne}, {Brown}, {Vallenari}, {Babusiaux}, {Bailer-Jones}, {Bastian},
  {Biermann}, {Evans}, \& et~al.}]{gaia16}
{Gaia Collaboration}, {Prusti}, T., {de Bruijne}, J.~H.~J., {et~al.} 2016,
  \aap, 595, A1, \dodoi{10.1051/0004-6361/201629272}

\bibitem[{{Gaia Collaboration} {et~al.}(2018){Gaia Collaboration}, {Brown},
  {Vallenari}, {Prusti}, {de Bruijne}, {Babusiaux}, {Bailer-Jones}, {Biermann},
  {Evans}, {Eyer}, \& et~al.}]{gaia18}
{Gaia Collaboration}, {Brown}, A.~G.~A., {Vallenari}, A., {et~al.} 2018, \aap,
  616, A1, \dodoi{10.1051/0004-6361/201833051}

\bibitem[{{Ginsburg} {et~al.}(2019){Ginsburg}, {Sip{\H{o}}cz}, {Brasseur},
  {Cowperthwaite}, {Craig}, {Deil}, {Guillochon}, {Guzman}, {Liedtke}, {Lian
  Lim}, {Lockhart}, {Mommert}, {Morris}, {Norman}, {Parikh}, {Persson},
  {Robitaille}, {Segovia}, {Singer}, {Tollerud}, {de Val-Borro}, {Valtchanov},
  {Woillez}, {Astroquery Collaboration}, \& {a subset of astropy
  Collaboration}}]{astroquery19}
{Ginsburg}, A., {Sip{\H{o}}cz}, B.~M., {Brasseur}, C.~E., {et~al.} 2019, \aj,
  157, 98, \dodoi{10.3847/1538-3881/aafc33}

\bibitem[{{Goldreich} \& {Tremaine}(1980)}]{gold80}
{Goldreich}, P., \& {Tremaine}, S. 1980, \apj, 241, 425, \dodoi{10.1086/158356}

\bibitem[{{Guerrero} {et~al.}(2021){Guerrero}, {Seager}, {Huang}, {Vanderburg},
  {Garcia Soto}, {Mireles}, {Hesse}, {Fong}, {Glidden}, {Shporer}, {Latham},
  {Collins}, {Quinn}, {Burt}, {Dragomir}, {Crossfield}, {Vanderspek},
  {Fausnaugh}, {Burke}, {Ricker}, {Daylan}, {Essack}, {G{\"u}nther}, {Osborn},
  {Pepper}, {Rowden}, {Sha}, {Villanueva}, {Yahalomi}, {Yu}, {Ballard},
  {Batalha}, {Berardo}, {Chontos}, {Dittmann}, {Esquerdo}, {Mikal-Evans},
  {Jayaraman}, {Krishnamurthy}, {Louie}, {Mehrle}, {Niraula}, {Rackham},
  {Rodriguez}, {Rowden}, {Sousa-Silva}, {Watanabe}, {Wong}, {Zhan},
  {Zivanovic}, {Christiansen}, {Ciardi}, {Swain}, {Lund}, {Mullally},
  {Fleming}, {Rodriguez}, {Boyd}, {Quintana}, {Barclay}, {Col{\'o}n},
  {Rinehart}, {Schlieder}, {Clampin}, {Jenkins}, {Twicken}, {Caldwell},
  {Coughlin}, {Henze}, {Lissauer}, {Morris}, {Rose}, {Smith}, {Tenenbaum},
  {Ting}, {Wohler}, {Bakos}, {Bean}, {Berta-Thompson}, {Bieryla}, {Bouma},
  {Buchhave}, {Butler}, {Charbonneau}, {Doty}, {Ge}, {Holman}, {Howard},
  {Kaltenegger}, {Kjeldsen}, {Kreidberg}, {Lin}, {Minsky}, {Narita}, {Paegert},
  {P{\'a}l}, {Palle}, {Sasselov}, {Spencer}, {Sozzetti}, {Stassun}, {Torres},
  {Udry}, \& {Winn}}]{guer21}
{Guerrero}, N.~M., {Seager}, S., {Huang}, C.~X., {et~al.} 2021, arXiv e-prints,
  arXiv:2103.12538.
\newblock \doarXiv{2103.12538}

\bibitem[{{Guillot} \& {Showman}(2002)}]{guil02}
{Guillot}, T., \& {Showman}, A.~P. 2002, \aap, 385, 156,
  \dodoi{10.1051/0004-6361:20011624}

\bibitem[{{Harris} {et~al.}(2020){Harris}, {Jarrod Millman}, {van der Walt},
  {Gommers}, {Virtanen}, {Cournapeau}, {Wieser}, {Taylor}, {Berg}, {Smith},
  {Kern}, {Picus}, {Hoyer}, {van Kerkwijk}, {Brett}, {Haldane}, {Fern{\'a}ndez
  del R{\'\i}o}, {Wiebe}, {Peterson}, {G{\'e}rard-Marchant}, {Sheppard},
  {Reddy}, {Weckesser}, {Abbasi}, {Gohlke}, \& {Oliphant}}]{harr20}
{Harris}, C.~R., {Jarrod Millman}, K., {van der Walt}, S.~J., {et~al.} 2020,
  arXiv e-prints, arXiv:2006.10256.
\newblock \doarXiv{2006.10256}

\bibitem[{{Hartman} \& {Bakos}(2016)}]{hart16}
{Hartman}, J.~D., \& {Bakos}, G.~{\'A}. 2016, Astronomy and Computing, 17, 1,
  \dodoi{10.1016/j.ascom.2016.05.006}

\bibitem[{{He} {et~al.}(2019){He}, {Ford}, \& {Ragozzine}}]{he19}
{He}, M.~Y., {Ford}, E.~B., \& {Ragozzine}, D. 2019, \mnras, 490, 4575,
  \dodoi{10.1093/mnras/stz2869}

\bibitem[{{Hellier} {et~al.}(2017){Hellier}, {Anderson}, {Collier Cameron},
  {Delrez}, {Gillon}, {Jehin}, {Lendl}, {Maxted}, {Neveu-VanMalle}, {Pepe},
  {Pollacco}, {Queloz}, {S{\'e}gransan}, {Smalley}, {Southworth}, {Triaud},
  {Udry}, {Wagg}, \& {West}}]{hell17}
{Hellier}, C., {Anderson}, D.~R., {Collier Cameron}, A., {et~al.} 2017, \mnras,
  465, 3693, \dodoi{10.1093/mnras/stw3005}

\bibitem[{{Hellier} {et~al.}(2019){Hellier}, {Anderson}, {Bouchy}, {Burdanov},
  {Collier Cameron}, {Delrez}, {Gillon}, {Jehin}, {Lendl}, {Nielsen}, {Maxted},
  {Pepe}, {Pollacco}, {Queloz}, {S{\'e}gransan}, {Smalley}, {Triaud}, {Udry},
  \& {West}}]{hell19}
{Hellier}, C., {Anderson}, D.~R., {Bouchy}, F., {et~al.} 2019, \mnras, 482,
  1379, \dodoi{10.1093/mnras/sty2741}

\bibitem[{{Hobson} {et~al.}(2021){Hobson}, {Brahm}, {Jord{\'a}n}, {Espinoza},
  {Kossakowski}, {Henning}, {Rojas}, {Schlecker}, {Sarkis}, {Trifonov},
  {Thorngren}, {Binnenfeld}, {Shahaf}, {Zucker}, {Ricker}, {Latham}, {Seager},
  {Winn}, {Jenkins}, {Addison}, {Bouchy}, {Bowler}, {Briegal}, {Bryant},
  {Collins}, {Daylan}, {Grieves}, {Horner}, {Huang}, {Kane}, {Kielkopf},
  {McLean}, {Mengel}, {Nielsen}, {Okumura}, {Plavchan}, {Shporer}, {Smith},
  {Tilbrook}, {Tinney}, {Twicken}, {Udry}, {Unger}, {West}, {Wittenmyer},
  {Wohler}, \& {Wright}}]{hobs21}
{Hobson}, M.~J., {Brahm}, R., {Jord{\'a}n}, A., {et~al.} 2021, arXiv e-prints,
  arXiv:2103.02685.
\newblock \doarXiv{2103.02685}

\bibitem[{{Hoffman} \& {Gelman}(2011)}]{hoff11}
{Hoffman}, M.~D., \& {Gelman}, A. 2011, arXiv e-prints, arXiv:1111.4246.
\newblock \doarXiv{1111.4246}

\bibitem[{{Huang} {et~al.}(2016){Huang}, {Wu}, \& {Triaud}}]{huan16}
{Huang}, C., {Wu}, Y., \& {Triaud}, A. H.~M.~J. 2016, \apj, 825, 98,
  \dodoi{10.3847/0004-637X/825/2/98}

\bibitem[{{Huang} {et~al.}(2020{\natexlab{a}}){Huang}, {Quinn}, {Vanderburg},
  {Becker}, {Rodriguez}, {Pozuelos}, {Gandolfi}, {Zhou}, {Mann}, {Collins},
  {Crossfield}, {Barkaoui}, {Collins}, {Fridlund}, {Gillon}, {Gonzales},
  {G{\"u}nther}, {Henry}, {Howell}, {James}, {Jao}, {Jehin}, {Jensen}, {Kane},
  {Lissauer}, {Matthews}, {Matson}, {Paredes}, {Schlieder}, {Stassun},
  {Shporer}, {Sha}, {Tan}, {Georgieva}, {Mathur}, {Palle}, {Persson}, {Eylen},
  {Ricker}, {Vanderspek}, {Latham}, {Winn}, {Seager}, {Jenkins}, {Burke},
  {Goeke}, {Rinehart}, {Rose}, {Ting}, {Torres}, \& {Wong}}]{huan20}
{Huang}, C.~X., {Quinn}, S.~N., {Vanderburg}, A., {et~al.} 2020{\natexlab{a}},
  \apjl, 892, L7, \dodoi{10.3847/2041-8213/ab7302}

\bibitem[{{Huang} {et~al.}(2020{\natexlab{b}}){Huang}, {Vanderburg}, {P{\'a}l},
  {Sha}, {Yu}, {Fong}, {Fausnaugh}, {Shporer}, {Guerrero}, {Vanderspek}, \&
  {Ricker}}]{huan20b}
{Huang}, C.~X., {Vanderburg}, A., {P{\'a}l}, A., {et~al.} 2020{\natexlab{b}},
  Research Notes of the American Astronomical Society, 4, 204,
  \dodoi{10.3847/2515-5172/abca2e}

\bibitem[{{Huang} {et~al.}(2020{\natexlab{c}}){Huang}, {Vanderburg}, {P{\'a}l},
  {Sha}, {Yu}, {Fong}, {Fausnaugh}, {Shporer}, {Guerrero}, {Vanderspek}, \&
  {Ricker}}]{huan20c}
---. 2020{\natexlab{c}}, Research Notes of the American Astronomical Society,
  4, 206, \dodoi{10.3847/2515-5172/abca2d}

\bibitem[{{Huang} {et~al.}(2019){Huang}, {Burt}, {Vanderburg}, {Gunther},
  {Shporer}, {Dittmann}, \& {Winn}}]{huan19}
{Huang}, X., {Burt}, J., {Vanderburg}, A., {et~al.} 2019, in American
  Astronomical Society Meeting Abstracts, Vol. 233, American Astronomical
  Society Meeting Abstracts \#233, 209.08

\bibitem[{{Hunter}(2007)}]{hunt07}
{Hunter}, J.~D. 2007, Computing in Science and Engineering, 9, 90,
  \dodoi{10.1109/MCSE.2007.55}

\bibitem[{{Ida} \& {Lin}(2008)}]{ida08}
{Ida}, S., \& {Lin}, D.~N.~C. 2008, \apj, 673, 487, \dodoi{10.1086/523754}

\bibitem[{{Ida} \& {Makino}(1992)}]{ida92}
{Ida}, S., \& {Makino}, J. 1992, \icarus, 96, 107,
  \dodoi{10.1016/0019-1035(92)90008-U}

\bibitem[{{Jackson} {et~al.}(2019){Jackson}, {Dawson}, \& {Zalesky}}]{jack19}
{Jackson}, J.~M., {Dawson}, R.~I., \& {Zalesky}, J. 2019, \aj, 157, 166,
  \dodoi{10.3847/1538-3881/ab09eb}

\bibitem[{{Jehin} {et~al.}(2011){Jehin}, {Gillon}, {Queloz}, {Magain},
  {Manfroid}, {Chantry}, {Lendl}, {Hutsem{\'e}kers}, \& {Udry}}]{trappsit11}
{Jehin}, E., {Gillon}, M., {Queloz}, D., {et~al.} 2011, The Messenger, 145, 2

\bibitem[{{Jenkins} {et~al.}(2002){Jenkins}, {Caldwell}, \& {Borucki}}]{jenk02}
{Jenkins}, J.~M., {Caldwell}, D.~A., \& {Borucki}, W.~J. 2002, \apj, 564, 495,
  \dodoi{10.1086/324143}

\bibitem[{{Jenkins} {et~al.}(2010){Jenkins}, {Caldwell}, {Chandrasekaran},
  {Twicken}, {Bryson}, {Quintana}, {Clarke}, {Li}, {Allen}, {Tenenbaum}, {Wu},
  {Klaus}, {Middour}, {Cote}, {McCauliff}, {Girouard}, {Gunter}, {Wohler},
  {Sommers}, {Hall}, {Uddin}, {Wu}, {Bhavsar}, {Van Cleve}, {Pletcher},
  {Dotson}, {Haas}, {Gilliland}, {Koch}, \& {Borucki}}]{jenk10}
{Jenkins}, J.~M., {Caldwell}, D.~A., {Chandrasekaran}, H., {et~al.} 2010,
  \apjl, 713, L87, \dodoi{10.1088/2041-8205/713/2/L87}

\bibitem[{{Jensen}(2013)}]{Jensen:2013}
{Jensen}, E. 2013, {Tapir: A web interface for transit/eclipse observability},
  Astrophysics Source Code Library.
\newblock \doeprint{1306.007}

\bibitem[{{Jones} {et~al.}(2003){Jones}, {Butler}, {Tinney}, {Marcy}, {Penny},
  {McCarthy}, \& {Carter}}]{jone03}
{Jones}, H.~R.~A., {Butler}, R.~P., {Tinney}, C.~G., {et~al.} 2003, \mnras,
  341, 948, \dodoi{10.1046/j.1365-8711.2003.06481.x}

\bibitem[{{Jord{\'a}n} {et~al.}(2020){Jord{\'a}n}, {Brahm}, {Espinoza},
  {Henning}, {Jones}, {Kossakowski}, {Sarkis}, {Trifonov}, {Rojas}, {Torres},
  {Drass}, {Nandakumar}, {Barbieri}, {Davis}, {Wang}, {Bayliss}, {Bouma},
  {Dragomir}, {Eastman}, {Daylan}, {Guerrero}, {Barclay}, {Ting}, {Henze},
  {Ricker}, {Vanderspek}, {Latham}, {Seager}, {Winn}, {Jenkins}, {Wittenmyer},
  {Bowler}, {Crossfield}, {Horner}, {Kane}, {Kielkopf}, {Morton}, {Plavchan},
  {Tinney}, {Addison}, {Mengel}, {Okumura}, {Shahaf}, {Mazeh}, {Rabus},
  {Shporer}, {Ziegler}, {Mann}, \& {Hart}}]{jord20}
{Jord{\'a}n}, A., {Brahm}, R., {Espinoza}, N., {et~al.} 2020, \aj, 159, 145,
  \dodoi{10.3847/1538-3881/ab6f67}

\bibitem[{{Kane} {et~al.}(2012){Kane}, {Ciardi}, {Gelino}, \& {von
  Braun}}]{kane12}
{Kane}, S.~R., {Ciardi}, D.~R., {Gelino}, D.~M., \& {von Braun}, K. 2012,
  \mnras, 425, 757, \dodoi{10.1111/j.1365-2966.2012.21627.x}

\bibitem[{{Kane} \& {Raymond}(2014)}]{kane14}
{Kane}, S.~R., \& {Raymond}, S.~N. 2014, \apj, 784, 104,
  \dodoi{10.1088/0004-637X/784/2/104}

\bibitem[{{Kempton} {et~al.}(2018){Kempton}, {Bean}, {Louie}, {Deming}, {Koll},
  {Mansfield}, {Christiansen}, {L{\'o}pez-Morales}, {Swain}, {Zellem},
  {Ballard}, {Barclay}, {Barstow}, {Batalha}, {Beatty}, {Berta-Thompson},
  {Birkby}, {Buchhave}, {Charbonneau}, {Cowan}, {Crossfield}, {de Val-Borro},
  {Doyon}, {Dragomir}, {Gaidos}, {Heng}, {Hu}, {Kane}, {Kreidberg}, {Mallonn},
  {Morley}, {Narita}, {Nascimbeni}, {Pall{\'e}}, {Quintana}, {Rauscher},
  {Seager}, {Shkolnik}, {Sing}, {Sozzetti}, {Stassun}, {Valenti}, \& {von
  Essen}}]{kemp18}
{Kempton}, E. M.~R., {Bean}, J.~L., {Louie}, D.~R., {et~al.} 2018, \pasp, 130,
  114401, \dodoi{10.1088/1538-3873/aadf6f}

\bibitem[{{Kipping} {et~al.}(2019){Kipping}, {Nesvorn{\'y}}, {Hartman},
  {Torres}, {Bakos}, {Jansen}, \& {Teachey}}]{kipp19}
{Kipping}, D., {Nesvorn{\'y}}, D., {Hartman}, J., {et~al.} 2019, \mnras, 486,
  4980, \dodoi{10.1093/mnras/stz1141}

\bibitem[{{Kipping}(2013)}]{exoplanet:kipping13}
{Kipping}, D.~M. 2013, \mnras, 435, 2152, \dodoi{10.1093/mnras/stt1435}

\bibitem[{{Kipping}(2014{\natexlab{a}})}]{kipp14a}
---. 2014{\natexlab{a}}, \mnras, 440, 2164, \dodoi{10.1093/mnras/stu318}

\bibitem[{{Kipping}(2014{\natexlab{b}})}]{kipp14b}
---. 2014{\natexlab{b}}, \mnras, 444, 2263, \dodoi{10.1093/mnras/stu1561}

\bibitem[{{Kipping} {et~al.}(2012){Kipping}, {Dunn}, {Jasinski}, \&
  {Manthri}}]{kipp12}
{Kipping}, D.~M., {Dunn}, W.~R., {Jasinski}, J.~M., \& {Manthri}, V.~P. 2012,
  \mnras, 421, 1166, \dodoi{10.1111/j.1365-2966.2011.20376.x}

\bibitem[{Kluyver {et~al.}(2016)Kluyver, Ragan-Kelley, P{\'e}rez, Granger,
  Bussonnier, Frederic, Kelley, Hamrick, Grout, Corlay, Ivanov, Avila, Abdalla,
  Willing, \& development team}]{kluy16}
Kluyver, T., Ragan-Kelley, B., P{\'e}rez, F., {et~al.} 2016, in Positioning and
  Power in Academic Publishing: Players, Agents and Agendas, ed. F.~Loizides \&
  B.~Scmidt (Netherlands: IOS Press), 87--90.
\newblock \url{https://eprints.soton.ac.uk/403913/}

\bibitem[{Kumar {et~al.}(2019)Kumar, Carroll, Hartikainen, \&
  Martin}]{Kumar2019}
Kumar, R., Carroll, C., Hartikainen, A., \& Martin, O. 2019, Journal of Open
  Source Software, 4, 1143, \dodoi{10.21105/joss.01143}

\bibitem[{{Lee} \& {Chiang}(2016)}]{lee16}
{Lee}, E.~J., \& {Chiang}, E. 2016, \apj, 817, 90,
  \dodoi{10.3847/0004-637X/817/2/90}

\bibitem[{{Lee} {et~al.}(2014){Lee}, {Chiang}, \& {Ormel}}]{lee14}
{Lee}, E.~J., {Chiang}, E., \& {Ormel}, C.~W. 2014, \apj, 797, 95,
  \dodoi{10.1088/0004-637X/797/2/95}

\bibitem[{{Lendl} {et~al.}(2014){Lendl}, {Triaud}, {Anderson}, {Collier
  Cameron}, {Delrez}, {Doyle}, {Gillon}, {Hellier}, {Jehin}, {Maxted},
  {Neveu-VanMalle}, {Pepe}, {Pollacco}, {Queloz}, {S{\'e}gransan}, {Smalley},
  {Smith}, {Udry}, {Van Grootel}, \& {West}}]{lend14}
{Lendl}, M., {Triaud}, A.~H.~M.~J., {Anderson}, D.~R., {et~al.} 2014, \aap,
  568, A81, \dodoi{10.1051/0004-6361/201424481}

\bibitem[{{Li} \& {Winn}(2016)}]{li16}
{Li}, G., \& {Winn}, J.~N. 2016, \apj, 818, 5,
  \dodoi{10.3847/0004-637X/818/1/5}

\bibitem[{{Lightkurve Collaboration} {et~al.}(2018){Lightkurve Collaboration},
  {Cardoso}, {Hedges}, {Gully-Santiago}, {Saunders}, {Cody}, {Barclay}, {Hall},
  {Sagear}, {Turtelboom}, {Zhang}, {Tzanidakis}, {Mighell}, {Coughlin}, {Bell},
  {Berta-Thompson}, {Williams}, {Dotson}, \& {Barentsen}}]{lightkurve18}
{Lightkurve Collaboration}, {Cardoso}, J.~V.~d.~M., {Hedges}, C., {et~al.}
  2018, {Lightkurve: Kepler and TESS time series analysis in Python},
  Astrophysics Source Code Library.
\newblock \doeprint{1812.013}

\bibitem[{{Lin} {et~al.}(1996){Lin}, {Bodenheimer}, \& {Richardson}}]{lin96}
{Lin}, D.~N.~C., {Bodenheimer}, P., \& {Richardson}, D.~C. 1996, \nat, 380,
  606, \dodoi{10.1038/380606a0}

\bibitem[{{Lin} \& {Papaloizou}(1986)}]{lin86}
{Lin}, D.~N.~C., \& {Papaloizou}, J. 1986, \apj, 309, 846,
  \dodoi{10.1086/164653}

\bibitem[{{Luger} {et~al.}(2019){Luger}, {Agol}, {Foreman-Mackey}, {Fleming},
  {Lustig-Yaeger}, \& {Deitrick}}]{exoplanet:luger18}
{Luger}, R., {Agol}, E., {Foreman-Mackey}, D., {et~al.} 2019, \aj, 157, 64,
  \dodoi{10.3847/1538-3881/aae8e5}

\bibitem[{{Mandel} \& {Agol}(2002)}]{mand02}
{Mandel}, K., \& {Agol}, E. 2002, \apjl, 580, L171, \dodoi{10.1086/345520}

\bibitem[{{Mireles} {et~al.}(2020){Mireles}, {Shporer}, {Grieves}, {Zhou},
  {G{\"u}nther}, {Brahm}, {Ziegler}, {Stassun}, {Huang}, {Nielsen}, {dos
  Santos}, {Udry}, {Bouchy}, {Ireland}, {Wallace}, {Sarkis}, {Henning},
  {Jord{\'a}n}, {Law}, {Mann}, {Paredes}, {James}, {Jao}, {Henry}, {Butler},
  {Rodriguez}, {Yu}, {Flowers}, {Ricker}, {Latham}, {Vanderspek}, {Seager},
  {Winn}, {Jenkins}, {Furesz}, {Hesse}, {Quintana}, {Rose}, {Smith},
  {Tenenbaum}, {Vezie}, {Yahalomi}, \& {Zhan}}]{mire20}
{Mireles}, I., {Shporer}, A., {Grieves}, N., {et~al.} 2020, \aj, 160, 133,
  \dodoi{10.3847/1538-3881/aba526}

\bibitem[{{Montalto} {et~al.}(2020){Montalto}, {Borsato}, {Granata},
  {Lacedelli}, {Malavolta}, {Manthopoulou}, {Nardiello}, {Nascimbeni}, \&
  {Piotto}}]{mont20}
{Montalto}, M., {Borsato}, L., {Granata}, V., {et~al.} 2020, \mnras, 498, 1726,
  \dodoi{10.1093/mnras/staa2438}

\bibitem[{{Moorhead} {et~al.}(2011){Moorhead}, {Ford}, {Morehead}, {Rowe},
  {Borucki}, {Batalha}, {Bryson}, {Caldwell}, {Fabrycky}, {Gautier}, {Koch},
  {Holman}, {Jenkins}, {Li}, {Lissauer}, {Lucas}, {Marcy}, {Quinn}, {Quintana},
  {Ragozzine}, {Shporer}, {Still}, \& {Torres}}]{moor11}
{Moorhead}, A.~V., {Ford}, E.~B., {Morehead}, R.~C., {et~al.} 2011, \apjs, 197,
  1, \dodoi{10.1088/0067-0049/197/1/1}

\bibitem[{{Naoz} {et~al.}(2012){Naoz}, {Farr}, \& {Rasio}}]{naoz12}
{Naoz}, S., {Farr}, W.~M., \& {Rasio}, F.~A. 2012, \apjl, 754, L36,
  \dodoi{10.1088/2041-8205/754/2/L36}

\bibitem[{{Neal}(2012)}]{neal12}
{Neal}, R.~M. 2012, arXiv e-prints, arXiv:1206.1901.
\newblock \doarXiv{1206.1901}

\bibitem[{{Newton} {et~al.}(2019){Newton}, {Mann}, {Tofflemire}, {Pearce},
  {Rizzuto}, {Vanderburg}, {Martinez}, {Wang}, {Ruffio}, {Kraus}, {Johnson},
  {Thao}, {Wood}, {Rampalli}, {Nielsen}, {Collins}, {Dragomir}, {Hellier},
  {Anderson}, {Barclay}, {Brown}, {Feiden}, {Hart}, {Isopi}, {Kielkopf},
  {Mallia}, {Nelson}, {Rodriguez}, {Stockdale}, {Waite}, {Wright}, {Lissauer},
  {Ricker}, {Vanderspek}, {Latham}, {Seager}, {Winn}, {Jenkins}, {Bouma},
  {Burke}, {Davies}, {Fausnaugh}, {Li}, {Morris}, {Mukai}, {Villase{\~n}or},
  {Villeneuva}, {De Rosa}, {Macintosh}, {Mengel}, {Okumura}, \&
  {Wittenmyer}}]{newt19}
{Newton}, E.~R., {Mann}, A.~W., {Tofflemire}, B.~M., {et~al.} 2019, \apjl, 880,
  L17, \dodoi{10.3847/2041-8213/ab2988}

\bibitem[{{Nielsen} {et~al.}(2019){Nielsen}, {Bouchy}, {Turner}, {Giles},
  {Mascare{\~n}o}, {Lovis}, {Marmier}, {Pepe}, {S{\'e}gransan}, {Udry},
  {Otegi}, {Ottoni}, {Stalport}, {Ricker}, {Vanderspek}, {Latham}, {Seager},
  {Winn}, {Jenkins}, {Kane}, {Wittenmyer}, {Bowler}, {Crossfield}, {Horner},
  {Kielkopf}, {Morton}, {Plavchan}, {Tinney}, {Zhang}, {Wright}, {Mengel},
  {Clark}, {Okumura}, {Addison}, {Caldwell}, {Cartwright}, {Collins},
  {Francis}, {Guerrero}, {Huang}, {Matthews}, {Pepper}, {Rose},
  {Villase{\~n}or}, {Wohler}, {Stassun}, {Howell}, {Ciardi}, {Gonzales},
  {Matson}, {Beichman}, \& {Schlieder}}]{niel19}
{Nielsen}, L.~D., {Bouchy}, F., {Turner}, O., {et~al.} 2019, \aap, 623, A100,
  \dodoi{10.1051/0004-6361/201834577}

\bibitem[{{Petrovich}(2015)}]{petr15}
{Petrovich}, C. 2015, \apj, 805, 75, \dodoi{10.1088/0004-637X/805/1/75}

\bibitem[{{Petrovich} \& {Tremaine}(2016)}]{petr16}
{Petrovich}, C., \& {Tremaine}, S. 2016, \apj, 829, 132,
  \dodoi{10.3847/0004-637X/829/2/132}

\bibitem[{{Petrovich} {et~al.}(2014){Petrovich}, {Tremaine}, \&
  {Rafikov}}]{petr14}
{Petrovich}, C., {Tremaine}, S., \& {Rafikov}, R. 2014, \apj, 786, 101,
  \dodoi{10.1088/0004-637X/786/2/101}

\bibitem[{{Queloz} {et~al.}(2010){Queloz}, {Anderson}, {Collier Cameron},
  {Gillon}, {Hebb}, {Hellier}, {Maxted}, {Pepe}, {Pollacco}, {S{\'e}gransan},
  {Smalley}, {Triaud}, {Udry}, \& {West}}]{quel10}
{Queloz}, D., {Anderson}, D.~R., {Collier Cameron}, A., {et~al.} 2010, \aap,
  517, L1, \dodoi{10.1051/0004-6361/201014768}

\bibitem[{{Rasio} \& {Ford}(1996)}]{rasi96}
{Rasio}, F.~A., \& {Ford}, E.~B. 1996, Science, 274, 954,
  \dodoi{10.1126/science.274.5289.954}

\bibitem[{{Ricker} {et~al.}(2015){Ricker}, {Winn}, {Vanderspek}, {Latham},
  {Bakos}, {Bean}, {Berta-Thompson}, {Brown}, {Buchhave}, {Butler}, {Butler},
  {Chaplin}, {Charbonneau}, {Christensen-Dalsgaard}, {Clampin}, {Deming},
  {Doty}, {De Lee}, {Dressing}, {Dunham}, {Endl}, {Fressin}, {Ge}, {Henning},
  {Holman}, {Howard}, {Ida}, {Jenkins}, {Jernigan}, {Johnson}, {Kaltenegger},
  {Kawai}, {Kjeldsen}, {Laughlin}, {Levine}, {Lin}, {Lissauer}, {MacQueen},
  {Marcy}, {McCullough}, {Morton}, {Narita}, {Paegert}, {Palle}, {Pepe},
  {Pepper}, {Quirrenbach}, {Rinehart}, {Sasselov}, {Sato}, {Seager},
  {Sozzetti}, {Stassun}, {Sullivan}, {Szentgyorgyi}, {Torres}, {Udry}, \&
  {Villasenor}}]{rick15}
{Ricker}, G.~R., {Winn}, J.~N., {Vanderspek}, R., {et~al.} 2015, Journal of
  Astronomical Telescopes, Instruments, and Systems, 1, 014003,
  \dodoi{10.1117/1.JATIS.1.1.014003}

\bibitem[{{Rodriguez} {et~al.}(2019){Rodriguez}, {Quinn}, {Huang},
  {Vanderburg}, {Penev}, {Brahm}, {Jord{\'a}n}, {Ikwut-Ukwa}, {Tsirulik},
  {Latham}, {Stassun}, {Shporer}, {Ziegler}, {Matthews}, {Eastman}, {Gaudi},
  {Collins}, {Guerrero}, {Relles}, {Barclay}, {Batalha}, {Berlind}, {Bieryla},
  {Bouma}, {Boyd}, {Burt}, {Calkins}, {Christiansen}, {Ciardi}, {Col{\'o}n},
  {Conti}, {Crossfield}, {Daylan}, {Dittmann}, {Dragomir}, {Dynes}, {Espinoza},
  {Esquerdo}, {Essack}, {Garcia Soto}, {Glidden}, {G{\"u}nther}, {Henning},
  {Jenkins}, {Kielkopf}, {Krishnamurthy}, {Law}, {Levine}, {Lewin}, {Mann},
  {Morgan}, {Morris}, {Oelkers}, {Paegert}, {Pepper}, {Quintana}, {Ricker},
  {Rowden}, {Seager}, {Sarkis}, {Schlieder}, {Sha}, {Tokovinin}, {Torres},
  {Vand erspek}, {Villanueva}, {Villase{\~n}or}, {Winn}, {Wohler}, {Wong},
  {Yahalomi}, {Yu}, {Zhan}, \& {Zhou}}]{rodr19}
{Rodriguez}, J.~E., {Quinn}, S.~N., {Huang}, C.~X., {et~al.} 2019, \aj, 157,
  191, \dodoi{10.3847/1538-3881/ab11d9}

\bibitem[{{Rodriguez} {et~al.}(2021){Rodriguez}, {Quinn}, {Zhou}, {Vanderburg},
  {Nielsen}, {Wittenmyer}, {Brahm}, {Reed}, {Huang}, {Vach}, {Ciardi},
  {Oelkers}, {Stassun}, {Hellier}, {Gaudi}, {Eastman}, {Collins}, {Bieryla},
  {Christian}, {Latham}, {Wright}, {Matthews}, {Gonzales}, {Ziegler},
  {Dressing}, {Howell}, {Tan}, {Wittrock}, {Plavchan}, {McLeod}, {Baker},
  {Wang}, {Radford}, {Schwarz}, {Esposito}, {Ricker}, {Vanderspek}, {Seager},
  {Winn}, {Jenkins}, {Addison}, {Anderson}, {Barclay}, {Beatty}, {Berlind},
  {Bouchy}, {Bowen}, {Bowler}, {Brasseur}, {Brice{\~n}o}, {Calkins},
  {Chaturvedi}, {Chaverot}, {Chimaladinne}, {Christiansen}, {Collins},
  {Crossfield}, {Eastridge}, {Espinoza}, {Esquerdo}, {Feliz}, {Fenske}, {Fong},
  {Gan}, {Gill}, {Gordon}, {Granados}, {Grieves}, {Guenther}, {Guerrero},
  {Henning}, {Henze}, {Hesse}, {Hobson}, {Horner}, {James}, {Jensen},
  {Jimenez}, {Jord{\'a}n}, {Kane}, {Kielkopf}, {Kim}, {Kuhn}, {Latouf}, {Law},
  {Levine}, {Lund}, {Mann}, {Mao}, {Matson}, {McDermott}, {Mengel}, {Mink},
  {Newman}, {O'Dwyer}, {Okumura}, {Palle}, {Pepper}, {Quintana}, {Sarkis},
  {Savel}, {Schlieder}, {Schnaible}, {Shporer}, {Sefako}, {Seidel}, {Siverd},
  {Skinner}, {Stalport}, {Stevens}, {Stibbard}, {Tinney}, {West}, {Yahalomi},
  \& {Zhang}}]{rodr21}
{Rodriguez}, J.~E., {Quinn}, S.~N., {Zhou}, G., {et~al.} 2021, arXiv e-prints,
  arXiv:2101.01726.
\newblock \doarXiv{2101.01726}

\bibitem[{Salvatier {et~al.}(2016)Salvatier, Wiecki, \&
  Fonnesbeck}]{exoplanet:pymc3}
Salvatier, J., Wiecki, T.~V., \& Fonnesbeck, C. 2016, PeerJ Computer Science,
  2, e55

\bibitem[{{Santerne} {et~al.}(2016){Santerne}, {Moutou}, {Tsantaki}, {Bouchy},
  {H{\'e}brard}, {Adibekyan}, {Almenara}, {Amard}, {Barros}, {Boisse},
  {Bonomo}, {Bruno}, {Courcol}, {Deleuil}, {Demangeon}, {D{\'{\i}}az},
  {Guillot}, {Havel}, {Montagnier}, {Rajpurohit}, {Rey}, \& {Santos}}]{sant16}
{Santerne}, A., {Moutou}, C., {Tsantaki}, M., {et~al.} 2016, \aap, 587, A64,
  \dodoi{10.1051/0004-6361/201527329}

\bibitem[{{Schlecker} {et~al.}(2020){Schlecker}, {Kossakowski}, {Brahm},
  {Espinoza}, {Henning}, {Carone}, {Molaverdikhani}, {Trifonov},
  {Molli{\`e}re}, {Hobson}, {Jord{\'a}n}, {Rojas}, {Klahr}, {Sarkis}, {Bakos},
  {Bhatti}, {Osip}, {Suc}, {Ricker}, {Vanderspek}, {Latham}, {Seager}, {Winn},
  {Jenkins}, {Vezie}, {Villase{\~n}or}, {Rose}, {Rodriguez}, {Rodriguez},
  {Quinn}, \& {Shporer}}]{schl20}
{Schlecker}, M., {Kossakowski}, D., {Brahm}, R., {et~al.} 2020, \aj, 160, 275,
  \dodoi{10.3847/1538-3881/abbe03}

\bibitem[{{Shabram} {et~al.}(2016){Shabram}, {Demory}, {Cisewski}, {Ford}, \&
  {Rogers}}]{shab16}
{Shabram}, M., {Demory}, B.-O., {Cisewski}, J., {Ford}, E.~B., \& {Rogers}, L.
  2016, \apj, 820, 93, \dodoi{10.3847/0004-637X/820/2/93}

\bibitem[{{Shallue} \& {Vanderburg}(2018)}]{shal18}
{Shallue}, C.~J., \& {Vanderburg}, A. 2018, \aj, 155, 94,
  \dodoi{10.3847/1538-3881/aa9e09}

\bibitem[{{Smith} {et~al.}(2014){Smith}, {Anderson}, {Armstrong}, {Barros},
  {Bonomo}, {Bouchy}, {Brown}, {Collier Cameron}, {Delrez}, {Faedi}, {Gillon},
  {G{\'o}mez Maqueo Chew}, {H{\'e}brard}, {Jehin}, {Lendl}, {Louden}, {Maxted},
  {Montagnier}, {Neveu-VanMalle}, {Osborn}, {Pepe}, {Pollacco}, {Queloz},
  {Rostron}, {Segransan}, {Smalley}, {Triaud}, {Turner}, {Udry}, {Walker},
  {West}, \& {Wheatley}}]{smit14}
{Smith}, A.~M.~S., {Anderson}, D.~R., {Armstrong}, D.~J., {et~al.} 2014, \aap,
  570, A64, \dodoi{10.1051/0004-6361/201424752}

\bibitem[{{Socrates} {et~al.}(2012){Socrates}, {Katz}, {Dong}, \&
  {Tremaine}}]{socr12}
{Socrates}, A., {Katz}, B., {Dong}, S., \& {Tremaine}, S. 2012, \apj, 750, 106,
  \dodoi{10.1088/0004-637X/750/2/106}

\bibitem[{{Stassun} \& {Torres}(2018)}]{stas18b}
{Stassun}, K.~G., \& {Torres}, G. 2018, \apj, 862, 61,
  \dodoi{10.3847/1538-4357/aacafc}

\bibitem[{{Stassun} {et~al.}(2018){Stassun}, {Oelkers}, {Pepper}, {Paegert},
  {De Lee}, {Torres}, {Latham}, {Charpinet}, {Dressing}, {Huber}, {Kane},
  {L{\'e}pine}, {Mann}, {Muirhead}, {Rojas-Ayala}, {Silvotti}, {Fleming},
  {Levine}, \& {Plavchan}}]{stas18}
{Stassun}, K.~G., {Oelkers}, R.~J., {Pepper}, J., {et~al.} 2018, \aj, 156, 102,
  \dodoi{10.3847/1538-3881/aad050}

\bibitem[{{Stassun} {et~al.}(2019){Stassun}, {Oelkers}, {Paegert}, {Torres},
  {Pepper}, {De Lee}, {Collins}, {Latham}, {Muirhead}, {Chittidi},
  {Rojas-Ayala}, {Fleming}, {Rose}, {Tenenbaum}, {Ting}, {Kane}, {Barclay},
  {Bean}, {Brassuer}, {Charbonneau}, {Ge}, {Lissauer}, {Mann}, {McLean},
  {Mullally}, {Narita}, {Plavchan}, {Ricker}, {Sasselov}, {Seager}, {Sharma},
  {Shiao}, {Sozzetti}, {Stello}, {Vanderspek}, {Wallace}, \& {Winn}}]{stas19}
{Stassun}, K.~G., {Oelkers}, R.~J., {Paegert}, M., {et~al.} 2019, \aj, 158,
  138, \dodoi{10.3847/1538-3881/ab3467}

\bibitem[{{Theano Development Team}(2016)}]{exoplanet:theano}
{Theano Development Team}. 2016, arXiv e-prints, abs/1605.02688.
\newblock \url{http://arxiv.org/abs/1605.02688}

\bibitem[{{Tsang} {et~al.}(2014){Tsang}, {Turner}, \& {Cumming}}]{tsan14a}
{Tsang}, D., {Turner}, N.~J., \& {Cumming}, A. 2014, \apj, 782, 113,
  \dodoi{10.1088/0004-637X/782/2/113}

\bibitem[{{Udry} {et~al.}(2003){Udry}, {Mayor}, \& {Santos}}]{udry03}
{Udry}, S., {Mayor}, M., \& {Santos}, N.~C. 2003, \aap, 407, 369,
  \dodoi{10.1051/0004-6361:20030843}

\bibitem[{{van der Walt} {et~al.}(2011){van der Walt}, {Colbert}, \&
  {Varoquaux}}]{vand11}
{van der Walt}, S., {Colbert}, S.~C., \& {Varoquaux}, G. 2011, Computing in
  Science and Engineering, 13, 22, \dodoi{10.1109/MCSE.2011.37}

\bibitem[{{Van Eylen} {et~al.}(2019){Van Eylen}, {Albrecht}, {Huang},
  {MacDonald}, {Dawson}, {Cai}, {Foreman-Mackey}, {Lundkvist}, {Silva Aguirre},
  {Snellen}, \& {Winn}}]{vinc19}
{Van Eylen}, V., {Albrecht}, S., {Huang}, X., {et~al.} 2019, \aj, 157, 61,
  \dodoi{10.3847/1538-3881/aaf22f}

\bibitem[{Virtanen {et~al.}(2020)Virtanen, Gommers, Oliphant, Haberland, Reddy,
  Cournapeau, Burovski, Peterson, Weckesser, Bright, {van der Walt}, Brett,
  Wilson, Millman, Mayorov, Nelson, Jones, Kern, Larson, Carey, Polat, Feng,
  Moore, {VanderPlas}, Laxalde, Perktold, Cimrman, Henriksen, Quintero, Harris,
  Archibald, Ribeiro, Pedregosa, {van Mulbregt}, \& {SciPy 1.0
  Contributors}}]{2020SciPy-NMeth}
Virtanen, P., Gommers, R., Oliphant, T.~E., {et~al.} 2020, Nature Methods, 17,
  261, \dodoi{10.1038/s41592-019-0686-2}

\bibitem[{{W}es {M}c{K}inney(2010)}]{mckinney-proc-scipy-2010}
{W}es {M}c{K}inney. 2010, in {P}roceedings of the 9th {P}ython in {S}cience
  {C}onference, ed. {S}t\'efan van~der {W}alt \& {J}arrod {M}illman, 56 -- 61,
  \dodoi{10.25080/Majora-92bf1922-00a}

\bibitem[{{Winn}(2010)}]{winn10}
{Winn}, J.~N. 2010, arXiv e-prints, arXiv:1001.2010.
\newblock \doarXiv{1001.2010}

\bibitem[{{Wittenmyer} {et~al.}(2010){Wittenmyer}, {O'Toole}, {Jones},
  {Tinney}, {Butler}, {Carter}, \& {Bailey}}]{witt10}
{Wittenmyer}, R.~A., {O'Toole}, S.~J., {Jones}, H.~R.~A., {et~al.} 2010, \apj,
  722, 1854, \dodoi{10.1088/0004-637X/722/2/1854}

\bibitem[{{Wu} \& {Lithwick}(2011)}]{wu11}
{Wu}, Y., \& {Lithwick}, Y. 2011, \apj, 735, 109,
  \dodoi{10.1088/0004-637X/735/2/109}

\bibitem[{{Yu} {et~al.}(2019){Yu}, {Vanderburg}, {Huang}, {Shallue},
  {Crossfield}, {Gaudi}, {Daylan}, {Dattilo}, {Armstrong}, {Ricker},
  {Vanderspek}, {Latham}, {Seager}, {Dittmann}, {Doty}, {Glidden}, \&
  {Quinn}}]{yu19}
{Yu}, L., {Vanderburg}, A., {Huang}, C., {et~al.} 2019, \aj, 158, 25,
  \dodoi{10.3847/1538-3881/ab21d6}

\end{thebibliography}

\end{document}